\documentclass{aa}

\usepackage{graphicx}
\usepackage{txfonts}
\usepackage{hyperref}
\usepackage{lscape}
\usepackage{subcaption}

\usepackage{natbib}
\bibpunct{(}{)}{;}{a}{}{,} 

\defcitealias{Perez-Diaz_2025}{Paper I}

\begin{document}

	\title{Chemical enrichment in LINERs from MaNGA}
    \subtitle{II. Characterizing the shape of their radial metallicity gradients}
	\titlerunning{LINERs from MaNGA. II. Metallicity gradient shapes}

	\author{Borja Pérez-Díaz\inst{\ref{inst1}}
		\and
		 José M. Vílchez\inst{\ref{inst1}} \and Enrique Pérez-Montero\inst{\ref{inst1}} \and Igor A. Zinchenko\inst{\ref{inst2a}, \ref{inst2}, \ref{inst3}} \and Brian Tapia-Contreras\inst{\ref{inst4}, \ref{inst5}} \and Patricia B. Tissera\inst{\ref{inst4}, \ref{inst5}}}

	\institute{Instituto de Astrofísica de Andalucía (IAA-CSIC), Glorieta de la Astronomía s/n, 18008 Granada, Spain\label{inst1} \\\email{borja.perezdiaz@inaf.it}\and Astronomisches Rechen-Institut, Zentrum f\"{u}r Astronomie der Universit\"{a}t Heidelberg, M\"{o}nchhofstra{\ss}e 12-14, D-69120 Heidelberg, Germany\label{inst2a}\and Faculty of Physics, Ludwig-Maximilians-Universit\"{a}t, Scheinerstr. 1, 81679 Munich, Germany\label{inst2} \and Main Astronomical Observatory, National Academy of Sciences of Ukraine, 27 Akad. Zabolotnoho St 03680 Kyiv, Ukraine\label{inst3} \and Institute of Astronomy, Pontificia Universidad Católica de Chile, Avenida Vicuña Mackena 4690, Santiago, Chile\label{inst4} \and Centro de Astro-Ingeniería, Pontificia Universidad Católica de Chile, Avenida Vicuña Mackena 4690, Santiago, Chile\label{inst5}}

	\date{Received MONTH DAY, YEAR; accepted MONTH DAY, YEAR}



\abstract
{Chemical abundance radial gradients provide key information on how the processes that affect chemical enrichment of the gas-phase interstellar medium (ISM) act at different galaxy scales. Whereas in the last decades there has been an increase in the number of galaxies studied with integral field spectroscopy, there is still not a clear picture on a subsequent characterization of the chemical abundance radial gradients in galaxies hosting Active Galactic Nuclei (AGNs). This lack of analysis is even more accentuated in the case of low-ionization nuclear emission-line regions (LINERs).}
{For the first time, we analyze the chemical abundance radial gradients in a sample of LINER-like galaxies, whose nuclear emission has been previously (Paper I) discussed.}
{We use a sample of 97 galaxies from the Mapping Nearby Galaxies at Apache Point Observatory (MaNGA), whose nuclear regions show LINER-like emission. We use the open-source code \textsc{HII-CHI-Mistry} to estimate the chemical abundance ratios 12+log(O/H) and log(N/O) in the HII regions across the disks in our sample, as well as in the nuclear parts where the LINER-like activity dominates. To fit the radial profiles we use a piecewise methodology which uses a non-fixed number of breaks to find the best fit for the data.}
{We obtain that majority of our sample of galaxies exhibits departures from the single linear gradient both in 12+log(O/H) and log(N/O) (as expected from the inside-out scenario). We investigate whether these departures are driven by galaxy properties (stellar mass, neutral gas mass, stellar velocity dispersion), finding not correlation at all. We also report that in most cases there is no correlation between the shape of the 12+log(O/H) and log(N/O) radial profiles. We propose a model in which AGN (feed)back, acting at different scales depending on the galaxy and its evolutionary stage, might be responsible for these departures.}
{}

\keywords{Galaxies: ISM --
	Galaxies: abundances --
	Galaxies: active -- Galaxies: nuclei}

\maketitle


\section{Introduction}
The metal content of the gas-phase ISM is a witness of the evolutionary processes that shape host galaxies. As the production of metals from stars that eventually eject them, incorporating them to the ISM, and as galaxy mass assembly is related to star formation, the analysis of the metal content of the gas-phase in galaxies allows us to shed light on their evolution \citep[e.g.][]{Tinsley_1980}. 

Studies analyzing the mean characteristic metallicity of the gas-phase ISM in galaxies have been performed over many decades \citep{Peimbert_1967, Lequeux_1979, Vilchez_1988, Thuan_1995, Hagele_2008, Pilyugin_2016}, using spectroscopic information from different regimes such as ultraviolet \citep{Feltre_2016, Perez-Montero_2017, Perez-Montero_2023}, optical \citep{Perez-Montero_2003, Belfiore_2015, Curti_2017, Curti_2020} and infrared \citep{Fernandez-Ontiveros_2021, Perez-Diaz_2022}. By means of the determination of physical properties of the gas-phase ISM, such as temperature and density \citep[direct method,][]{Osterbrock_book}, photoionization models \citep{Perez-Montero_2014, Thomas_2018} or strong line calibrations based on bright emission lines \citep[see Table 1 from][]{Maiolino_2019}. Although most studies of  the metal content of the gas-phase ISM are focused on star-forming galaxies (SFGs), its study has also recently extended to Active Galactic Nuclei (AGN) \citep[e.g.][]{Perez-Montero_2019, Dors_2019, Carvalho_2020, Perez-Diaz_2021}.

Oxygen, whose relative abundance is usually expressed as 12+log(O/H), is generally used as the main  tracer of the metal content \citep[see][for a review]{Maiolino_2019} as a consequence of its bright emission lines easily detected in almost all spectral ranges (ultraviolet, optical and infrared) and because it is the most abundant element in mass in the gas-phase ISM \citep{Peimbert_2007}. However, oxygen expressed in relative terms to the hydrogen content of the gas-phase ISM is susceptible of changes in the gas composition due to inflows \citep[e.g.][]{Perez-Diaz_2024}, outflows \citep[e.g.][]{Villar-Martin_2024} or depletion into dust \citep[e.g.][]{Calura_2008}. On the other hand, information on other chemical species can be provided to have a better understanding on the ISM chemical enrichment history. For instance, the nitrogen-to-oxygen ratio (log(N/O)), involves an \ensuremath{\alpha}-element (O), mainly produced by massive stars,  with nitrogen, which has an extra channel of production by means of CNO cycles inside stars of intermediate mass \citep[e.g.][]{Henry_2000}. Thus, under the requirement of O already present in the ISM from which stars were born, the simultaneous study of O/H and N/O allows us to quantify the several effects leading to a dilution of metallicity in the gas-phase of galaxies. 

All these observational studies of metallicity in the gas-phase of galaxies can be complemented with theoretical works based on chemical evolution models \citep{Pagel_1975, Koppen_1999, Spitoni_2019, Sharda_2021}, simulations of individual/interacting galaxies \citep{Montuori_2010, Rupke_2010, Perez_2011} or cosmological simulations \citep{Mosconi_2001, Lia_2002, Kobayashi_2007}, in which the many astrophysical processes that enrich the ISM (supernovae feedback, outflow enrichment, pAGB stars, etc) are incorporated in the sub-grid modeling, and later on, compared to the observational results to constrain and predict the different ways in which galaxies evolve. In this context, the predicted timescales of the different processes that enrich the ISM, later affecting  the overall content of gas, stars and dust, are not  only relevant for the galaxy as a whole, but also at smaller scales \citep{Sharda_2021, Sharda_2024, Tissera_2022}. Therefore, the study of both O/H and N/O in the gas-phase ISM across different regions in the galaxy (including radial metallicity gradients) is also crucial in understanding the chemical evolution of galaxies. 

Radial metallicity gradients have largely benefit in the last decades from to the advent of large surveys acquired with integral field spectroscopy (IFS) such as CALIFA \citep{Sanchez_2013}, MaNGA \citep{Bundy_2015} or SAMI \citep{Poetrodjojo_2018} for low-redshift galaxies (z<0.2). Several studies \citep[e.g.][]{Vila-Costas_1992, Rich_2012, Sanchez_2014, Sanchez-Menguiano_2016, Zinchenko_2019} found evidence that most galaxies present a negative oxygen abundance gradient from the inner to the outer most parts of galaxies with discs.  The analysis of the radial metallicity gradient by means of log(N/O) reveals a similar trend \citep[e.g.][]{Pilyugin_2004, Perez-Montero_2016, Zurita_2021, Zinchenko_2021}. These results, also supported by the analysis of stellar populations within the disc \citep[e.g.][]{Taylor_2005, Munoz-Mateos_2007} as well as the star formation histories \citep[SFHs,][]{Sanchez-Blazquez_2009}, reinforces the scenario in which gas accretion moves from outer to inner parts in the galaxy, reaching higher density and triggering star formation in the innermost parts, leading to its faster enrichment \citep{Matteucci_1989}. This is the so-called inside-out growth of galaxies.  

However, the above scenario is challenged by other set of observations. First of all, some high-redshift galaxies also exhibit positive or flattened radial gradients \citep{Cresci_2010, Carton_2018}, attributed in many cases to the infall of metal-poor gas \citep[e.g.][]{Bresolin_2012}. Bars are thought to be an efficient mechanism in gas migration within galaxies \citep{Athanassoula_1992, Friedli_1994}, but no consensus is found if either bars affect radial metallicity gradients flattening them \citep{Vila-Costas_1992, Zaristky_1994} or remain unchanged \citep{Sanchez_2014, Zinchenko_2019, Zurita_2021b}. Likewise, contrary results are reported on whether radial gradients change with morphological type \citep{Vila-Costas_1992} or they are independent \citep{Zaristky_1994}. Many other effects add complexity to the picture: galactic fountains that causes gas being ejected from the disc due to supernovas and capture again due to the gravitational potential \citep{Spitoni_2013}; interaction and merger between galaxies \citep{Rupke_2010, Sillero_2017}; stellar mass biasing gas accretion \citep{Spitoni_2021, Camps_2023}; stellar age differentiation \citep{Zinchenko_2021}; strong-metal rich galactic outflows \citep[e.g.][]{Tissera_2022}; azimuthal variations \citep{Vogt_2017, Spitoni_2019}; or spiral arms \citep{Spitoni_2021}. Moreover, it has been reported that the idea of a single slope radial metallicity gradient might not be accurate in many cases \citep[e.g.][]{Sanchez-Menguiano_2018, Tapia-Contreras_2025}. To asses the impact of some of these processes, the analysis of log(N/O) has been extensively used \citep[e.g.][]{Perez-Montero_2016, Zurita_2021, Zinchenko_2021}, finding similar results to those reported for the 12+log(O/H) radial gradient.

Another important aspect that needs to be considered in the picture of chemical enrichment within galaxies, is the effect of a possible  presence of Active Galactic Nuclei (AGNs). As the realm of chemical studies of the Narrow Line Region (NLR) in AGNs is scarce, much less studies can be found on the radial metallicity gradient in galaxies with a significant fraction of AGN activity.
\citet{Taylor_2017}, using cosmological simulations accounting for AGN feedback, they found that galaxies with strong AGN activity cannot recover from previous merger events and their metallicity radial profiles tend to flatten. However, it has been reported observationally that metallicity radial gradients might a decrease in the inner parts. Indeed, \citet{Nascimento_2022} reported from the analysis of radial chemical abundance gradient in Seyfert galaxies that nuclear abundances are even lower than the expected extrapolation, supporting this scenario. Other studies, such us those analyzing 12+log(O/H) and log(N/O) abundances in the nuclear region of AGN-hosting galaxies \citep{Perez-Diaz_2021, Perez-Diaz_2022, Oliveira_2024} support this scenario as log(N/O) mostly remains solar or suprasolar (an advanced chemical enrichment history), whereas 12+log(O/H) spread over a wide range of metallicities, from subsolar to solar, as a possible consequence of infalls of gas. More recently, \citet{Amiri_2024} found not only that AGN nuclear abundances are lower than the extrapolated value, but also that there is an inverse radial metallicity gradient within the regions dominated by AGN activity.

However, the realm of low-luminosity AGNs (LLAGNs), a tag under which LINERs can be accounted for, is poorly explored. Indeed, only the galaxy UGC 4805, with IFS data from MaNGA has been analyzed \citep{Krabbe_2021}, reporting that the nuclear abundances are in consonance with the extrapolation from the radial metallicity gradient. The regime of LLAGN is key in order to understand the differences in evolution between AGNs and SFGs, as the AGN emission is much weaker and, thus, its effects can be constrained to the nuclear region, producing lower deviations from the expected radial behavior. Furthermore, LLAGNs represent a predominant fraction of active galaxies in the local Universe \citep[e.g.][]{Ho_2008}, so a proper analysis of the radial metallicity gradients in these objects is necessary to have a clear picture on how AGNs affect the evolution of galaxies. As AGN feed and feedback is intrinsically associated to gas hydrodynamics, a simultaneous analysis of N and O radial profiles is essential to disentangle their effects.

Exploiting the capabilities of IFS data from SDSS IV - MaNGA survey \citep{Bundy_2015, Blanton_2017}, in \citet[][hereinafter Paper I]{Perez-Diaz_2025}, we presented an analysis of  105 galaxies classified as LINERs. By means of \textsc{HCm} \citep{Perez-Montero_2014}, we obtained independent estimations for O/H and N/O chemical abundances in the nuclear region of these galaxies, concluding that their chemical composition is well explained under the AGN ionizing scenario. This second paper is devoted to the analysis of the radial metallicity gradients in the same sample. In section \ref{s4.2}, the selection of the galaxy sample hosting LINER-like emission is explained as well as other different criteria used in this study. In section \ref{s4.3} we discuss the methodology employed to estimate chemical abundances in our sample. In section \ref{s4.4} we present the main results of this study followed up by a discussion in section \ref{s4.5}. In section \ref{s4.6} we summarize our main conclusions.

\section{Sample selection}
\label{s4.2}
\subsection{MaNGA data and emission line measurements}
\label{ss4.2.1}

The Mapping Nearby Galaxies at Apache Point Observatory \citep[MaNGA;][]{Bundy_2015} is part of the Sloan Digital Sky Survey IV \citep[SDSS IV;][]{Blanton_2017}. For this work, we used data release 17 (DR17, \citealt{Abdurro_2022}). We used individual spaxels ensuring that their size was significantly smaller than that of the point spread function (PSF) in the MaNGA datacubes. Overall, the spatial resolution of these cubes has a median full width at half maximum (FWHM) of 2.54 arcsec \citep{Law_2016}. 

We examined the MaNGA spectra following the same procedure as in \citet{Zinchenko_2016, Zinchenko_2021}. Briefly, we used the code STARLIGHT \citep{Cid-Fernandes_2005, Mateus_2006, Asari_2007} to fit the stellar background throughout all spaxels, adapting it for parallel datacube processing. We used simple stellar population (SSP) spectra from \citet{Bruzual_2003} evolutionary synthesis models for stellar fitting, and we subtracted them from the observed spectrum to obtain a pure gas spectrum. After that, we fitted emission lines using our ELF3D code. Each emission line was fitted with a single-Gaussian profile. For each spectrum, we measured the fluxes of the [\ion{O}{II}]\ensuremath{\lambda,\lambda}3726,3729\ensuremath{\AA} (hereinafter [\ion{O}{II}]\ensuremath{\lambda}3727\ensuremath{\AA}), [\ion{Ne}{III}]\ensuremath{\lambda}3868\ensuremath{\AA}, H\ensuremath{_{\beta}}, [\ion{O}{III}]\ensuremath{\lambda}4959\ensuremath{\AA}, [\ion{O}{III}]\ensuremath{\lambda}5007\ensuremath{\AA}, [\ion{N}{II}]\ensuremath{\lambda}6548\ensuremath{\AA}, H\ensuremath{_{\alpha}}, [\ion{N}{II}]\ensuremath{\lambda}6584\ensuremath{\AA}, and [\ion{S}{II}]\ensuremath{\lambda,\lambda}6717,6731\ensuremath{\AA} lines with a signal-to-noise ratio, S/N, \ensuremath{>} 3.

\subsection{Sample of galaxies}
\label{ss4.2.2}
Sample selection is described in more detail in \citetalias{Perez-Diaz_2025}. In short, we selected from the MaNGA survey \citep{Bundy_2015, Blanton_2017} a sample of galaxies whose nuclear emission is dominated by LINER-like emission according to the diagnostic diagrams \citep{Baldwin_1981, Kauffmann_2003, Kewley_2006}. After imposing that galaxies must be classified as LINERs in all diagnostic diagrams, we came up with a sample of 329 galaxies.

From that sample, we imposed the criterion that LINER-like emission comes mainly from the nuclear region (\ensuremath{<} 2kpc), i.e., the emission line ratios from spaxels retrieving emission from the outer regions are mainly classified as HII regions according to the same diagnostic diagrams. By doing this, we ensure to have enough HII regions to properly analyze radial metallicity gradients and compare the HII regions properties with those in central spaxels. After this filter, we came up with a sample of 105 LINER-like galaxies, classified according to the WHAN diagram \citep{Cid-Fernandes_2010, Cid-Fernandes_2011} as weak-AGNs (wAGNs, 57) and retired galaxies (RG, 48). Hereinafter, we will not distinguish between RGs and wAGNs as: i) the results from \citetalias{Perez-Diaz_2025} showed that their ionizing spectra is likely identical; and, ii) the statistical analysis of their radial metallicity gradients showed similar distributions.

In order to avoid systematic errors in the deprojection effects of the highly inclined galaxies in our sample, we imposed a minimum value for the axis ratio b/a > 0.3 from the r-band Sersic profile fit. This criterion excludes 8 objects, leading to a final sample of 97 galaxies, divided as 55 wAGNs (56.7\%) and 42 RGs (43.3\%).

HII regions in each galaxy were selected from the remaining spaxels that verify the diagnostic criteria for the BPT diagrams \citep{Baldwin_1981, Kauffmann_2003, Kewley_2006}. Particularly, we selected only those regions either classified as "SFG" (equivalently HII regions) simultaneously in the three classical BPT diagrams or those regions that fall in the "composite" region \citep{Kewley_2006}.  As the composite region can accommodate HII regions with high log(N/O) ratios, which are expected in the central parts according to the inside-out scenario \citep[e.g.][]{Perez-Montero_2016, Zurita_2021}, we considered them as well in our analysis.

\subsection{Ancillary data}
\label{ss4.2.3}
We retrieved complementary data on the physical properties of the host galaxies from the NASA-Sloan Atlas (NSA) catalog\footnote{\url{https://www.sdss4.org/dr17/manga/manga-target-selection/nsa/}.}. Particularly, we retrieved the stellar mass (M\ensuremath{_*}) as estimated from a K-correction fit to the elliptical Petrosian fluxes assuming an initial mass function (IMF) from \citet{Chabrier_2003} and stellar population models from \citet{Bruzual_2003}, as well as \ensuremath{R_{e}} radii, Sersic 50\% light radius along major axis across the r-band.

Additionally, we retrieved the chemical abundances estimations for the nuclear region of our sample of LINERs as reported in \citetalias{Perez-Diaz_2025}. Since the authors provided the estimations based on different ionizing sources from the HII-CHI-Mistry\footnote{The code is publicly available at \url{http://home.iaa.csic.es/~epm/HII-CHI-mistry.html}.} code \citep{Perez-Montero_2014}, we took four estimations as representative of all the considered scenarios:
AGN characterized by \ensuremath{\alpha_{OX} = -1.6}\footnote{We note that very recently, \citet{Perez-Montero_2025} obtain that LINERs are characterized by slightly higher slopes (\ensuremath{\alpha_{OX} \approx -1.4}). As demonstrated in \citetalias{Perez-Diaz_2025}, results under both assumptions are compatible.}; hot old stellar populations dominated by post-asymptotic giant branch (pAGB) stars and characterized by two different effective temperatures T\ensuremath{_{eff} = 10^{5}} K and T\ensuremath{_{eff} = 1.5\cdot10^{5}} K; and advection-dominated accretion flow (ADAF) model for the AGN. Details on the considered models and how they were implemented can be found in \citetalias{Perez-Diaz_2025}.

\section{Methodology}
\label{s4.3}
In this section we present a detailed explanation on the methodology used to estimate chemical abundances as well as other physical parameters from the nebular emission of HII regions in our sample of galaxies. We also give details on the calculation of the corresponding  radial gradients estimated for those properties.

\subsection{HII-CHI-Mistry}
\label{ss4.3.1}
To ensure consistency with the results reached in \citetalias{Perez-Diaz_2025}, as well as between the estimations of chemical abundances in the nuclear region (dominated by the LINER-like emission) and the rest or regions in the disk of each galaxy in our sample, we used \textsc{HII-CHI-Mistry} v5.5 (hereinafter \textsc{HCm}). \textsc{HCm} uses a grid of photoionization models (to be selected among the available ones or introduced by the user)  with three free parameters, the chemical properties of the gas-phase ISM 12+log(O/H) and log(N/O) as well as the ionization parameter log(U). The code compares the emission-line fluxes predicted by the grid of models with the observed (input) emission line ratios sensitive to those parameters. Firstly, the code estimates log(N/O) which is used to constrain the grid of models in the later iterations. Secondly, the code performs an estimation of 12+log(O/H) and log(U). 

In the case of HII regions, we selected as ionizing source a young stellar cluster with an age of 1 Myr, as taken from \textsc{POPSTAR} \citep{Molla_2009} synthesis code for an initial mass function (IMF) that follows the trend reported by \citet{Chabrier_2003}. The density of the gas is assumed to be constant with a value of 100 cm\ensuremath{^{-3}}. This grid of models was calculated using \textsc{Cloudy} v17 \citep{Ferland_2017}. Due to the lack of measurements of the auroral line [\ion{O}{iii}]\ensuremath{\lambda }4363\ensuremath{\AA} in our MaNGA sample, we used an additional  constrain to the grid of models, consisting of  the relation between 12+log(O/H) and log(U) reported by \citet{Perez-Montero_2014}. The estimation of log(N/O) remains completely independent, thus no fixed relation with 12+log(O/H) was assumed in this work.

We used as input for \textsc{HCm} the emission line ratios  [\ion{O}{ii}]\ensuremath{\lambda }3727\ensuremath{\AA}, [\ion{Ne}{iii}]\ensuremath{\lambda }3868\ensuremath{\AA}, [\ion{O}{iii}]\ensuremath{\lambda }5007\ensuremath{\AA}, [\ion{N}{ii}]\ensuremath{\lambda }6584\ensuremath{\AA} and [\ion{S}{ii}]\ensuremath{\lambda }6717\ensuremath{\AA}+[\ion{S}{ii}]\ensuremath{\lambda }6731\ensuremath{\AA}, referred to H\ensuremath{_{\beta}} emission. All emission line ratios have been corrected for reddening assuming Case B photoionization and an expected ratio between H\ensuremath{_{\alpha}} and H\ensuremath{_{\beta}} of 2.86 for standard conditions found in HII regions, that is, an electron density n\ensuremath{_{e} \sim 100} cm\ensuremath{^{-3}} and an electron temperature T\ensuremath{_{e} \sim 10^{4}} K \citep{Osterbrock_book}. We assumed the extinction curve from \citet{Howarth_1983} for R\ensuremath{_{V} = 3.1}.

\subsection{Radial metallicity gradients}
\label{ss4.3.2}
For each galaxy, we used the estimated chemical abundances (12+log(O/H) and log(N/O)) from \textsc{HCm} in the HII regions and their distance to the galaxy center in terms of \ensuremath{R_{e}} to characterize their radial metallicity gradient up to 4\ensuremath{R_{e}}. In order to fit the observed trends, we tested three different methodologies to reproduce most of the scenarios explored in the literature:
\begin{itemize}
	\item Single linear fit with no restrictions in the slope neither on the Intersect.
	\item Double linear fit accounting for a break\footnote{Hereinafter we will refer to the point where the linear fit changes as break.} at exactly \ensuremath{R=R_{e}}, and again with no restrictions on the slopes or intersects.
	\item Piecewise fit allowing for several breaks at no fix positions but ensuring continuity in the estimation through all intervals \citep[see][for more details on the methodology]{Tapia-Contreras_2025}.
\end{itemize}

To test the goodness of the fits, we used the Square Root Error (RSE), which can be applied to any fit:
\begin{equation}
	\label{RSE} \mathrm{RSE} = \frac{\sqrt{\sum_{i}^{n}\left( p_{i} - y_{i} \right) ^{2}}}{n-2}
\end{equation}
here \ensuremath{p} are the values predicted from the fit, \ensuremath{y} are the estimated values and \ensuremath{n} the number of points (i.e. number of HII regions). As piecewise methodology accounts for a non-fixed number of breaks, we also assessed the goodness of the fit accounting for the number of breaks obtained from the piecewise fit.

\begin{figure*}
	\centering
	\includegraphics[width=17cm]{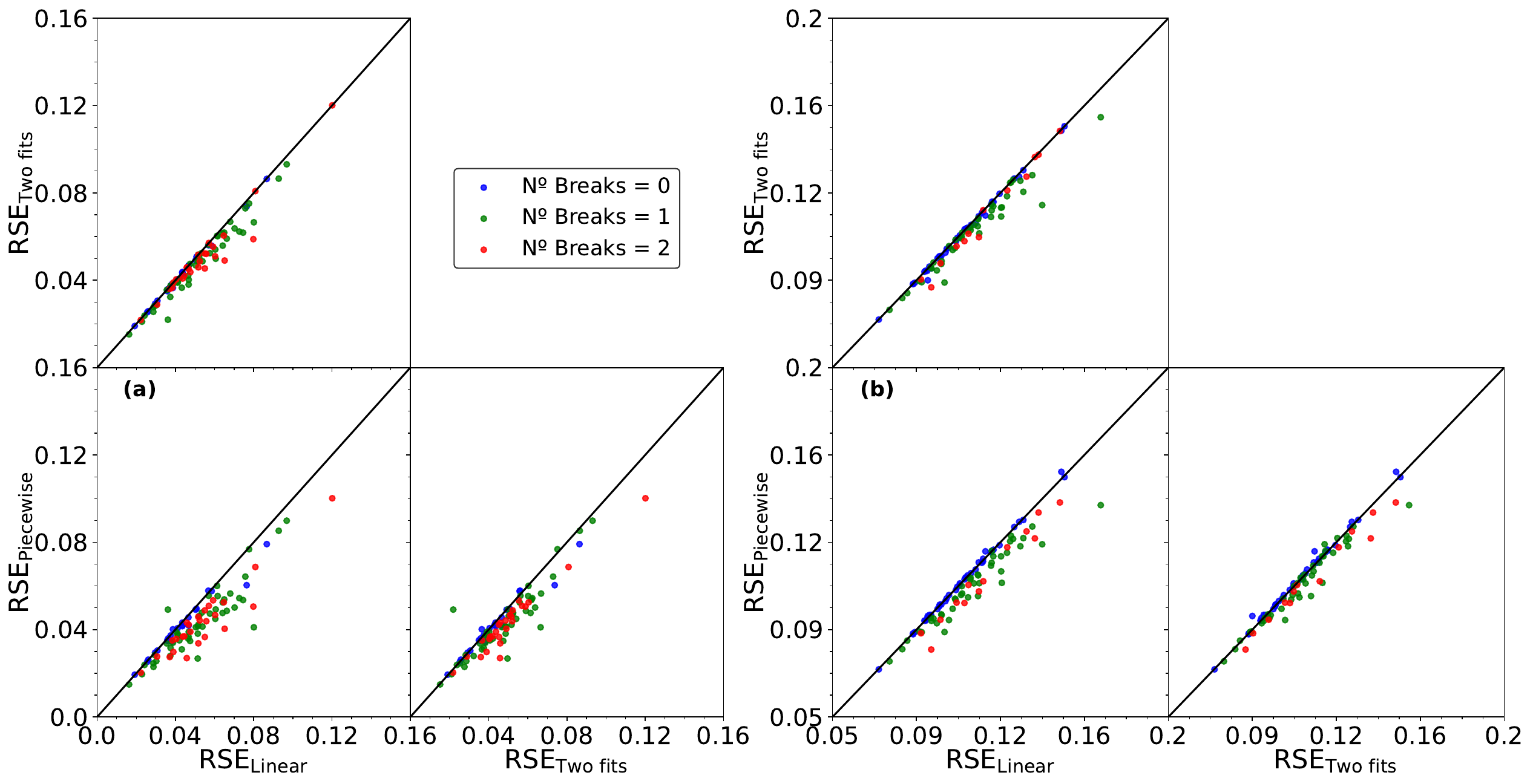}
	\caption[Root square error for different fit techniques of the radial metallicity gradients]{Root Square Error (RSE) for the different techniques used to fit the  radial metallicity gradients of 12+log(O/H) (a) and log(N/O) (b) in our sample of LINERs. The sample is segregated according to the number of breaks used by piecewise fit for each chemical abundance ratio.}
	\label{RSE_plot}
\end{figure*}

We show in Fig. \ref{RSE_plot} (a) the goodness of the different fit techniques for the 12+log(O/H) radial gradient. We conclude that single linear fit offers the highest RSE, with exception of those galaxies where piecewise methodology finds a fit without any break. When comparing  piecewise methodology with double linear fit, we can see that the majority of the fits offer a lower RSE for the former one. A similar result is obtained when analyzing the fit for the radial gradient of log(N/O) (see Fig. \ref{RSE_plot} (b)). 

We also explored the distribution of the scatter subtracting the values from the obtained fits to the individual abundances in each galaxy. On average, we observe a median scatter within the range 0.01 - 0.03 dex when considering the piecewise methodology, for both1 12+log(O/H) and log(N/O), a similar range to that reported by other authors using linear profiles in other samples \citep[e.g.][]{Sanchez_2015, Zinchenko_2016, Kreckel_2019, Grasha_2022}. If a single linear fit is considered, then the median scatter increases up to 0.1 dex. Thus, we only consider thereafter for our analysis the results obtained from piecewise fit.

\section{Results}
\label{s4.4}
\subsection{Trends in the radial metallicity gradients}
\label{ss4.4.1}
We show in Appendix \ref{A1} the radial metallicity gradient for each galaxy both for 12+log(O/H) and log(N/O). From Fig. \ref{Gradients_1} we can conclude that our sample of LINER-like galaxies shows a variety of trends\footnote{The binned values shown in Fig. \ref{Gradients_1} are drawn as visual guides for the reader, but they are not used in the piecewise fit.}. We summarize these trends for both 12+log(O/H) and log (N/O) radial gradients in Table \ref{OH_fit}.
\subsubsection{Breaks in the fits}
\label{sss4.4.1.1}
Around half of our sample (49.49\%) exhibits a 12+log(O/H) radial gradient with one single break, whereas the rest of galaxies are almost split between none (23.71\%) or two breaks (26.80\%). In the case of log(N/O) radial gradient we found a different scenario: galaxies are almost equally distributed between none or one break (\ensuremath{\sim}43-44\%), and only a small group of galaxies (13.4\%) shows two breaks. These results highlight the importance of using fit profiles more complex than a simple linear fit for radial metallicity gradients in galaxies, specially for 12+log(O/H), as highlighted by the higher residuals obtained when assuming a simple linear fit.

\begin{table*}
	\caption[Radial metallicity gradients statistics in the LINER sample from MaNGA]{Statistics of the radial metallicity gradients in our sample as obtained using piecewise method. Column (1) and (2) show the number of breaks needed to fit 12+log(O/H) and log(N/O) radial gradients respectively. Column (3) shows the total number of galaxies for each group. Column (4) shows the relative number of galaxies with respect to the group. Column (5) shows the relative number of galaxies with respect to the whole sample.}
	\label{OH_fit}
	\centering
	\fontsize{9.25pt}{10.00pt}\selectfont 
	\begin{tabular}{c | c | c  c c } 
		\hline\hline
		\textbf{N. Breaks (O/H)} & \textbf{N. Breaks (N/O)} & \textbf{N. Gal} & \textbf{Perc. (\%)} & \textbf{Perc. tot. (\%) } \\ 
		\textbf{(1)} & \textbf{(2)} & \textbf{(3)} & \textbf{(4)} & \textbf{(5)} \\ \hline
		& 0 & 9 & 39.13 & 9.28 \\
		0 & 1 & 12 & 52.17 & 12.37 \\
		& 2 & 2 & 8.70 & 2.07 \\ \hline
		& 0 & 18 & 37.50 & 18.56 \\
		1 & 1 & 22 & 45.83 & 22.68 \\
		& 2 & 8 & 16.67 & 8.25 \\ \hline
		& 0 & 14 & 53.84 & 14.43 \\
		2 & 1 & 9 & 34.62 & 9.28 \\
		& 2 & 3 & 11.54 & 3.10 \\
	\end{tabular}
\end{table*}

When comparing between them the results from 12+log(O/H) and log(N/O) radial metallicity gradients, we observe that there is no correlation between the number of breaks in each one of them in each galaxy. Regardless of the fit obtained for  12+log(O/H), the best fit for log(N/O) radial gradient has either a none or a single break (\ensuremath{>} 80\%). Moreover, for a given particular scenario of the 12+log(O/H) profile, there is a higher probability that the log(N/O) profile does not follow the same trend (\ensuremath{>} 50\%). This result might be interpreted as a decoupling of 12+log(O/H) and log(N/O) radial gradients. 

\begin{table}
	\caption[Radial metallicity 12+log(O/H) gradient breaks statistics for the LINER sample from MaNGA]{Statistics of the breaks in the obtained radial metallicity gradients of 12+log(O/H) in terms of R\ensuremath{_{e}}. Column (1) shows the number of breaks. Column (2) shows the total number of galaxies. Columns (3) and (4) show the median and standard deviation distances in terms of R\ensuremath{_{e}}.}
	\label{OH_breaks}
	\centering
	\fontsize{9.25pt}{10.00pt}\selectfont
	\begin{tabular}{c | c c c c } 
		\hline\hline
		&  & \textbf{12+log(O/H)} &  &  \\ \hline
		\textbf{N. Breaks} & \textbf{N. Gal.} & \textbf{R}\boldmath{$_{median}$} & \textbf{R}\boldmath{$_{std}$} & \textbf{Range} \\
		\textbf{(1)} & \textbf{(2)} & \textbf{(3)} & \textbf{(4)} & \textbf{(5)} \\ \hline
		1 & 48 & 0.87 & 0.49 & [0.25, 2.97] \\
		2 (Inner) & 26 & 0.61 & 0.32 & [0.05, 1.56] \\
		2 (Outer) & 26 & 1.09 & 0.6 & [0.09, 2.93] \\
	\end{tabular}
\end{table}
\begin{table}
	\caption[Radial metallicity log(N/O) gradient breaks statistics for the LINER sample from MaNGA]{Same as Table \ref{OH_breaks} but for log(N/O).}
	\label{NO_breaks}
	\centering
	\fontsize{9.25pt}{10.00pt}\selectfont
	\begin{tabular}{c | c c c c } 
		\hline\hline
		&  & \textbf{log(N/O))} &  &  \\ \hline
		\textbf{N. Breaks} & \textbf{N. Gal.} & \textbf{R}\boldmath{$_{median}$} & \textbf{R}\boldmath{$_{std}$} & \textbf{Range} \\
		\textbf{(1)} & \textbf{(2)} & \textbf{(3)} & \textbf{(4)} & \textbf{(5)} \\ \hline
		1 & 43 & 0.83 & 0.53 & [0.08, 2.98] \\
		2 (Inner) & 13 & 0.76 & 0.43 & [0.18, 1.64] \\
		2 (Outer) & 13 & 1.16 & 0.68 & [0.34, 3.17] \\
	\end{tabular}
\end{table}

\begin{figure}
	\centering
	\resizebox{\hsize}{!}{\includegraphics{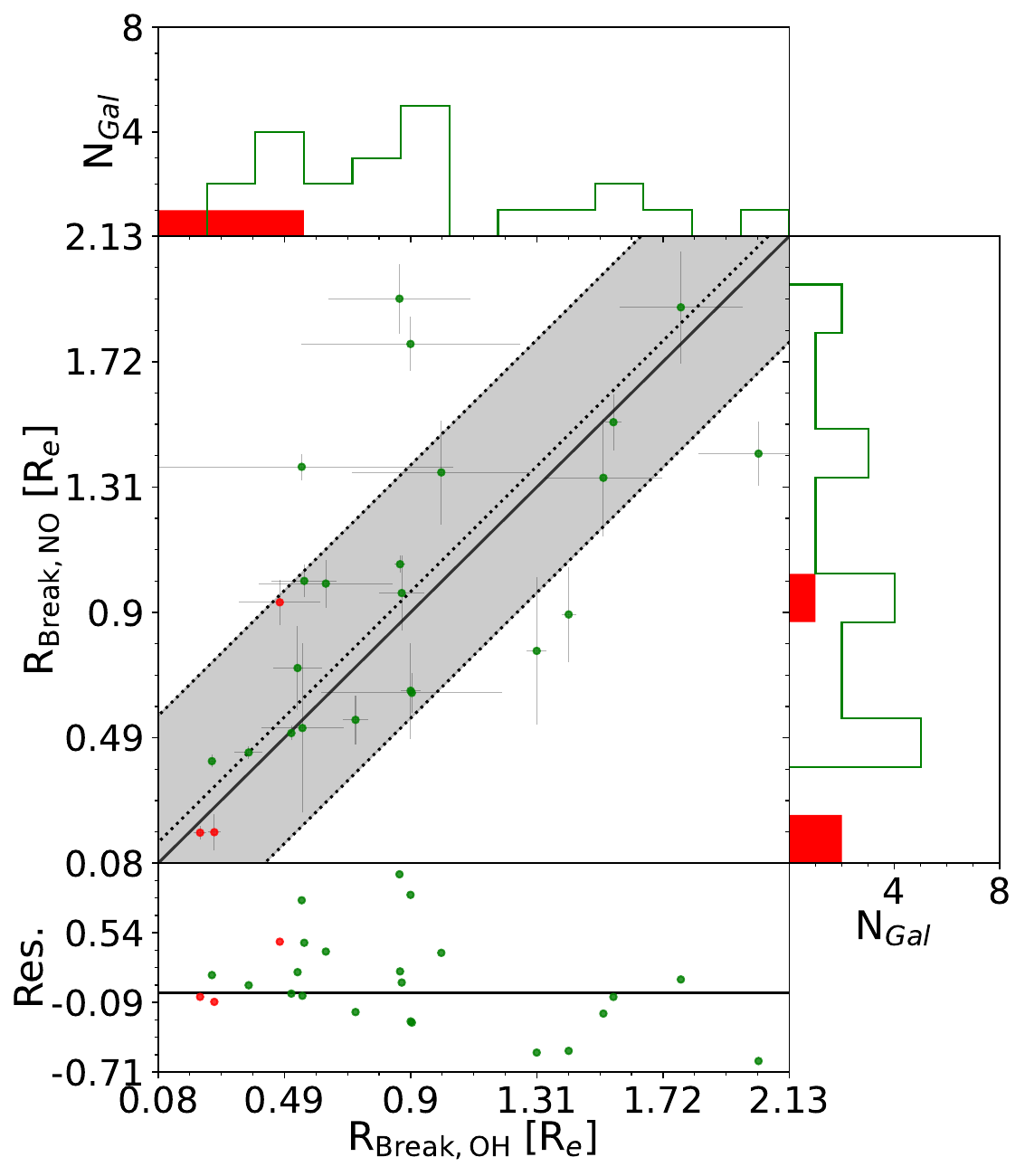}}
	\caption[Comparison between 12+log(O/H) and log(N/O) breaks in the LINER sample from MaNGA]{Comparison between the breaks found for the log(N/O) radial gradients (y-axis) and 12+log(O/H) radial gradients (x-axis) in our sample. Green dots correspond to LINERs showing one break in both gradients, whereas red dots correspond to LINERs showing two breaks. The solid black lines represent the one-to-one relation, the dotted lines represent the median offset, and the shaded gray area the deviation.}
	\label{Breaks}
\end{figure}

We also present in Tables \ref{OH_breaks} and \ref{NO_breaks} the statistics of the breaks in the obtained radial metallicity gradients both for 12+log(O/H) and log(N/O), respectively. When comparing the breaks obtained for 12+log(O/H) and log(N/O) radial gradients, we observe that, although the statistics might reflect similar values, there is not any correspondence, as observed in Fig. \ref{Breaks}. This result reinforces the idea that 12+log(O/H) and log(N/O) radial gradients are essentially decoupled. 


As the fitting technique yields a deterministic result, we compute uncertainties in the breaking points from 1,000 bootstrap fittings per galaxy, drawing random HII regions. In most cases, the uncertainty correlates with the number of HII regions present in the breaking point. This is the case of GAL 8981-6101, GAL 10518-6103 or GAL 11746-9102, where the uncertainties are high although there is clear change in the slope of the radial metallicity gradient.

For the subsequent analysis, we consider  the part of the gradient that goes from the outer most region to the more external break as \textit{outer} part. In the case of a galaxy presenting two breaks, we consider the region that goes between them as \textit{middle} part. Finally, in both cases, the \textit{inner} part is defined as the one that goes from the nuclear region to the break closer to the center. In the case of galaxies with two breaks, we consider the one closer to the center of the galaxy as \textit{inner break}, and the other as \textit{outer break}. No specification is made when the corresponding radial fit only presents one break.

\subsubsection{Slopes of the radial gradients}
\label{sss4.4.1.2}
\begin{figure*}
	\centering
	\includegraphics[width=17cm]{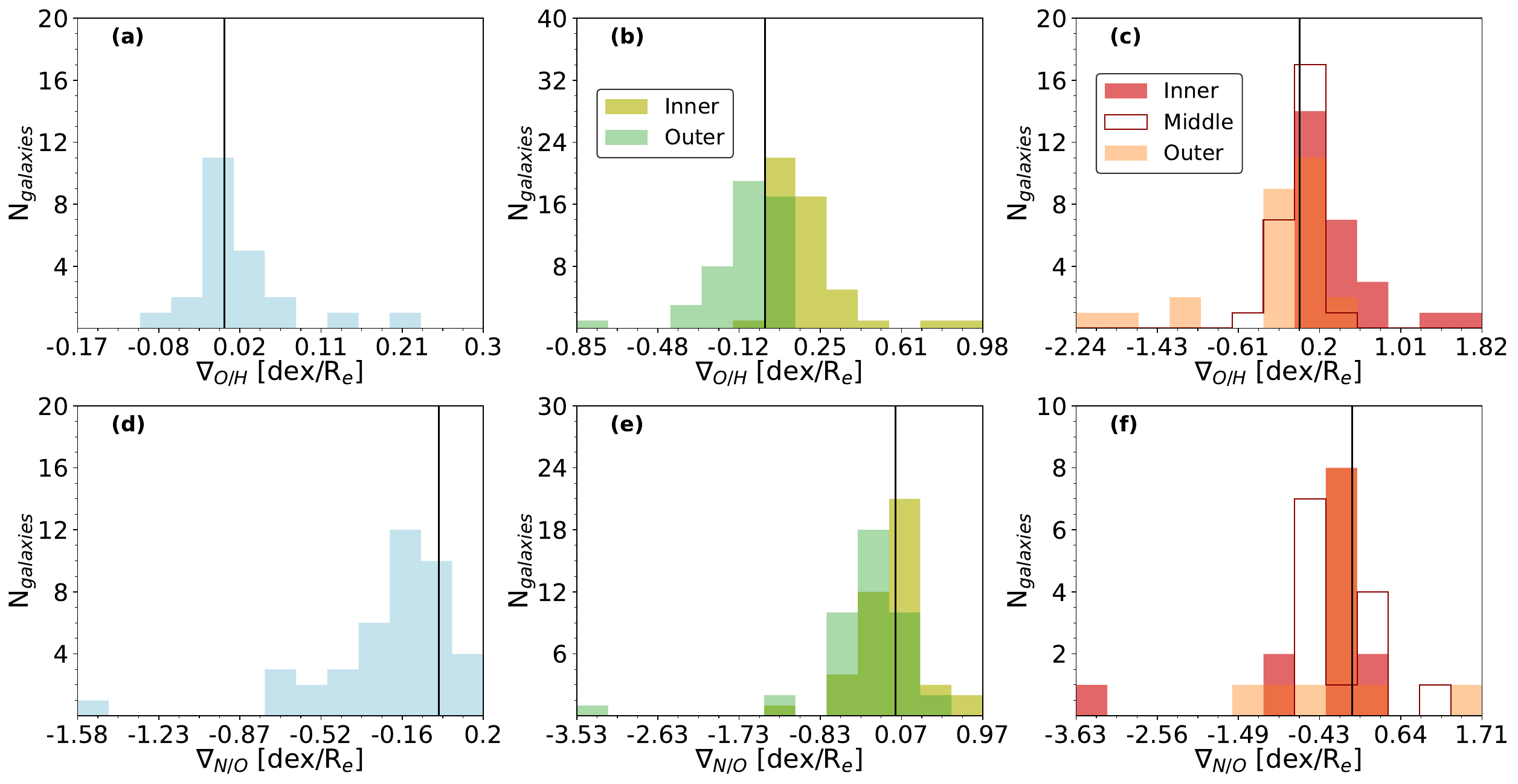}
	\caption[Radial metallicity gradient slopes for the LINER sample from MaNGA]{Histogram of the obtained gradient slopes in the fits for both 12+log(O/H) and log(N/O) for different categories attending to the number of obtained breaks. Slopes for 12+log(O/H): a) galaxies with no breaks, b) galaxies with one single break, and c) galaxies with two breaks. Lower plots d), e) and f) are similar to the above ones but for log(N/O) radial gradient. For all plots, solid black lines represent the flatten profile (\ensuremath{\nabla = 0}).}
	\label{Slopes}
\end{figure*}

We show in Fig. \ref{Slopes} the slopes of the fits in our sample of LINER-like galaxies, simultaneously for 12+log(O/H) and log(N/O). We distinguish among galaxies with no breaks (panels (a) and (d)), galaxies with one break (panels (b) and (e)) and galaxies with two breaks (panels (c) and (f)). 

As it can be seen, galaxies presenting no breaks in the fit of the 12+log(O/H) radial gradient (panel (a)) show a flattened behavior, as the slope is close to \ensuremath{\nabla_{O/H} = 0.0}. On the contrary, galaxies with no breaks in the log(N/O) radial fit mainly show negative gradients\footnote{We notice the strong negative gradient for galaxy 8942-12702, which is explained by the lack of HII regions at larger radii (see Fig. \ref{Gradients_1} in the supplementary material).} (panel(b)), in concordance with the inside-out growth scenario. If we consider the group of galaxies with only one break (panels (b) and (d)), we observe that for both 12+log(O/H) and log(N/O) the outer slope is mainly negative or flattened, in contrast to the inner slope which is mainly positive for 12+log(O/H), with some cases (five galaxies) for log(N/O) as well.

Finally, the analysis of the slopes in galaxies with two breaks reinforces the previous results obtained for 12+log(O/H): the inner  part shows positive slopes while the outer part shows negative or close to zero slopes (see panel (e)). The middle parts of the galaxies tend to show almost flattened gradients. Regarding log(N/O), apart from a galaxy that shows a strong negative slope in the inner part\footnote{Again, this outlier corresponds to galaxy 8243-9102, whose fit to the radial  gradient can be visually inspected  in Fig. \ref{Gradients_1} (supplementary material), the analysis of the slopes  reveals that there is not enough HII regions to properly trace the inner gradient.}, both the inner and middle slopes are mainly negative whereas the outer parts are close to \ensuremath{\nabla_{N/O} = 0.0}.

\subsubsection{Intersects of the radial gradients}
\label{sss4.4.4.3}
\begin{figure*}
	\centering
	\includegraphics[width=15cm]{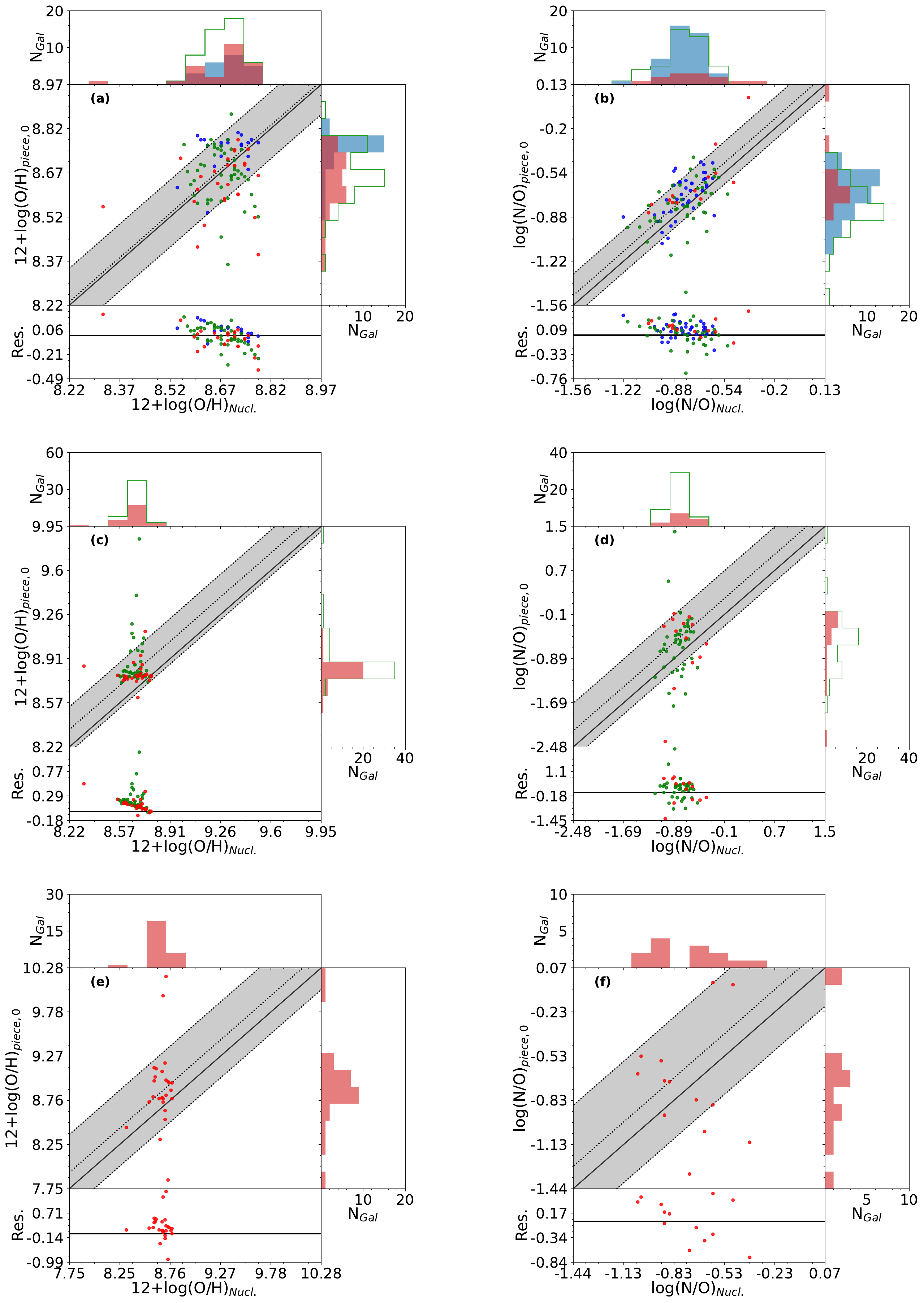}
	\caption[Comparison of the intersects from the radial metallicity gradients fits and nuclear estimations in the LINER sample from MaNGA]{Comparison of the abundances obtained as intersects from the fits  with the abundances estimated in the nuclear regions both for 12+log(O/H) (left column) and log(N/O) (right column).  Plots a) and b) represent the extrapolation of the innermost radial fits, while plots c) and d) represent the extrapolation of the middle radial fits, and  plots e) and f) the extrapolations of the outer radial fits. For all plots, Blue dots represent galaxies with no breaks in their corresponding fits, green dots galaxies with one break, and red dots galaxies with  two breaks. The solid black lines represent the one-to-one relation, the dotted lines represent the median offset, and the shaded gray areas the deviation.}
	\label{Intersects}
\end{figure*}

Lastly, we analyzed the intersects of the metallicity radial gradients, i.e., the extrapolation of the fits to the galaxy nuclei (we considered different intersects according to the shape and breaks of their gradients). We show the results in Fig. \ref{Intersects}, for which we considered as reference in each galaxy the value in the nuclear region as estimated in \citetalias{Perez-Diaz_2025}. These abundances were calculated assuming AGN models, which are the ones found to better reproduce most of the scaling relations in the studied sample \citepalias{Perez-Diaz_2025}.

Firstly, we analyzed whether the extrapolation of the innermost radial metallicity gradients match the estimations obtained from the nuclear regions (as traced by the metallicity of the AGN). From Fig. \ref{Intersects} (a) we conclude that, on average, the extrapolations of the 12+log(O/H) fits are in agreement with the nuclear estimations (median offset of 0.02 dex), although there is a wide scatter. On the other hand, we obtained that the extrapolations of the log(N/O) fit  (panel (b)) are also, on average, in agreement (median offset 0.07 dex), but with less scatter. This is a similar result as that reported in \citetalias{Perez-Diaz_2025}, where all ionizing scenarios were tested.

When we account for the extrapolations of the middle radial gradient (for galaxies with two breaks) or the outer gradient (for galaxies with one break), we obtained that the extrapolations from that fit lead to a clear overestimation of the nuclear abundance in both cases (panels (c) and (d)). We observe that the offset from the log(N/O) radial gradient (\ensuremath{\sim 0.27} dex/$R_{e}$) is higher that that found for the 12+log(O/H) (\ensuremath{\sim 0.14} dex/$R_{e}$). Interestingly, we observe that the extrapolations from the 12+log(O/H) middle radial gradient are systematically above the predictions for the nuclear region, whereas in the case of the log(N/O) extrapolations there is a non-negligible group of galaxies for which their middle radial fits predict lower abundances.

Finally, we analyzed the extrapolations from the outer radial fits in the case of LINERs with two breaks. The results shown in panels (e) and (f) reveal a similar picture to that observed in the extrapolations from the middle radial fit. However, in this case we also report that the deviation is much higher and that the relative number of galaxies whose radial fits predict lower abundances compared to the nuclear estimations is higher as well.

\subsection{Exploring the connection of galaxy properties with gradient shapes}
\label{ss4.4.2}
Several authors have discussed the role played by different galaxy properties, such as stellar mass or star formation rate, in the shape of the radial metallicity gradients observed in disk galaxies \citep[e.g.][]{Sanchez-Menguiano_2018, Cardoso_2025}. In this section we discuss whether the variety of the metallicity radial gradient profiles found in our sample are connected to the properties of the host galaxies and/or the nature of their nuclear region. We separately explore the three characteristics of the radial profiles: the number of breaks, the slopes and the intersects.


Firstly, we analyzed whether the number of found breaks in 12+log(O/H) and log(N/O) radial fits correlate with the stellar mass or HI mass. We found similar distributions for the whole sample and for each group of galaxies, implying that neither the stellar nor the HI mass are responsible for the number of breaks observed in our sample (see Fig. \ref{Hist_ste} and \ref{Hist_mhi}). We also explored whether the number of breaks correlate with any physical property of the nuclear part, including their derived chemical abundances, the stellar velocity dispersion (which directly correlates with the mass of the Super Massive Black Hole (SMBH) in case of AGN activity) or the equivalent width of H\ensuremath{_{\alpha}} (whose strength allows us to discriminate the type of nuclear activity). In all cases, we found again that there is no correlation at all with the number of breaks (see Fig. \ref{Hist_vel}).

\begin{figure*}
	\centering
	\includegraphics[width=17cm]{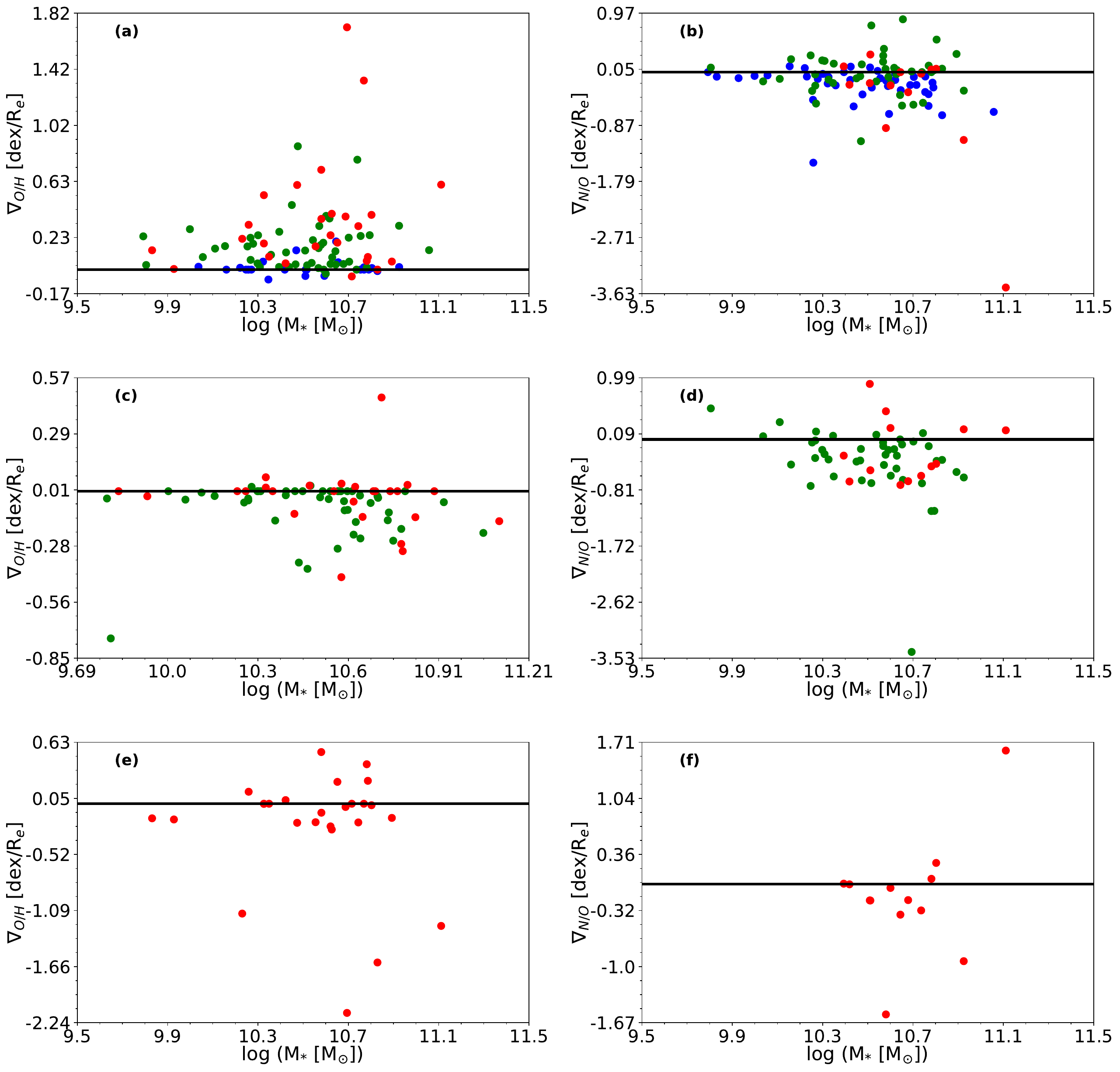}
	\caption[Radial metallicity gradient slopes as a function of stellar mass in the LINER sample from MaNGA]{Relation between stellar mass and the  slopes for the radial fits of 12+log(O/H) (left column) and log(N/O) (right column).  Panels a) and b) represent the slopes of the inner radial fits, panels c) and d) the slopes of the middle radial fits, and  panels e) and f) the slopes of the outer radial fits. For all plots, Blue dots represent galaxies with no breaks,  green dots with one break, and red dots with two breaks. Solid black lines represent the flatten profile (\ensuremath{\nabla = 0}).}
	\label{Slopes_mass}
\end{figure*}

\begin{figure*}
	\centering
	\includegraphics[width=17cm]{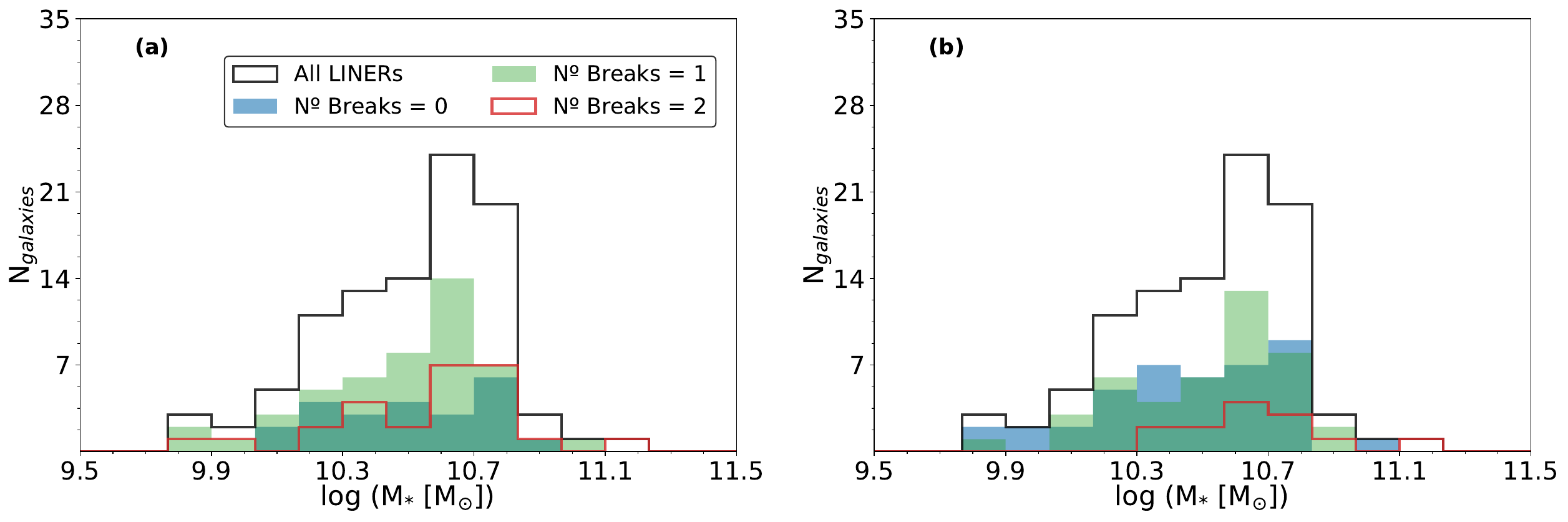}
	\caption[Histogram of stellar masses in the LINER sample from MaNGA]{Histogram of the stellar masses for our sample of LINERs. (a) The sample is segregated according to the number of breaks shown in the 12+log(O/H) radial profile. (b) The segregation is done according to the number of breaks showed in the log(N/O) radial profile.}
	\label{Hist_ste}
\end{figure*}

\begin{figure*}
	\centering
	\includegraphics[width=17cm]{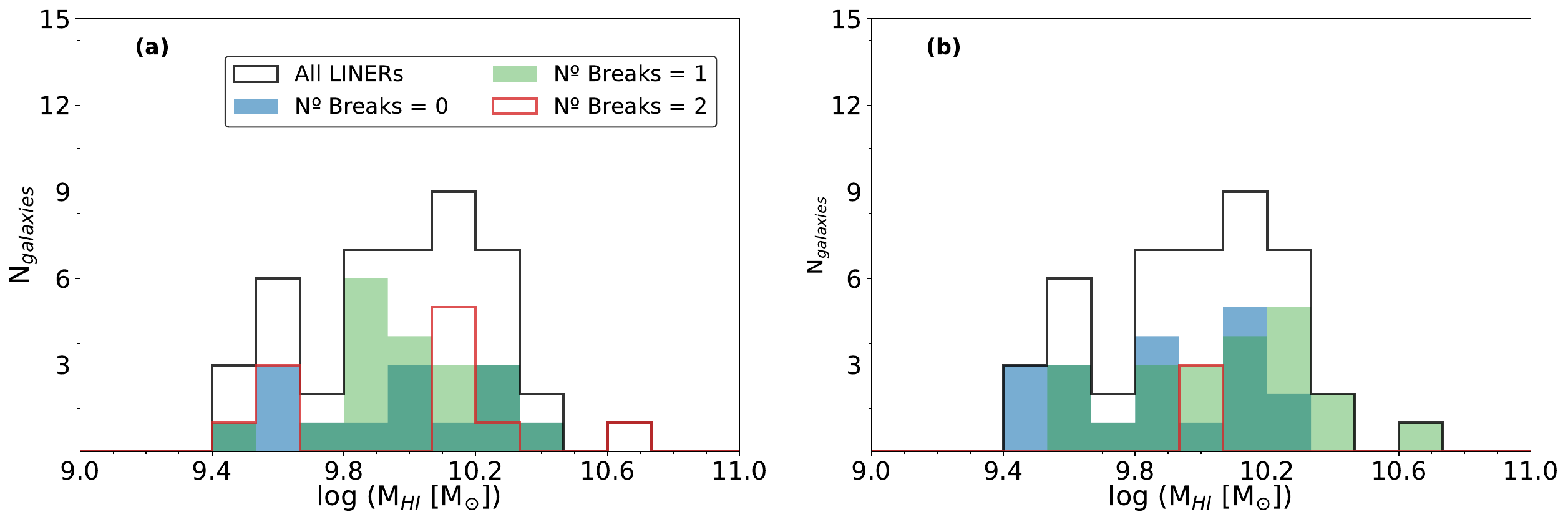}
	\caption[Histogram of HI masses in the LINER sample from MaNGA]{Same as Fig. \ref{Hist_ste} but for the HI mass.}
	\label{Hist_mhi}
\end{figure*}

\begin{figure*}
	\centering
	\includegraphics[width=17cm]{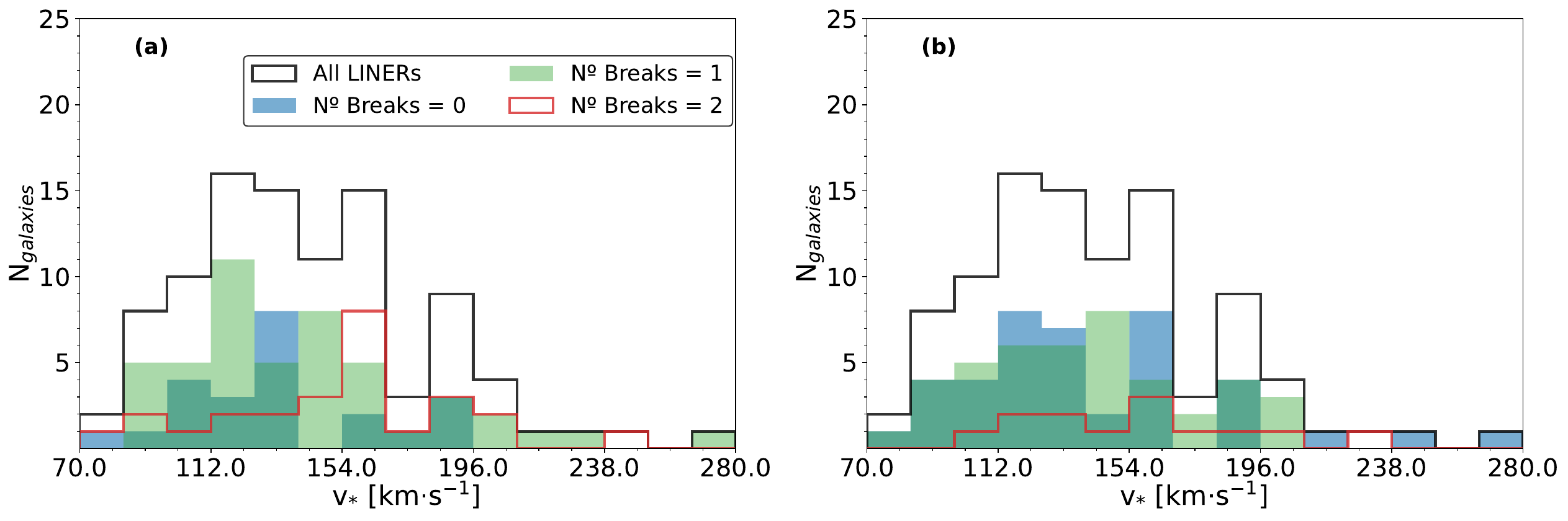}
	\caption[Histogram of nuclear stellar velocity dispersion in the LINER sample from MaNGA]{Same as Fig. \ref{Hist_ste} but for the stellar velocity dispersion in the nuclear regions.}
	\label{Hist_vel}
\end{figure*}

We then analyzed whether the slopes of the radial fits (\ensuremath{\nabla}) for both 12+log(O/H) and log(N/O) correlate with the stellar mass. Our results are shown in Fig. \ref{Slopes_mass}. In the case of 12+log(O/H), we obtained that: i) galaxies with no breaks tend to show flatten profiles (blue dots in panel (a)), whereas the rest of galaxies show positive slopes; ii) galaxies with one (and two breaks) tend to show negative slopes in the outer (middle) part of their profiles, being more  negative when stellar mass increases (panel (c)); and, iii) the slope of outer profile in galaxies with two breaks does not correlate with stellar mass (panel (e)).

On the other hand, when analyzing log(N/O), we obtained that i) there is no correlation between stellar mass and the slope of the inner radial fit for all types of galaxies; ii) the slope of the outer fit in galaxies with one  break is meanly negative and shows a strong anti-correlation with stellar mass (a similar result is obtained for the slope in the middle part of galaxies with two breaks); and, iii) similarly to 12+log(O/H), no particular trend is found for the slope of the outer profile of galaxies with two breaks. 

Finally, we inspected whether intersects correlate with any of the other studied properties. In particular, we analyzed the relation between the host galaxy properties with  the extrapolations of the metallicity radial fits to the nuclei (\ensuremath{R = 0}) (i.e. the intersects of the inner radial fits). First of all, we checked if there is a correlation with stellar mass, as it might be suggested by the so-called mass-metallicity relations (MZR) and the equivalent one for log(N/O) (MNOR). Our results are similar to those reported in \citetalias{Perez-Diaz_2025} for AGN models (see Fig. 6 and 7 from that work). 

\begin{figure*}
	\centering
	\includegraphics[width=17cm]{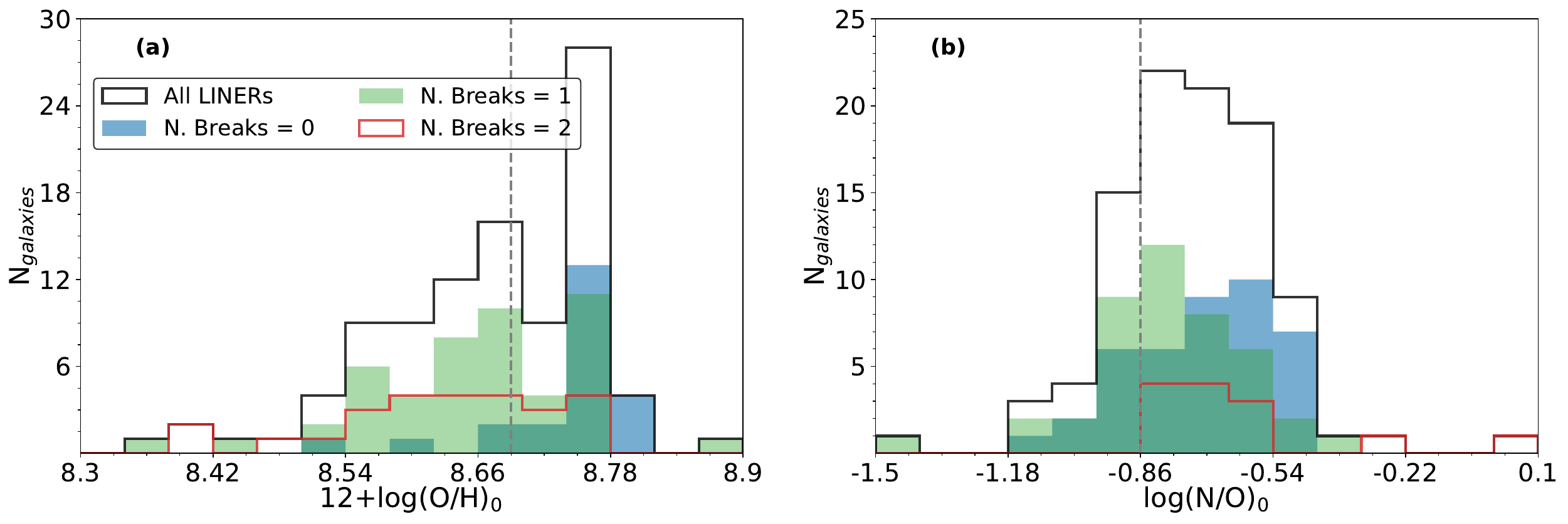}
	\caption[Histogram of radial metallicity gradient intersects in the LINER sample from MaNGA]{Histogram of the intersects of the inner parts of the metallicity radial fits for (a) 12+log(O/H), and (b) for log(N/O). Sample of LINERs is segregated into three groups according to the number of breaks shown in the abundance radial gradient. For both plots, gray dashed lines correspond to the solar value.}
	\label{Hist_Intersects}
\end{figure*}

We also explored whether the number of breaks introduces differences in the intersects, as shown in Fig. \ref{Hist_Intersects}. Only for galaxies with no breaks in the 12+log(O/H) radial fit we obtained that they tend to solar abundances (12+log(O/H) \ensuremath{\sim} 8.69), but in the rest of cases we do not see any relation between the number of breaks and the extrapolated abundances in the nuclear part.


\subsection{The log(N/O) vs 12+log(O/H) relation for individual HII regions}
\label{ss4.4.3}
We show in Appendix \ref{A2} the log(N/O) vs 12+log(O/H) diagram for all the selected HII regions in each galaxy, colored by their distance to the galactic center. For the majority of our sample, we obtain that HII regions mainly follow the log(N/O) vs 12+log(O/H) relation (see Fig. \ref{NOOH_1}) within the scatter, and that the HII regions located closer to the nuclear parts exhibit higher log(N/O) ratios, concluding that there is a decreasing trend between log(N/O) and 12+log(O/H), It is worth to note that 12 galaxies (\ensuremath{\sim} 13\% of the sample) exhibit a decrease of 12+log(O/H) in the closer parts towards the nuclear region, while the log(N/O) remains high (solar or suprasolar). Namely, these galaxies are 10518-12705, 11013-6104, 12078-12703, 7495-12704, 8249-12704, 8259-9102, 8320-9102, 8562-9102, 8563-12705, 8492-12702, 8983-3703 and 8997-12704.

\begin{figure*}
	\centering
	\includegraphics[width=17cm]{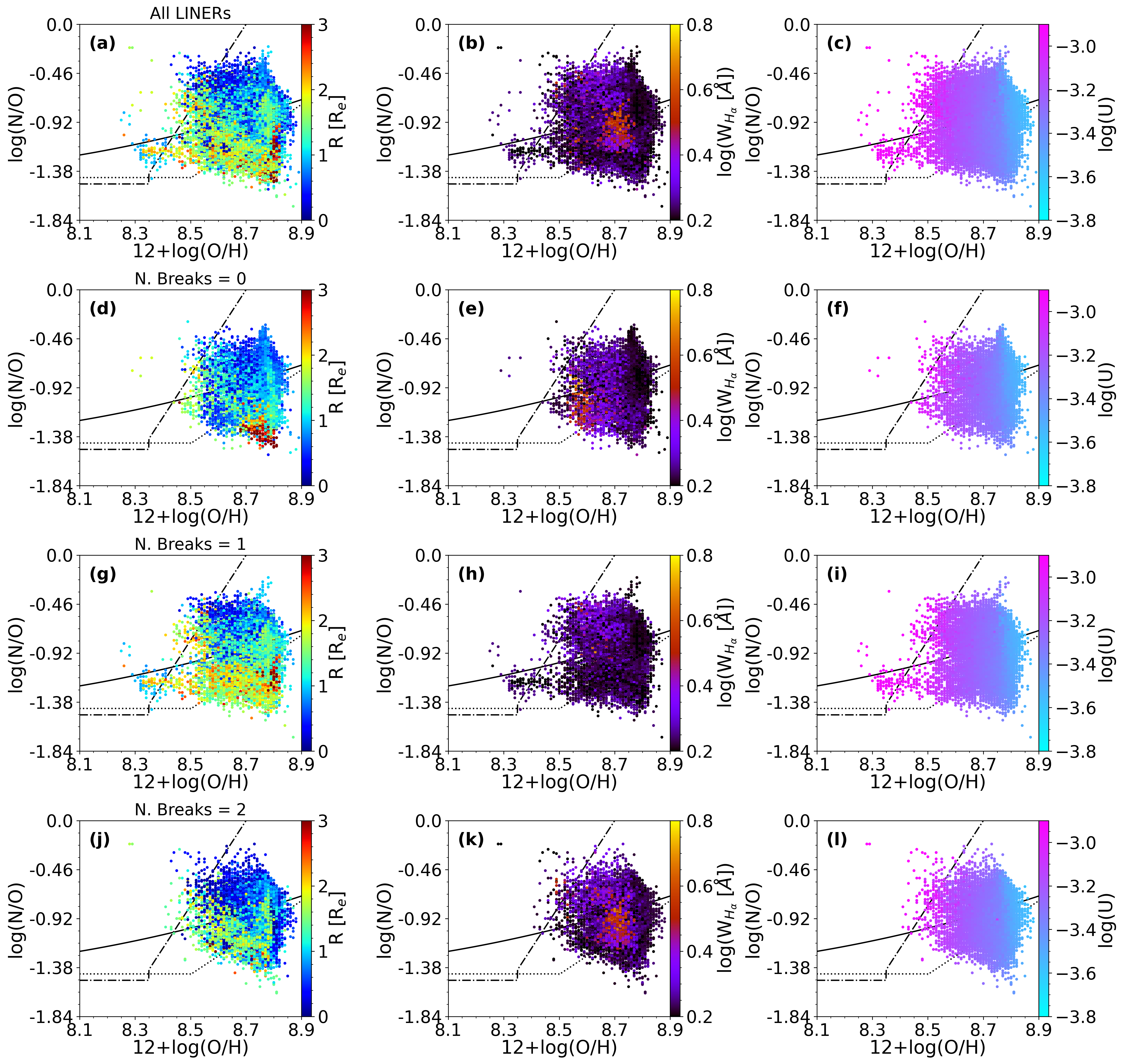}
	\caption[N/O vs O/H relation for all HII regions in the LINER sample from MaNGA]{log(N/O) vs 12+log(O/H) diagram for the HII regions in our sample of LINERs. Top to bottom: first row shows diagrams for all HII regions in all LINERs; second row shows diagrams for those HII regions in LINERs with no breaks in the 12+log(O/H) radial fit; third row shows diagrams for HII regions in LINERs with one break; and forth row shows diagrams for HII regions in LINERs with two breaks. Each column shows different colorbared properties: normalized distance to the galaxy centers (left), equivalent width of H\ensuremath{_{\alpha}} (middle) and ionization parameter (right). For all plots, the solid back line represents the fit provided by \citet{Coziol_1999}, the dotted line shows the fit by \citet{Andrews_2013}, and the dash-dotted line shows the fit by \citep{Belfiore_2015}.}
	\label{NOOH_global}
\end{figure*}

We show in Fig. \ref{NOOH_global} the results for all HII regions in our sample. First of all, looking at the ionization parameter (right column of the plot), we can check on the robustness of the methodology: the lack of measurements of the auroral line [\ion{O}{iii}]\ensuremath{\lambda }4363\ensuremath{\AA} forces the assumption on the log(U) vs 12+log(O/H) relation to break the degeneracy. Nonetheless, it is clearly shown that this assumption introduces no dependence at all in the log(N/O) ratio by log(U). 

Considering the information on the distance of the HII regions to the galactic centers, probes that the primary N production is mainly located in the outer HII regions (R \ensuremath{\geq} 1.5R\ensuremath{_{e}}), as shown in Fig. \ref{NOOH_global} panel (a). In addition, the HII regions in galaxies without any break in the radial fits are mainly located in the regime of secondary N production, with increasing log(N/O) with 12+log(O/H), which is consistent with previous studies (see panel (d)). Interestingly, those outer HII regions with primary N production are found in galaxies that exhibit a 12+log(O/H) gradient with one break (panel (g)). Overall, the position of the large majority of HII regions is well reproduced by the scatter reported by different relations in the literature.

Finally, the analysis of the position of HII regions in the log(N/O) vs 12+log(O/H) relation based on the equivalent width of H\ensuremath{_{\alpha}} complements the picture that emerged from their distance to the galactic centers. We only found a particular trend in galaxies with no breaks in their radial fits (see panel (e)). Particularly, we obtained that HII regions with oxygen abundances close to the solar value \citep[(12+log(O/H)\ensuremath{_{\odot}} = 8.69;][]{Asplund_2009} exhibit the highest values of W\ensuremath{_{H_{\alpha}}}, as well as some of the closest distances to the galactic centers. This might be explained by the contamination of their spectra from AGN activity.

In summary, a majority of HII regions in our sample falls between the reported relations for log(N/O) and 12+log(O/H). Therefore, even though we observe some deviations from the reported relation, they are not introducing a departure.

\subsection{A representative value for the metallicity in galaxies}
\label{ss4.4.4}
A key aspect when analyzing chemical enrichment of galaxies is to define their characteristic metallicities. Several ideas have been proposed: the abundance ratios at 0.4R\ensuremath{_{25}} (being R\ensuremath{_{25}} the isophotal radius, i.e., the radius at which the surface brightness equals 25 mag/arcsec\ensuremath{^{2}}, \citealp{Zaristky_1994}); the central (extrapolation to the nucleus) abundances \citep{Ryder_1995}; or, the chemical abundance ratios at the effective radius (i.e. encompassing 50\% of the light coming from the disk component) \citep[e.g.][]{Sanchez_2013, Cresci_2019, Alvarez-Hurtado_2022, Sanchez-Menguiano_2024}. This is even more critical in the case of AGNs, as their activity affects the nuclear parts and, hence, the representative value might account or not for the AGN role.

\begin{figure*}
	\centering
	\includegraphics[width=17cm]{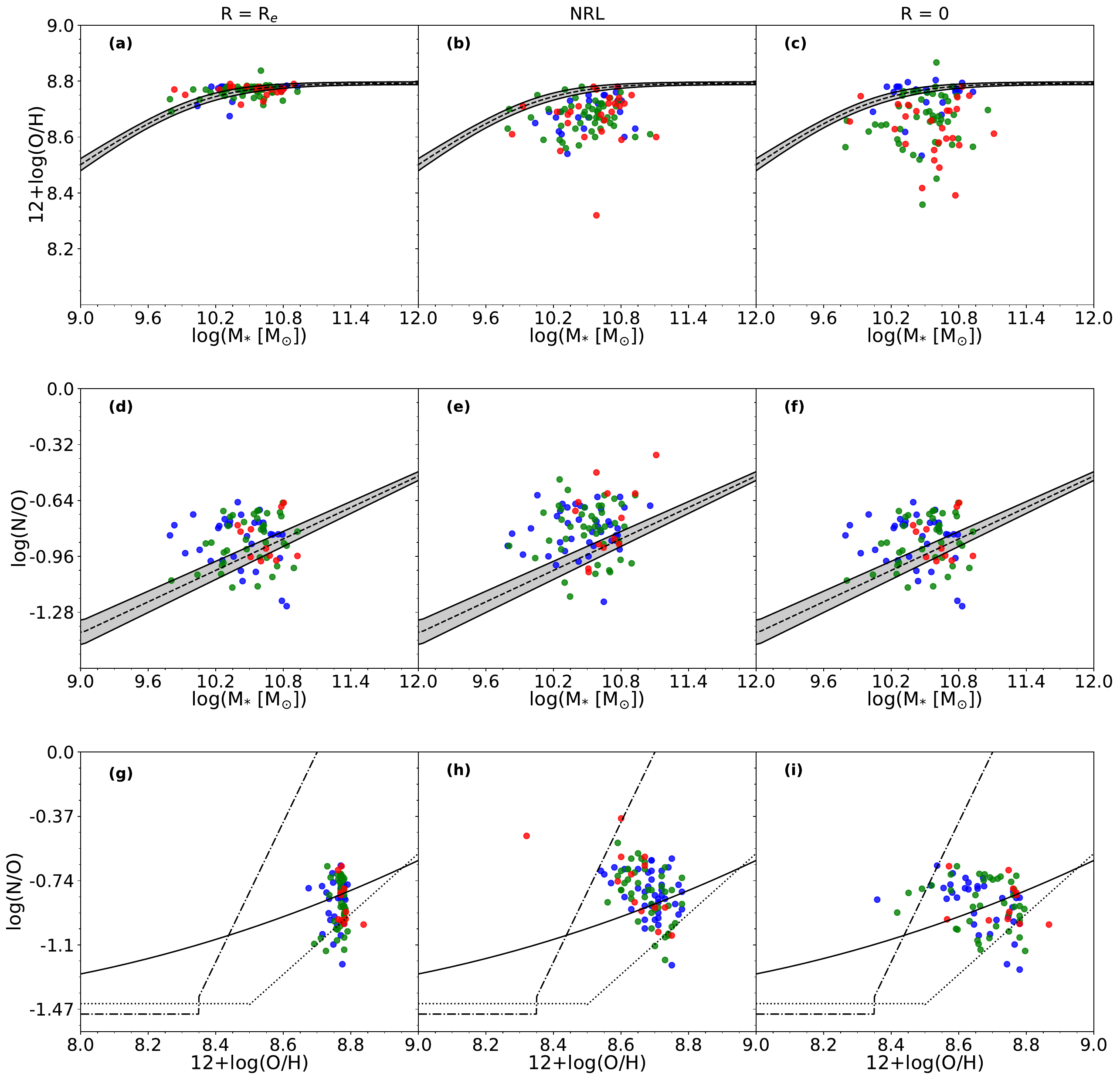}
	\caption[Scaling relations for different characteristic abundance ratios in the LINER sample from MaNGA]{Scaling relations for the characteristics abundance ratios in our sample of LINER-like galaxies. Panels in the first row show the mass-metallicity relation (MZR), and the continuous line represents the fit obtained by \citet{Curti_2020}. Panels in the second row show the mass-NO relation (MNOR), with the continuous line representing the fit obtained by \citet{Andrews_2013}. Panels in the third row show the log(N/O) vs 12+log(O/H relation), with the lines representing the different fits explained in Fig. \ref{NOOH_global}. Panels in the first column show the chemical abundance ratios estimated at the effective radius. Panels in the second column show the estimations of nuclear abundances assuming AGN models with \ensuremath{\alpha_{OX} = -1.6} (abundance of the Narrow Line Region, NLR). Panels in the third column show the intersects of the metallicity radial fits. For all plots, blue dots represent galaxies with no breaks in their corresponding radial fits, green dots have one break, and red dots represent galaxies with two breaks.}
	\label{Charac_abun}
\end{figure*}

We explored three different values of the chemical abundance ratios for each galaxy: the expected value at the effective radius (R\ensuremath{_{e}}), the abundance ratios derived for the Narrow Line Region of the AGN (NLR) and the extrapolated abundance to the nucleus (R=0). Our results are presented in Fig. \ref{Charac_abun}, comparing them in different scaling relations such as the MZR, the mass-NO relation (MNOR) or the N/O vs O/H diagram.

We found that the metallicities of the ISM at the effective radius are the ones that best reproduce the MZR \citep{Curti_2020} (panel (a) in Fig. \ref{Charac_abun}), whereas the MNOR \citep{Andrews_2013} seems to not be well reproduced in any of the considered cases. Finally, the log(N/O) vs 12+log(O/H) relation tells complementary stories depending on which chemical abundance ratio is used. If we account for chemical abundance ratios derived in the regions ionized by the AGN, we observe that there is a wide scatter, but mostly reproduced by the observing trends in literature. If we instead consider chemical abundances as extrapolated to the nuclear part from the radial fits, we observe almost an anti-correlation, which might be indicative of hydrodynamical processes affecting the oxygen abundance. Finally, the chemical abundances at the effective radius report galaxies clustering slightly above the solar oxygen abundance, but with a large spread of log(N/O) values. 

\section{Discussion}
\label{s4.5}
As a general remark, we note the lack of studies analyzing metallicity radial gradients in galaxies with nuclear emission dominated by LLAGNs. Therefore, in this section we discuss our results with previous studies that are targeting different types of objects from those analyzed here.

\subsection{On the shape of the O/H metallicity gradient in LINERs}
\label{ss4.5.1}
The inside-out growth scenario for galaxies, the formation of an exponential disk, due to the accretion from the outer parts, implies higher gas densities in the inner parts, triggering more star formation, leading to the result of galaxies showing negative radial gradients in both the metallicity of the gas-phase ISM (as traced by 12+log(O/H)) and the metallicity of the stellar populations \citep{Rich_2012, Sanchez_2014, Gonzalez-Delgado_2015, Sanchez-Menguiano_2016, Goddard_2017, Zinchenko_2019}. This is also supported by chemical evolution models \citep{Matteucci_1989}.

Recent observational results from IFS surveys such as MaNGA \citep{Bundy_2015} or CALIFA \citep{Sanchez_2013} have allowed not only to increase the statistics of metallicity radial gradients at low redshift, but also to refine chemical evolution models and their predictions. The study from \citet{Belfiore_2017} (MaNGA) showed that not only the negative metallicity gradient is observed in most galaxies, but that the steepness of the gradient depends on stellar mass. The study from \citet{Perez-Montero_2016} (CALIFA) also reported a similar trend, although the authors warned about the statistical significance of such trend. Indeed, both studies obtained that galaxies with moderate low stellar masses (\ensuremath{10^{9.5}-10^{10.5}}M\ensuremath{_{\odot}}) show steeper gradients (normalized to a characteristic size) than more massive galaxies. This result has also theoretical background from chemical evolution models. Low-stellar mass galaxies (considered as progenitors of more massive systems) have initially star formation mainly located on the central regions. As star formation occurs in the outer parts, the pollution from stars in the outskirts of galaxies helps flattening the metallicity profile, while the inner parts, already rich in primary metals,  do not increase their metallicity at the same rate (for a constant value of the metal yield).

Recently, \citet{Sanchez-Menguiano_2018} published a study of 102 spiral galaxies observed with MUSE. Their analysis of the metallicity radial gradients shows that only 55 galaxies exhibit the expected negative gradient, while 37 galaxies show inner drops and 26 a flat profile. This implies that a significant number of galaxies deviate from the negative gradient predicted from the inside-out scenario. In the same way, \citet{Pilyugin_2024} found that spiral galaxies can be divided into two main categories: galaxies with a single linear radial gradient (called S-galaxies) and galaxies with a flat inner gradient and a negative outer gradient (called LS-galaxies). Cosmological simulations also report results that add more complexity to the inside-out scenario: instead of a correlation between the slope of the metallicity gradient and stellar mass, \citet{Tissera_2022} obtained no correlation at all at low redshifts if all morphological types (not only disk-dominated galaxies) are considered.

In the realm of AGN-dominated galaxies, the picture of gas-phase abundance gradients is more uncertain due to the scarcity of studies analyzing them. \citet{Amiri_2024} observed that the Seyfert-host galaxy NGC 7130 exhibits an inverse metallicity gradient (i.e. metallicity of the gas-phase ISM increases with radius), and they conclude that the AGN is the responsible for the shape of the metallicity radial gradient. On the other hand, \citet{Nascimento_2022} found that the majority of metallicity radial gradients in Seyfert-like galaxies show almost flattened profiles and they report that there is a significant drop in metallicity in the parts closest to the AGN, implying that accretion of metal-poor gas is the responsible for the dilution.

Our results show that LINERs are not characterized by a single unique 12+log(O/H) radial gradient shape. Out of our sample of 97 LINER-like galaxies, we obtain that only 23 galaxies (23.7\%) can be approximated with the single linear metallicity profile. For these galaxies, we obtained that the slopes (\ensuremath{\nabla_{O/H}}) are close to zero dex/R\ensuremath{_{e}} (Fig. \ref{Slopes} panel (a)) and in some rare cases they are positive. This is in agreement with the results obtained by \citet{Nascimento_2022} for Seyfert-like galaxies, as we obtain the same reported scenarios. 

We also report a significant group of LINER-like galaxies that exhibit a metallicity radial gradient profile with a break. In total, 48 galaxies (49.5\%) show this metallicity radial profile characterized by a positive 12+log(O/H) gradient in the inner parts (as reported in systems dominated by gas inflows), whereas the outer parts exhibits either a flattened or negative profile characteristic of the inside-out scenario (Fig. \ref{Slopes} panel (b)). This group of galaxies (which is the majority of our sample) follows the same trend reported by \citet{Amiri_2024} for the Seyfert galaxy NGC 7130: the inner parts of the galaxy (dominated by AGN emission) exhibits a positive gradient, whereas the outer parts (dominated by star formation) are characterized by a mildly negative gradient.

Finally, we also report a non-negligible group of galaxies, 26 out of 97 (26.8\%) that exhibit two breaks in the 12+log(O/H) metallicity radial profile. As in the previous case, the inner and middle parts are characterized by positive or almost flattened gradients, whereas the outer part is mainly characterized by a negative or flattened metallicity radial profile (Fig. \ref{Slopes} panel (c)). Only the recent work by \citet{Tapia-Contreras_2025}, based on cosmological simulations using the  ChemodynamIc propErties of gaLaxies and the cOsmic web project \citep[CIELO, ][]{Tissera_2025}, has reported a similar result for some galaxies, in which effects from mergers, satellites, galactic fountains and cold gas inflows of gas lead to these varied profiles.

The relation between the slope(s) of the 12+log(O/H) gradients and the stellar mass gives more insights on the processes that are shaping the metallicity radial gradient. For galaxies with no breaks, the gradient remains flat for almost all stellar mass (see Fig. \ref{Slopes_mass} panel (a)), in contrast to the expected behavior reported by \citet{Belfiore_2017}. This might be explained due to the fact that these galaxies might have experienced a faster evolution, already reaching the characteristic flat profile of massive galaxies. Galaxies showing at least one break exhibit positive slopes, and reach higher values for massive galaxies, which might be explained by the fact the gravitational potential of these galaxies is more effective in retaining the metals and/or favoring and capturing gas towards the inner parts, leading to a infall-dominated scenario.

The slopes in the outer parts of 12+log(O/H) radial gradient (for galaxies with at least one break) as a function of stellar mass revealed that there is a hint of anti-correlation, as reported for star-forming galaxies, although there are many galaxies with \ensuremath{\nabla_{O/H} \sim 0} (see Fig. \ref{Slopes_mass} panel (c)). This result reinforces the idea that the outer parts (or middle parts) behave accordingly to the inside-out scenario, although some galaxies seem to be already evolved. It is also important to mention that gas captured from the inner parts (galactic fountains) or from satellite gas also introduces significant changes in the shape of the metallicity profile in the outer parts \citep{Perez_2011, Sillero_2017, Tapia-Contreras_2025}

Finally, the outer parts of galaxies with two breaks present a wide dispersion of values of the slope with respect to the stellar mass (Fig. \ref{Slopes_mass} panel (e)). This can be interpreted as due to the fact that some galaxies might be experiencing galactic fountains that do not reach the very outer parts, leading to strong negative slopes, whereas others might have experienced merger events which could be stripping part of the gas or enhancing metal production due to the star-formation in the outermost parts.

\subsection{On the shape of the N/O metallicity radial gradient}
\label{ss4.5.2}
In the inside-out scenario, chemical evolution models predict that the log(N/O) radial gradient should be steeper than that observed for 12+log(O/H), as the time delay between nitrogen and oxygen production increases the difference between the inner and outer parts \citep[e.g.][]{Matteucci_1989}. Moreover, as the nitrogen production is also affected by the star-formation efficiency \citep[e.g.][]{Molla_2006}, and the inner parts are characterized by lower star formation efficiencies \citep{Spindler_2021}, this would increase even more the difference between log(N/O) and 12+log(O/H) radial profiles \citep{Vincenzo_2016}.

As stated by several authors \citep[e.g.][]{Amorin_2010, Perez-Montero_2009, Vincenzo_2016, Belfiore_2017, Perez-Diaz_2024}, measuring log(N/O) is essential to complement the picture of chemical enrichment (either for galaxies as whole or at different distances). Moreover, having a prior determination of log(N/O) is essential to use the information of nitrogen emission lines to properly measure oxygen abundances without adding biases. Indeed, several works have been published relying on estimators which are mainly tracing either the log(N/O) abundance ratio (such as O3N2, \citealp{Sanchez_2014, Sanchez-Menguiano_2018, Nascimento_2022, Amiri_2024}) or the 12+log(N/H) abundance ratio (such as N2, \citealt{Nascimento_2022}). In contrast, our methodology allows in both the SF-dominated and AGN-dominated regions to independently determine log(N/O) and 12+log(O/H) abundances, allowing us to  simultaneously explore both radial profiles.

\citet{Pilyugin_2004} published an analysis of the metallicity radial gradients (12+log(O/H) and log(N/O)) for a sample of 54 nearby spiral galaxies. Their results were in agreement with the inside-out scenario: mainly negative gradients for all the considered chemical abundance ratios, and with log(N/O) radial gradients being steeper than those reported from 12+log(O/H). A similar conclusion was reported by \citet{Perez-Montero_2016} from the analysis of metallicity radial gradients in CALIFA, obtaining that on average the slopes of the radial gradients of log(N/O) point towards steeper metallicity profiles than those obtained for 12+log(O/H). Moreover, they also analyzed the possible dependence of the slope (\ensuremath{\nabla_{N/O}}) with stellar mass, obtaining an slight dependence for less-massive systems but not statistically significant. 

Later on, \citet{Belfiore_2017} also found slightly steeper gradients in the log(N/O), but they do report an anti-correlation between the slopes and stellar mass, similar to that found in the 12+log(O/H) radial profile. They also reported higher log(N/O) ratios in the outer parts of many galaxies, concluding that pollution in the form of galactic winds from the inner to the outer parts might be responsible for this behavior. Finally, \citet{Zinchenko_2021} found not only similar results to previous studies, but also found a correlation between the median slopes for the log(N/O) profile and the stellar ages (traced by the D4000 index): galaxies with older populations (D4000 \ensuremath{>} 1.2) tend to have steeper gradients that the others. This result is in agreement with the general scheme for nitrogen production and the time delay between oxygen and nitrogen production.

The previous results only cover galaxies whose nuclear activity is dominated by star-formation, and we do not have a picture on how the log(N/O) gradient behaves in galaxies with AGN activity. Our study uses a robust methodology for the independent estimation of 12+log(O/H) and log(N/O) abundances across the HII regions to the nuclear AGN-dominated region, allowing us to study of log(N/O) abundance gradients for the first time in galaxies hosting LLAGNs.

Our results show that 41 LINER-like galaxies (42.3\%) show a single linear profile in the log(N/O) metallicity radial gradient, with the majority of the slopes being negative or close to 0. We do report very few cases (three) with clear positive slopes (see Fig. \ref{Slopes} panel (d)). Interestingly, only nine galaxies show simultaneously a single linear profile in both their 12+log(O/H) and log(N/O) metallicity radial gradients, whereas the large majority show at least one break in the 12+log(O/H). This might imply the effect of infalls of gas that mainly affects the 12+log(O/H) (also explaining the positive slopes found in the inner parts) but do not significantly affect the log(N/O) ratio due to the non-linear relation between them.

Focusing our attention on galaxies with one break, we obtained that they represent the 44.3\% of the total sample (43 galaxies). As in the case of galaxies with one break in the 12+log(O/H) radial fit, our results also indicate that the inner parts tend to present positive slopes whereas the outer parts are biased towards negative slopes. However, contrary to the clear distinction obtained for the 12+log(O/H) radial profile, we observe a clear overlap between both distributions (see Fig. \ref{Slopes} (e)). We discuss the possible causes for these profiles together with the information from the 12+log(O/H) radial gradient:

\begin{itemize}
	\item \textbf{One break in the log(N/O) radial profile but no breaks in the 12+log(O/H) radial profile}. Positive slopes in the inner parts might be indicative of the effects of AGN-driven outflows in the inner parts: if outflows are efficient at gas removal. This decrease in the inner parts does not only affect to  12+log(O/H), but also to log(N/O), as the gas removal would lead to a quenching on star formation and, thus, stars responsible for the production of N by means of secondary production would not be born at the same rate as in the inside-out scenario. Negative slopes can be interpreted as the effects of AGN-driven outflows with lower impact, dilution from galaxy inflows or a simultaneous combination of both.
	\item \textbf{One break in both log(N/O) and 12+log(O/H) radial profiles}. This is the scenario for 22 galaxies in our sample. We observe that some of them present clear negative slopes in the inner parts and less prominent or even close to flat slopes in the outer parts, which might be interpreted as the effects of those outer regions not having already undergo the secondary production N. On the other hand, there is also a group of galaxies for which we observe the contrary scenario, an almost flat profile in the inner parts, whereas there is a clear negative profile in the outer parts. This might be interpreted as the effects of quenching from AGN feed and/or feedback, simultaneously lowering the O and N abundances due to the suppression of star formation, whereas the outer part resembles the inside-out growth scenario.
\end{itemize}

Finally, we detected 13 galaxies (13.4\%) that present two breaks in the log(N/O) radial profile, the majority of them found in galaxies with one break in the 12+log(O/H) profile. In most of them, a closer look to their log(N/O) profiles (see Fig. \ref{Gradients_1}) reveals that the lack of HII regions in some parts forces this profile, and that a single linear fit or imposing just one break could fit the data with similar residuals. Thus, these cases could be also likely explained by the same scenarios proposed for those galaxies with just one break in the log(N/O) profile.  

Lastly, we also analyzed the possible effects of stellar mass in the slopes \ensuremath{\nabla_{N/O}}. For the innermost parts and for galaxies with no breaks (Fig. \ref{Slopes_mass} (b)), we did not obtain any correlation, which is in agreement with the results from \citet{Perez-Montero_2016}. In the case of the outer parts in galaxies with one break, we do not observe any significant correlation, although they are mainly negative, which again might be indicative of the standard inside-out scenario or contamination from outflows which is captured by the galaxy at the break \citep[see for instance the effects of galactic fountains, ][]{Spitoni_2013}.

\subsection{The relation between N/O and O/H}
\label{ss4.5.3}
The log(N/O) vs 12+log(O/H) diagram is a powerful tool to discriminate the processes that might shape the chemical enrichment history of galaxies. While O has a primary production origin mainly from massive stars, N can have either a primary production origin from the same massive stars and/or a production from intermediate-mass stars \citep[4-7M\ensuremath{_{\odot }}][]{Kobayashi_2020} by means of CNO cycles \citep[secondary production;][]{Henry_2000}. As a consequence, the relation between these two elements has  an almost constant ratio of log(N/O) for low oxygen abundances \citep[12+log(O/H) \ensuremath{\lesssim} 8.5][]{Andrews_2013, Vincenzo_2016} as both species are produced in the primary process of the nucleosynthesis in massive stars. On the other hand, when there is already enough O (12+log(O/H) \ensuremath{\gtrsim} 8.5) in the ISM from which stars were born, then the CNO cycles contribute to an extra enrichment of N leading to an increasing relation between log(N/O) and 12+log(O/H). However, several factors affect the expected relation between log(N/O) and 12+log(O/H) leading to a non-negligible scatter in the relation \citepalias[see the discussion provided in][]{Perez-Diaz_2025}. Hence, in the inside-out growth scenario, this is translated into a decreasing log(N/O) and 12+log(O/H) as radii increases, reaching the highest values in the regions closer to the nuclear part of the galaxies \citep[e.g.][]{Zurita_2021, Zinchenko_2021}. 

However, there is no such a characterization for galaxies hosting AGNs, specially for LLAGNs. \citet{Perez-Diaz_2021} obtained that the nuclear regions of nearby LLAGNs from the Palomar Survey do not follow the expected relation between log(N/O) and 12+log(O/H), although later on \citet{Perez-Diaz_2025} analyzing LLAGNs from MaNGA obtained that the abundances in nuclear regions are consistent with the relation within the scatter if they are derived assuming that they are  ionized by an AGN. \citep{Oliveira_2024} obtained for a sample of retired galaxies from MaNGA that, if the nuclear regions are ionized by hot, old stellar populations, the trend between log(N/O) and 12+log(O/H) is inverted, i.e., log(N/O) decreases with 12+log(O/H). Moreover, still no characterization has been provided for the chemical abundance radial gradients in those galaxies.

Our results show that the majority of HII regions from the disks of our sample of LINER-like galaxies follow the expected trend between log(N/O) and 12+log(O/H), and the scatter is similar to that observed for the nuclear parts (see \citetalias{Perez-Diaz_2025}). In some cases, the innermost HII regions exhibit log(N/O) abundances which are characteristic of the solar and suprasolar metallicity regimes, but their 12+log(O/H) ratios are even lower than those reported for the outer parts, which can be interpreted as the effect of outflows carrying metal-rich gas \citep[e.g.][]{Villar-Martin_2024} to the outer parts before nitrogen pollution from intermediate massive stars. In other cases, the inner most regions with slightly depressed O/H abundances are similar to the outermost ones, which can be interpreted as the effect of gas inflows from the outer to the inner parts \citep[e.g.][]{Bresolin_2012}, or the effects of galactic fountains \citep[e.g.][]{Spitoni_2013, Tapia-Contreras_2025}.

\subsection{A galaxy model accounting for AGN feed(back)}
\label{ss4.5.4}
Among the different observational scenarios for AGNs, LLAGNs provide a useful laboratory as we can address the effects of the AGN activity in the innermost parts as they are mostly inefficient AGNs \citep{Marquez_2017}. It is beyond the scope of this paper to either provide a theoretical framework or a hydrodynamical simulation that account for the AGNs. However, given the complementary information that we can infer from the simultaneous analysis of 12+log(O/H) and log(N/O) we can propose a general and simplified scheme that might reproduce and explain the observed trends.

The bathtub model to explain galaxy mass assembly and chemical enrichment in the whole galaxy provides feasible explanation for observed properties in large samples of galaxies such as the mass-metallicity relation (MZR), the fundamental metallicity relation (FMR) and the mass-stellar metallicity relation \citep[e.g.][]{Bouche_2010, Lilly_2013, Peng_2014}. Latter on, \citet{Belfiore_2019} brought together the bathtub model and the inside-out scenario \citep[following the approach provided by ][]{Matteucci_1989} to provide a theoretical framework to explain the metallicity gradients of the gas-phase ISM observed in galaxies. They came up with a model with four free parameters: the infall time scale and radial dependence, the star formation efficiency at the center of the galaxy and the outflow mass loading factor.

\begin{figure*}
	\centering
	\includegraphics[width=15cm]{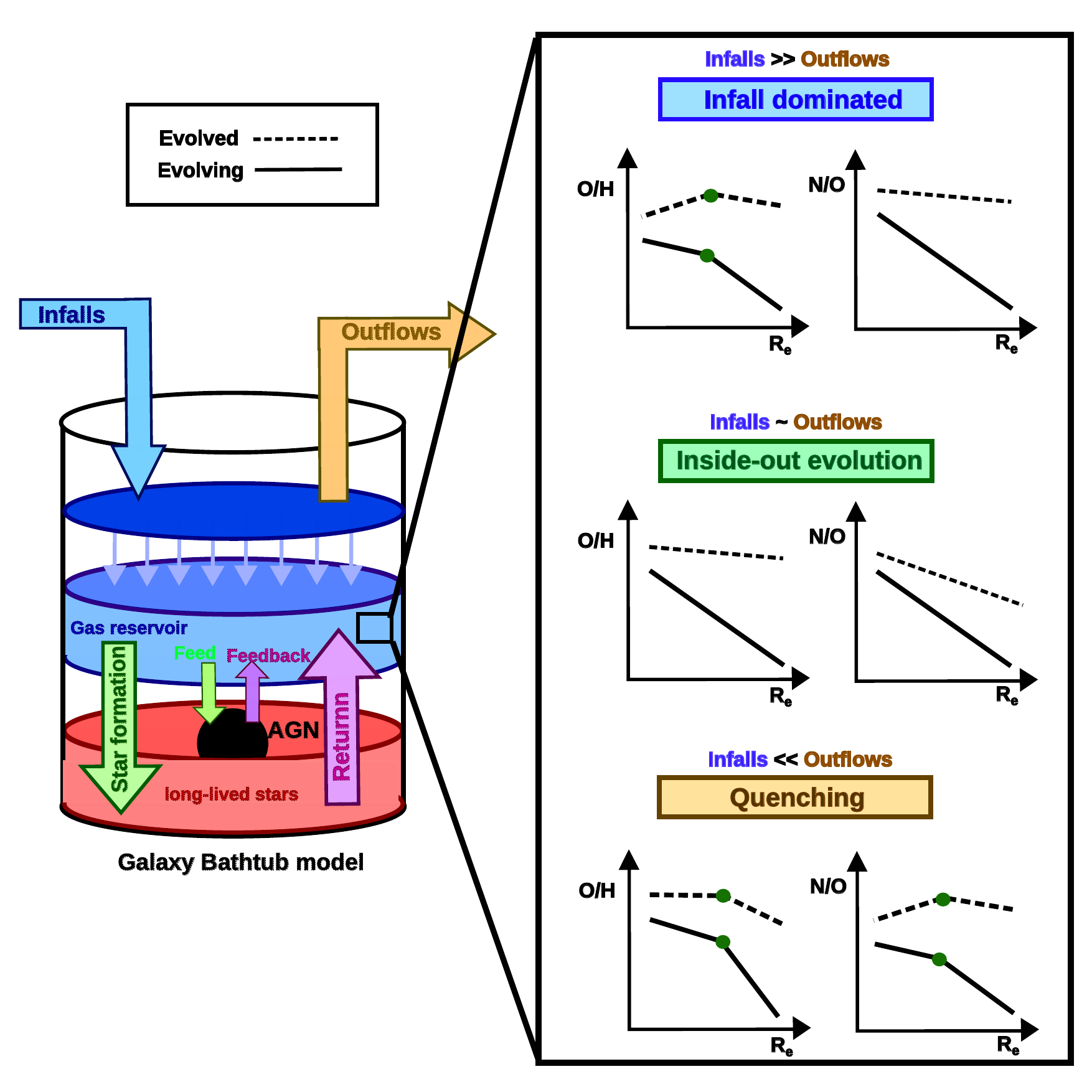}
	\caption[Adaption of the bathtub model to explain radial metallicity gradients in the LINER sample from MaNGA]{Adapted scheme from the bathtub model proposed by \citet{Lilly_2013}. Left part shows the scheme of galaxy flows within the galaxy. Right part shows the expected gas-phase ISM metallicity radial gradients for different scenarios. Continuous tracks of the gradients shows evolving systems, whereas dashed tracks shows the expected behavior for already evolved systems.}
	\label{Bathtub}
\end{figure*}

Theoretical models just by accounting for pure star formation processes are used by  \citet{Belfiore_2019} to reproduce the 12+log(O/H) radial gradients and trends observed by \citet{Belfiore_2017}. Indeed, they were able to reproduce the slight drop found in massive galaxies (\ensuremath{>10^{10.5}} M\ensuremath{_{\odot }}) in the innermost parts, and also positive gradients for low mass galaxies (\ensuremath{<10^{9}} M\ensuremath{_{\odot }}). However, they also warned against the degeneracy of the free parameters of their models as well as on the effects of the calibration used to estimate chemical abundances on the gas-phase ISM.

More recently, based on hydrodynamical simulations, \citet{Tapia-Contreras_2025} obtained departures from the 12+log(O/H) single linear radial profile. By just accounting for the same effects (inflows, outflows and star formation efficiency), they report that some systems experience breaks depending on the dominant mechanism for gas dynamics at a different scales. They detect some inner parts with significant steeper negative slopes than in the outer parts, and they concluded this is due to an increase of the star formation activity due to past infalls of gas. On the other hand, they also report that some galaxies experience an inner drop, i.e., a change in the slope from negative (outer) to positive (inner). They conclude that they are driven by metal-rich outflows, and that they have shorter life times. They also find that galactic fountains as well as gas-inflows might introduce changes in the outer parts of the 12+log(O/H) profile.

Both approaches, the theoretical model by \citet{Belfiore_2019} as well as the predictions from hydrodynamical simulations \citep{Tapia-Contreras_2025} evidence that several factors compete in the shape of metallicity radial gradients, and that the general assumption of a single, negative, profile as predicted by the inside-out scenario might not be accurate in several cases \citep[e.g.][]{Sanchez-Menguiano_2018, Pilyugin_2024}. However, both approaches ignore the influence of AGN activity in their evolutionary scenarios. 

Using as starting point the bathtub model from \citet{Lilly_2013}, we just added an extra component representing the AGN. The expected behavior for the AGN would be similar to the effects of long-lived stars in the closed-box models: a sinking point. As the AGN feeds from gas, that gas would be captured in the closest parts such as the Narrow Line Region (NLR) and the Broad Line Region (BLR), and from there they will eventually feed the SMBH. Part of the gas captured might eventually be ejected to the AGN-driven outflows, similar to the outflows expected in the star-formation that would lead to the long-lived stars. In short, a fraction of the gas would be captured by the AGN (inflows) and part would be returned (outflows). We represent this small re-adaption on Fig. \ref{Bathtub}. 

Therefore, considering all these factors, we propose three main scenarios,
distinguishing in each of them between evolving and already evolved systems, which essentially imply low-mass or massive galaxies:
\begin{itemize}
	\item \textbf{Inside-out}: This scenario is reached when there is equilibrium between the different hydrodynamical processes that affect the gas and allows its secular evolution.
	\begin{itemize}
		\item For evolving systems we expect that both O/H and N/O radial profiles show clear negative slopes. We do not detect any clear example of this scenario in our sample of galaxies, indicating that most of them are already evolved systems.
		\item For already evolved systems we expect a mildly negative or mostly flatten 12+log(O/H) profile for our sample, whereas the log(N/O) gradient shows a steeper profile. In this case, the HII regions in the log(N/O) vs 12+log(O/H) diagram should be located in a very tight range of oxygen abundances, with an increasing N/O as they are located closer to the nuclear part. This is the case of galaxies such as GAL 7990-12704, GAL 8331-6102 or GAL 8984-12705
	\end{itemize}
	In all these cases, we report that the abundances in the NLR of the AGN might not reflect the expected trend as the gas might have been captured at a different epoch (e.g. GAL 7990-12704) or there are hints of inflows that have not yet perturbed 12+log(O/H) (e.g. GAL 8331-6102). 
	\item \textbf{Infall dominated}: This is the expected scenario when inflows dominate the gas dynamics in the inner parts. We also report that it could be the case in which the inflow might shape the whole radial profile as it is the case in GAL 10510-6103 and GAL 8141-6102. 
	\begin{itemize}
		\item For evolving systems we expect a drastic change in the slope of the 12+log(O/H) radial profile, presenting negative slopes in the outer parts and positive slopes in the inner parts. 
		At the same time, the log(N/O) radial profile should reflect the expected behavior from the inside-out scenario or, some small changes in the slope due to the mixing. The position of the HII regions in the log(N/O) vs 12+log(O/H) displays an inverted "c" with the lower tail populated by middle/outer regions and the upper tail populated by inner regions. This is the case for GAL 11838-3794, GAL 11746-9102, GAL 8134-9102 or GAL 8549-3703.
		\item For already evolved systems, we expect that the inner parts of the 12+log(O/H) radial profile shows a positive slope whereas the outer part is mainly characterized by a flat profile. The log(N/O) profile reflects again a mildly negative or almost flat profile. In this scenario, the log(N/O) vs 12+log(O/H) diagram for HII regions shows a vertical relation for the middle/outer regions, and an horizontal branch for the inner regions, with high N/O ratios and a wide range of values for 12+log(O/H). This is the case for GAL 10842-12704, GAL 11945-3704, GAL 12078-12703 or GAL 7495-12704 (among many others).
	\end{itemize}
	Contrary to the inside-out scenario, there seems to be a better agreement between the trends and the abundances estimated in the NLR region of AGN. This would reinforce our proposed scenario, as the infall would feed the AGN and, thus, the metallicity in the NLR should follow the trend.
	\item \textbf{Quenching}: This is the expected scenario for an outflow dominated system. Due to the galactic winds driven by the AGN, extreme star formation processes or both, gas is removed from the inner parts and expelled to the halo (in less massive systems) or the outskirts of the galaxy (for massive systems), and thus, preventing the chemical enrichment of the inner parts due to the suppression of star formation. As the secondary production of nitrogen is delayed with respect to oxygen, it is expected that the effects are less notorious in the log(N/O) gradient as those stars can contaminate the remaining gas.
	\begin{itemize}
		\item For evolving systems, this scenario predicts metallicity gradients with at least one break for both 12+log(O/H) and log(N/O). The slopes for both cases change from negative (outer parts) to positive (inner parts), as the quenching process is preventing the chemical enrichment in the inner parts, whereas the outer parts can be enriched by this material or remain unaffected following the inside-out scenario. However, due to the time delay in nitrogen production, the profile for the log(N/O) gradient might just reflect a change in the absolute value of the slope rather than in the sign. With respect to the log(N/O) vs 12+log(O/H) diagram, HII regions from inner parts and outer parts are expected to coexist in the same range of O values, and with some differentiation in the N/O values (being higher for the inner parts). This is the case for GAL 104519-9102, GAL 1190-6103 or GAL 8258-12704.
		\item For evolved systems, we expect a similar behavior in the log(N/O) radial profile to the evolving system scenario, but 12+log(O/H) should reflect a change from an almost flat gradient (outer parts) to a positive slope (inner parts). The expected behavior in the log(N/O) vs 12+log(O/H) diagram will depend on how much nitrogen can be produced after the gas removal. If little is produced, then a clustering of HII regions in similar positions within the diagram is expected, and with independence of the distance. On the contrary, if the remaining gas (which can be more easily polluted as it is less abundant) is contaminated with nitrogen, then an anti-correlation (or flat distribution) in the log(N/O) vs 12+log(O/H) diagram is expected: inner HII regions would reflect moderate N/O ratios but low O/H abundances, and outer parts would reflect higher O/H values and moderate to low N/O values. This is the case for GAL 10839-12795, GAL 11013-6104, GAL 7964-9102 or GAL 8080-12703 (among many others).
	\end{itemize}
	In this case, the metallicity of the NLR of the AGN might be a footprint of the gas-phase metallicity of the ISM that was expelled, but also it can be driven away by the same AGN-driven outflows, lowering its metallicity.
\end{itemize}

Although we understand that many physical processes at micro- and macro-physical scales play a role in the evolution of metallicity radial gradients, our adapted model helps providing a physical explanation to the observed results in the case of our sample of LINER-like galaxies. The strength of the different processes and the properties of the host galaxy are important actors in determining the scenario. For instance, the quenching (or outflow dominated) scenario might lead to galactic fountains as gas might be captured and accreted again into the galaxy. We suspect this is the case for GAL 10215-3703 or GAL 8078-12703. Both galaxies exhibit in the most outer parts values of 12+log(O/H) and log(N/O) that mimic the inner parts of their galaxies (even the values derived for the NLR of the AGN).

The processes that are shaping the galactic radial gradients, specifically in the inner parts, are mainly inflows and outflows. Evidences for outflows driven by LLAGNs have been reported in the literature \citep{Masegosa_2011, Cazzoli_2018, Cazzoli_2022, Hermosa_2022, Hermosa_2024}. Hot accretion flows are expected to power LLAGNs \citep[e.g.][]{Ho_2008, Marquez_2017}, but reported evidences for such inflows have been provided in counted cases \citep[see the case for M87,][]{Yuan_2022}. On the other hand, powerful inflows and outflows can also be observed in star-forming dominated systems. For instance, \citet{Perez-Diaz_2024} provided indirect evidences of gas inflows diluting metallicity and enhancing star formation in interacting galaxies. Powerful outflows are also expected and reported in galaxies classified as starbursts \citep[see reviews by ][and references therein]{Veilleux_2005, Veilleux_2020}. Very recently, \citet{Tapia-Contreras_2025} have shown that this hydrodynamical processes driven by star formation can drastically affect the shape of metallicity radial gradients finding similar results to ours. Overall, we cannot rule out that the star formation processes are also contributing to the AGN-driven hydrodynamical processes that shapes of metallicity gradients.

\section{Conclusions}
\label{s4.6}
We studied the gas-phase metallicity radial gradients (12+log(O/H) and log(N/O)) in a sample of 97 LINERs from SSDS-IV MaNGA, whose central abundances were previously analyzed in \citetalias{Perez-Diaz_2025}. We select galaxies whose nuclear regions are dominated by LINER-like activity, while the rest of analyzed active regions through their disks are powered by star formation (as also demonstrated by means of the diagnostic diagrams). We used \textsc{HCm} to estimate chemical abundances both in the nuclear and disk regions, allowing us to perform a consistent analysis while keeping track of their differences in the ionizing sources. We applied a piecewise methodology to fit those metallicity gradients allowing breaks as a consequence of changes in the slopes. Our results are summarized as follows:
\begin{enumerate}
	\item The majority of our sample of galaxies exhibit departures from the single linear radial gradient both in 12+log(O/H) and log(N/O) (as would be expected from the inside-out scenario). Particularly, for 12+log(O/H) gradients we found that only 23 galaxies (23.7\%) show the characteristic single linear fit, whereas 48 (49.5\%) exhibit one break and 26 galaxies (26.8\%) exhibit two. In the case of log(N/O) gradient profile, we obtained that 41 (42.3\%) galaxies are well reproduced by the single linear profile, 43 galaxies (44.3\%) exhibit one break and 13 galaxies (13.4\%) two breaks.
	\item We obtained that for a given galaxy, the 12+log(O/H) and log(N/O) radial profiles do not follow the same trend and the positions found for the breaks are different. Moreover, for a given 12+log(O/H) radial profile, it is more likely that the log(N/O) one presents a different shape. This is in general agreement with the expected time delay between oxygen an nitrogen production.
	\item We did not obtain any correlation between the general shape of the metallicity radial profiles  (breaks, slopes, intersects) and stellar mass of galaxies. Only in the outer (middle) parts of galaxies exhibiting one (two) break(s) that are characterized by negative gradients we obtained a very weak anti-correlation as previously reported, although it is not statistically significant.
	\item We propose an adaptation of the bathtub model (accounting for AGN feed and feedback) as a driver for the departures from the inside-out scenario. Infalls supplying gas to the AGN and innermost parts might be the dominant mechanism for galaxies with a broken profile in the 12+log(O/H) radial profile and with an almost unaltered log(N/O) one. On the other hand, outflows removing gas from the inner parts and favoring quenching might explain the broken profiles seen simultaneously in log(N/O) and 12+log(O/H). 
	\item The observational effects of the imbalance between inflows and outflows in the metallicity radial gradients might depend on the evolutionary stage of the galaxy: being mild in galaxies already chemically evolved.
	\item The chemical content of the gas-phase ISM surrounding the AGN (NLR) seems to be more representative of the chemical footprints of the inner parts of the galaxy, whereas the chemical content as derived from the metallicity gradient at the effective radius seems to be more representative of the galaxy as a whole when accounting for global properties such as the mass-metallicity relation.
\end{enumerate}

Our studies employing a robust methodology to analyze a sample of galaxies, in which the AGN activity is not dominant at all scales, show that there is a wide variety of trends for the chemical abundance gradients. Although we propose a toy model accounting for AGN activity to explain the observed behaviors, a comparative analysis with a control sample of galaxies in the same mass range is needed to assess whether AGNs alter or not the shape of the radial metallicity gradients. Such study will be published as a forthcoming work within this series of papers. Additionally, this sample can be later on explored with high-resolution spectroscopic data of the inner parts of the galaxies to provide a more precise quantification on the effects of AGNs at different scales.

\section*{Data availability}
Supplementary online material for this work is available on Zenodo at \url{https://zenodo.org/records/17900866}.

\begin{acknowledgements}
We thank the anonymous referee for the constructive report that improved this manuscript. BPD, EPM and JVM acknowledge support from the Spanish MINECO grant PID2022-136598NB-C32. We also acknowledge financial support from the Severo Ochoa grant CEX2021-001131-S funded by MICIU/AEI/ 10.13039/501100011033. BPD acknowledges support from the Spanish FPI-grant PRE2020-092550. IAZ acknowledges funding from the Deutsche Forschungsgemeinschaft (DFG; German Research Foundation)---project-ID 550945879. PBT acknowledges partial funding by Fondecyt-ANID 1240465/2024, and ANID Basal Project FB210003. This research made use of \textsc{Astropy}, which is a community-developed core Python package for Astronomy \citep{Astropy_2013, Astropy_2018, Astropy_2022}, and other software and packages: \textsc{Numpy} \citep{Walt_2011}, and \textsc{Scipy} \citep{Virtanen_2020}. The plots for this research were created using \textsc{Matplotlib} \citep{Hunter_2007}. We acknowledge the fruitful discussions with our research team.  E.P.M. acknowledges the assistance from his guide dog, Rocko, without whose daily help this work would have been much more difficult.
\end{acknowledgements}

\bibliographystyle{bibtex/aa} 
\bibliography{hcm}

\begin{thebibliography}{137}
\expandafter\ifx\csname natexlab\endcsname\relax\def\natexlab#1{#1}\fi

\bibitem[{{Abdurro'uf} {et~al.}(2022){Abdurro'uf}, {Accetta}, {Aerts}, {Silva
  Aguirre}, {Ahumada}, {Ajgaonkar}, {Filiz Ak}, {Alam}, {Allende Prieto},
  {Almeida}, {Anders}, {Anderson}, {Andrews}, {Anguiano}, {Aquino-Ort{\'\i}z},
  {Arag{\'o}n-Salamanca}, {Argudo-Fern{\'a}ndez}, {Ata}, {Aubert},
  {Avila-Reese}, {Badenes}, {Barb{\'a}}, {Barger}, {Barrera-Ballesteros},
  {Beaton}, {Beers}, {Belfiore}, {Bender}, {Bernardi}, {Bershady}, {Beutler},
  {Bidin}, {Bird}, {Bizyaev}, {Blanc}, {Blanton}, {Boardman}, {Bolton},
  {Boquien}, {Borissova}, {Bovy}, {Brandt}, {Brown}, {Brownstein}, {Brusa},
  {Buchner}, {Bundy}, {Burchett}, {Bureau}, {Burgasser}, {Cabang}, {Campbell},
  {Cappellari}, {Carlberg}, {Wanderley}, {Carrera}, {Cash}, {Chen}, {Chen},
  {Cherinka}, {Chiappini}, {Choi}, {Chojnowski}, {Chung}, {Clerc}, {Cohen},
  {Comerford}, {Comparat}, {da Costa}, {Covey}, {Crane}, {Cruz-Gonzalez},
  {Culhane}, {Cunha}, {Dai}, {Damke}, {Darling}, {Davidson}, {Davies},
  {Dawson}, {De Lee}, {Diamond-Stanic}, {Cano-D{\'\i}az}, {S{\'a}nchez},
  {Donor}, {Duckworth}, {Dwelly}, {Eisenstein}, {Elsworth}, {Emsellem},
  {Eracleous}, {Escoffier}, {Fan}, {Farr}, {Feng}, {Fern{\'a}ndez-Trincado},
  {Feuillet}, {Filipp}, {Fillingham}, {Frinchaboy}, {Fromenteau}, {Galbany},
  {Garc{\'\i}a}, {Garc{\'\i}a-Hern{\'a}ndez}, {Ge}, {Geisler}, {Gelfand},
  {G{\'e}ron}, {Gibson}, {Goddy}, {Godoy-Rivera}, {Grabowski}, {Green},
  {Greener}, {Grier}, {Griffith}, {Guo}, {Guy}, {Hadjara}, {Harding},
  {Hasselquist}, {Hayes}, {Hearty}, {Hern{\'a}ndez}, {Hill}, {Hogg},
  {Holtzman}, {Horta}, {Hsieh}, {Hsu}, {Hsu}, {Huber}, {Huertas-Company},
  {Hutchinson}, {Hwang}, {Ibarra-Medel}, {Chitham}, {Ilha}, {Imig}, {Jaekle},
  {Jayasinghe}, {Ji}, {Johnson}, {Jones}, {J{\"o}nsson}, {Katkov}, {Khalatyan},
  {Kinemuchi}, {Kisku}, {Knapen}, {Kneib}, {Kollmeier}, {Kong}, {Kounkel},
  {Kreckel}, {Krishnarao}, {Lacerna}, {Lane}, {Langgin}, {Lavender}, {Law},
  {Lazarz}, {Leung}, {Leung}, {Lewis}, {Li}, {Li}, {Lian}, {Liang}, {Lin},
  {Lin}, {Lin}, {Lintott}, {Long}, {Longa-Pe{\~n}a}, {L{\'o}pez-Cob{\'a}},
  {Lu}, {Lundgren}, {Luo}, {Mackereth}, {de la Macorra}, {Mahadevan},
  {Majewski}, {Manchado}, {Mandeville}, {Maraston}, {Margalef-Bentabol},
  {Masseron}, {Masters}, {Mathur}, {McDermid}, {Mckay}, {Merloni},
  {Merrifield}, {Meszaros}, {Miglio}, {Di Mille}, {Minniti}, {Minsley}, \&
  {Monachesi}}]{Abdurro_2022}
{Abdurro'uf}, {Accetta}, K., {Aerts}, C., {et~al.} 2022, \apjs, 259, 35

\bibitem[{{Alvarez-Hurtado} {et~al.}(2022){Alvarez-Hurtado},
  {Barrera-Ballesteros}, {S{\'a}nchez}, {Colombo}, {L{\'o}pez-S{\'a}nchez}, \&
  {Aquino-Ort{\'\i}z}}]{Alvarez-Hurtado_2022}
{Alvarez-Hurtado}, P., {Barrera-Ballesteros}, J.~K., {S{\'a}nchez}, S.~F.,
  {et~al.} 2022, \apj, 929, 47

\bibitem[{{Amiri} {et~al.}(2024){Amiri}, {Knapen}, {Comer{\'o}n}, {Marconi}, \&
  {Lehmer}}]{Amiri_2024}
{Amiri}, A., {Knapen}, J.~H., {Comer{\'o}n}, S., {Marconi}, A., \& {Lehmer},
  B.~D. 2024, arXiv e-prints, arXiv:2407.12158

\bibitem[{{Amor{\'\i}n} {et~al.}(2010){Amor{\'\i}n}, {P{\'e}rez-Montero}, \&
  {V{\'\i}lchez}}]{Amorin_2010}
{Amor{\'\i}n}, R.~O., {P{\'e}rez-Montero}, E., \& {V{\'\i}lchez}, J.~M. 2010,
  \apjl, 715, L128

\bibitem[{{Andrews} \& {Martini}(2013)}]{Andrews_2013}
{Andrews}, B.~H. \& {Martini}, P. 2013, \apj, 765, 140

\bibitem[{{Asari} {et~al.}(2007){Asari}, {Cid Fernandes}, {Stasi{\'n}ska},
  {Torres-Papaqui}, {Mateus}, {Sodr{\'e}}, {Schoenell}, \&
  {Gomes}}]{Asari_2007}
{Asari}, N.~V., {Cid Fernandes}, R., {Stasi{\'n}ska}, G., {et~al.} 2007,
  \mnras, 381, 263

\bibitem[{{Asplund} {et~al.}(2009){Asplund}, {Grevesse}, {Sauval}, \&
  {Scott}}]{Asplund_2009}
{Asplund}, M., {Grevesse}, N., {Sauval}, A.~J., \& {Scott}, P. 2009, \araa, 47,
  481

\bibitem[{{Astropy Collaboration} {et~al.}(2022){Astropy Collaboration},
  {Price-Whelan}, {Lim}, {Earl}, {Starkman}, {Bradley}, {Shupe}, {Patil},
  {Corrales}, {Brasseur}, {N{"o}the}, {Donath}, {Tollerud}, {Morris},
  {Ginsburg}, {Vaher}, {Weaver}, {Tocknell}, {Jamieson}, {van Kerkwijk},
  {Robitaille}, {Merry}, {Bachetti}, {G{"u}nther}, {Aldcroft},
  {Alvarado-Montes}, {Archibald}, {B{'o}di}, {Bapat}, {Barentsen}, {Baz{'a}n},
  {Biswas}, {Boquien}, {Burke}, {Cara}, {Cara}, {Conroy}, {Conseil}, {Craig},
  {Cross}, {Cruz}, {D'Eugenio}, {Dencheva}, {Devillepoix}, {Dietrich},
  {Eigenbrot}, {Erben}, {Ferreira}, {Foreman-Mackey}, {Fox}, {Freij}, {Garg},
  {Geda}, {Glattly}, {Gondhalekar}, {Gordon}, {Grant}, {Greenfield}, {Groener},
  {Guest}, {Gurovich}, {Handberg}, {Hart}, {Hatfield-Dodds}, {Homeier},
  {Hosseinzadeh}, {Jenness}, {Jones}, {Joseph}, {Kalmbach}, {Karamehmetoglu},
  {Ka{l}uszy{'n}ski}, {Kelley}, {Kern}, {Kerzendorf}, {Koch}, {Kulumani},
  {Lee}, {Ly}, {Ma}, {MacBride}, {Maljaars}, {Muna}, {Murphy}, {Norman},
  {O'Steen}, {Oman}, {Pacifici}, {Pascual}, {Pascual-Granado}, {Patil},
  {Perren}, {Pickering}, {Rastogi}, {Roulston}, {Ryan}, {Rykoff}, {Sabater},
  {Sakurikar}, {Salgado}, {Sanghi}, {Saunders}, {Savchenko}, {Schwardt},
  {Seifert-Eckert}, {Shih}, {Jain}, {Shukla}, {Sick}, {Simpson},
  {Singanamalla}, {Singer}, {Singhal}, {Sinha}, {Sip{H{o}}cz}, {Spitler},
  {Stansby}, {Streicher}, {{{S}}umak}, {Swinbank}, {Taranu}, {Tewary},
  {Tremblay}, {Val-Borro}, {Van Kooten}, {Vasovi{'c}}, {Verma}, {de Miranda
  Cardoso}, {Williams}, {Wilson}, {Winkel}, {Wood-Vasey}, {Xue}, {Yoachim},
  {Zhang}, {Zonca}, \& {Astropy Project Contributors}}]{Astropy_2022}
{Astropy Collaboration}, {Price-Whelan}, A.~M., {Lim}, P.~L., {et~al.} 2022,
  \apj, 935, 167

\bibitem[{{Astropy Collaboration} {et~al.}(2018){Astropy Collaboration},
  {Price-Whelan}, {Sip{\H{o}}cz}, {G{\"u}nther}, {Lim}, {Crawford}, {Conseil},
  {Shupe}, {Craig}, {Dencheva}, {Ginsburg}, {Vand erPlas}, {Bradley},
  {P{\'e}rez-Su{\'a}rez}, {de Val-Borro}, {Aldcroft}, {Cruz}, {Robitaille},
  {Tollerud}, {Ardelean}, {Babej}, {Bach}, {Bachetti}, {Bakanov}, {Bamford},
  {Barentsen}, {Barmby}, {Baumbach}, {Berry}, {Biscani}, {Boquien}, {Bostroem},
  {Bouma}, {Brammer}, {Bray}, {Breytenbach}, {Buddelmeijer}, {Burke},
  {Calderone}, {Cano Rodr{\'\i}guez}, {Cara}, {Cardoso}, {Cheedella}, {Copin},
  {Corrales}, {Crichton}, {D'Avella}, {Deil}, {Depagne}, {Dietrich}, {Donath},
  {Droettboom}, {Earl}, {Erben}, {Fabbro}, {Ferreira}, {Finethy}, {Fox},
  {Garrison}, {Gibbons}, {Goldstein}, {Gommers}, {Greco}, {Greenfield},
  {Groener}, {Grollier}, {Hagen}, {Hirst}, {Homeier}, {Horton}, {Hosseinzadeh},
  {Hu}, {Hunkeler}, {Ivezi{\'c}}, {Jain}, {Jenness}, {Kanarek}, {Kendrew},
  {Kern}, {Kerzendorf}, {Khvalko}, {King}, {Kirkby}, {Kulkarni}, {Kumar},
  {Lee}, {Lenz}, {Littlefair}, {Ma}, {Macleod}, {Mastropietro}, {McCully},
  {Montagnac}, {Morris}, {Mueller}, {Mumford}, {Muna}, {Murphy}, {Nelson},
  {Nguyen}, {Ninan}, {N{\"o}the}, {Ogaz}, {Oh}, {Parejko}, {Parley}, {Pascual},
  {Patil}, {Patil}, {Plunkett}, {Prochaska}, {Rastogi}, {Reddy Janga},
  {Sabater}, {Sakurikar}, {Seifert}, {Sherbert}, {Sherwood-Taylor}, {Shih},
  {Sick}, {Silbiger}, {Singanamalla}, {Singer}, {Sladen}, {Sooley},
  {Sornarajah}, {Streicher}, {Teuben}, {Thomas}, {Tremblay}, {Turner},
  {Terr{\'o}n}, {van Kerkwijk}, {de la Vega}, {Watkins}, {Weaver}, {Whitmore},
  {Woillez}, {Zabalza}, \& {Astropy Contributors}}]{Astropy_2018}
{Astropy Collaboration}, {Price-Whelan}, A.~M., {Sip{\H{o}}cz}, B.~M., {et~al.}
  2018, \aj, 156, 123

\bibitem[{{Astropy Collaboration} {et~al.}(2013){Astropy Collaboration},
  {Robitaille}, {Tollerud}, {Greenfield}, {Droettboom}, {Bray}, {Aldcroft},
  {Davis}, {Ginsburg}, {Price-Whelan}, {Kerzendorf}, {Conley}, {Crighton},
  {Barbary}, {Muna}, {Ferguson}, {Grollier}, {Parikh}, {Nair}, {Unther},
  {Deil}, {Woillez}, {Conseil}, {Kramer}, {Turner}, {Singer}, {Fox}, {Weaver},
  {Zabalza}, {Edwards}, {Azalee Bostroem}, {Burke}, {Casey}, {Crawford},
  {Dencheva}, {Ely}, {Jenness}, {Labrie}, {Lim}, {Pierfederici}, {Pontzen},
  {Ptak}, {Refsdal}, {Servillat}, \& {Streicher}}]{Astropy_2013}
{Astropy Collaboration}, {Robitaille}, T.~P., {Tollerud}, E.~J., {et~al.} 2013,
  \aap, 558, A33

\bibitem[{{Athanassoula}(1992)}]{Athanassoula_1992}
{Athanassoula}, E. 1992, \mnras, 259, 345

\bibitem[{{Baldwin} {et~al.}(1981){Baldwin}, {Phillips}, \&
  {Terlevich}}]{Baldwin_1981}
{Baldwin}, J.~A., {Phillips}, M.~M., \& {Terlevich}, R. 1981, \pasp, 93, 5

\bibitem[{{Belfiore} {et~al.}(2015){Belfiore}, {Maiolino}, {Bundy}, {Thomas},
  {Maraston}, {Wilkinson}, {S{\'a}nchez}, {Bershady}, {Blanc}, {Bothwell},
  {Cales}, {Coccato}, {Drory}, {Emsellem}, {Fu}, {Gelfand}, {Law}, {Masters},
  {Parejko}, {Tremonti}, {Wake}, {Weijmans}, {Yan}, {Xiao}, {Zhang}, {Zheng},
  {Bizyaev}, {Kinemuchi}, {Oravetz}, \& {Simmons}}]{Belfiore_2015}
{Belfiore}, F., {Maiolino}, R., {Bundy}, K., {et~al.} 2015, \mnras, 449, 867

\bibitem[{{Belfiore} {et~al.}(2017){Belfiore}, {Maiolino}, {Tremonti},
  {S{\'a}nchez}, {Bundy}, {Bershady}, {Westfall}, {Lin}, {Drory}, {Boquien},
  {Thomas}, \& {Brinkmann}}]{Belfiore_2017}
{Belfiore}, F., {Maiolino}, R., {Tremonti}, C., {et~al.} 2017, \mnras, 469, 151

\bibitem[{{Belfiore} {et~al.}(2019){Belfiore}, {Vincenzo}, {Maiolino}, \&
  {Matteucci}}]{Belfiore_2019}
{Belfiore}, F., {Vincenzo}, F., {Maiolino}, R., \& {Matteucci}, F. 2019,
  \mnras, 487, 456

\bibitem[{{Blanton} {et~al.}(2017){Blanton}, {Bershady}, {Abolfathi},
  {Albareti}, {Allende Prieto}, {Almeida}, {Alonso-Garc{\'\i}a}, {Anders},
  {Anderson}, {Andrews}, {Aquino-Ort{\'\i}z}, {Arag{\'o}n-Salamanca},
  {Argudo-Fern{\'a}ndez}, {Armengaud}, {Aubourg}, {Avila-Reese}, {Badenes},
  {Bailey}, {Barger}, {Barrera-Ballesteros}, {Bartosz}, {Bates}, {Baumgarten},
  {Bautista}, {Beaton}, {Beers}, {Belfiore}, {Bender}, {Berlind}, {Bernardi},
  {Beutler}, {Bird}, {Bizyaev}, {Blanc}, {Blomqvist}, {Bolton}, {Boquien},
  {Borissova}, {van den Bosch}, {Bovy}, {Brandt}, {Brinkmann}, {Brownstein},
  {Bundy}, {Burgasser}, {Burtin}, {Busca}, {Cappellari}, {Delgado Carigi},
  {Carlberg}, {Carnero Rosell}, {Carrera}, {Chanover}, {Cherinka}, {Cheung},
  {G{\'o}mez Maqueo Chew}, {Chiappini}, {Choi}, {Chojnowski}, {Chuang},
  {Chung}, {Cirolini}, {Clerc}, {Cohen}, {Comparat}, {da Costa}, {Cousinou},
  {Covey}, {Crane}, {Croft}, {Cruz-Gonzalez}, {Garrido Cuadra}, {Cunha},
  {Damke}, {Darling}, {Davies}, {Dawson}, {de la Macorra}, {Dell'Agli}, {De
  Lee}, {Delubac}, {Di Mille}, {Diamond-Stanic}, {Cano-D{\'\i}az}, {Donor},
  {Downes}, {Drory}, {du Mas des Bourboux}, {Duckworth}, {Dwelly}, {Dyer},
  {Ebelke}, {Eigenbrot}, {Eisenstein}, {Emsellem}, {Eracleous}, {Escoffier},
  {Evans}, {Fan}, {Fern{\'a}ndez-Alvar}, {Fernandez-Trincado}, {Feuillet},
  {Finoguenov}, {Fleming}, {Font-Ribera}, {Fredrickson}, {Freischlad},
  {Frinchaboy}, {Fuentes}, {Galbany}, {Garcia-Dias},
  {Garc{\'\i}a-Hern{\'a}ndez}, {Gaulme}, {Geisler}, {Gelfand},
  {Gil-Mar{\'\i}n}, {Gillespie}, {Goddard}, {Gonzalez-Perez}, {Grabowski},
  {Green}, {Grier}, {Gunn}, {Guo}, {Guy}, {Hagen}, {Hahn}, {Hall}, {Harding},
  {Hasselquist}, {Hawley}, {Hearty}, {Gonzalez Hern{\'a}ndez}, {Ho}, {Hogg},
  {Holley-Bockelmann}, {Holtzman}, {Holzer}, {Huehnerhoff}, {Hutchinson},
  {Hwang}, {Ibarra-Medel}, {da Silva Ilha}, {Ivans}, {Ivory}, {Jackson},
  {Jensen}, {Johnson}, {Jones}, {J{\"o}nsson}, {Jullo}, {Kamble}, {Kinemuchi},
  {Kirkby}, {Kitaura}, {Klaene}, {Knapp}, {Kneib}, {Kollmeier}, {Lacerna},
  {Lane}, {Lang}, {Law}, {Lazarz}, {Lee}, {Le Goff}, {Liang}, {Li}, {Li},
  {Lian}, {Lima}, {Lin}, {Lin}, {Bertran de Lis}, {Liu}, {de Icaza Lizaola},
  {Long}, {Lucatello}, {Lundgren}, {MacDonald}, {Deconto Machado}, {MacLeod},
  {Mahadevan}, {Geimba Maia}, {Maiolino}, {Majewski}, {Malanushenko},
  {Malanushenko}, {Manchado}, {Mao}, {Maraston}, {Marques-Chaves}, {Masseron},
  {Masters}, {McBride}, {McDermid}, {McGrath}, {McGreer}, {Medina Pe{\~n}a},
  {Melendez}, {Merloni}, {Merrifield}, {Meszaros}, {Meza}, {Minchev},
  {Minniti}, {Miyaji}, {More}, {Mulchaey}, {M{\"u}ller-S{\'a}nchez}, {Muna},
  {Munoz}, {Myers}, {Nair}, {Nandra}, {Correa do Nascimento}, {Negrete},
  {Ness}, {Newman}, {Nichol}, {Nidever}, {Nitschelm}, {Ntelis}, {O'Connell},
  {Oelkers}, {Oravetz}, {Oravetz}, {Pace}, {Padilla}, {Palanque-Delabrouille},
  {Alonso Palicio}, {Pan}, {Parejko}, {Parikh}, {P{\^a}ris}, {Park}, {Patten},
  {Peirani}, {Pellejero-Ibanez}, {Penny}, {Percival}, {Perez-Fournon},
  {Petitjean}, {Pieri}, {Pinsonneault}, {Pisani}, {Poleski}, {Prada},
  {Prakash}, {Queiroz}, {Raddick}, {Raichoor}, {Barboza Rembold}, {Richstein},
  {Riffel}, {Riffel}, {Rix}, {Robin}, {Rockosi}, {Rodr{\'\i}guez-Torres},
  {Roman-Lopes}, {Rom{\'a}n-Z{\'u}{\~n}iga}, {Rosado}, {Ross}, {Rossi}, {Ruan},
  {Ruggeri}, {Rykoff}, {Salazar-Albornoz}, {Salvato}, {S{\'a}nchez}, {Aguado},
  {S{\'a}nchez-Gallego}, {Santana}, {Santiago}, {Sayres}, {Schiavon}, {da Silva
  Schimoia}, {Schlafly}, {Schlegel}, {Schneider}, {Schultheis}, {Schuster},
  {Schwope}, {Seo}, {Shao}, {Shen}, {Shetrone}, {Shull}, {Simon}, {Skinner},
  {Skrutskie}, {Slosar}, {Smith}, {Sobeck}, {Sobreira}, {Somers}, {Souto},
  {Stark}, {Stassun}, {Stauffer}, {Steinmetz}, {Storchi-Bergmann},
  {Streblyanska}, {Stringfellow}, {Su{\'a}rez}, {Sun}, {Suzuki}, {Szigeti},
  {Taghizadeh-Popp}, {Tang}, {Tao}, {Tayar}, {Tembe}, {Teske}, {Thakar},
  {Thomas}, {Thompson}, {Tinker}, {Tissera}, {Tojeiro}, {Hernandez Toledo}, {de
  la Torre}, {Tremonti}, {Troup}, {Valenzuela}, {Martinez Valpuesta},
  {Vargas-Gonz{\'a}lez}, {Vargas-Maga{\~n}a}, {Vazquez}, {Villanova}, {Vivek},
  {Vogt}, {Wake}, {Walterbos}, {Wang}, {Weaver}, {Weijmans}, {Weinberg},
  {Westfall}, {Whelan}, {Wild}, {Wilson}, {Wood-Vasey}, {Wylezalek}, {Xiao},
  {Yan}, {Yang}, {Ybarra}, {Y{\`e}che}, {Zakamska}, {Zamora}, {Zarrouk},
  {Zasowski}, {Zhang}, {Zhao}, {Zheng}, {Zheng}, {Zhou}, {Zhou}, {Zhu},
  {Zoccali}, \& {Zou}}]{Blanton_2017}
{Blanton}, M.~R., {Bershady}, M.~A., {Abolfathi}, B., {et~al.} 2017, \aj, 154,
  28

\bibitem[{{Bouch{\'e}} {et~al.}(2010){Bouch{\'e}}, {Dekel}, {Genzel}, {Genel},
  {Cresci}, {F{\"o}rster Schreiber}, {Shapiro}, {Davies}, \&
  {Tacconi}}]{Bouche_2010}
{Bouch{\'e}}, N., {Dekel}, A., {Genzel}, R., {et~al.} 2010, \apj, 718, 1001

\bibitem[{{Bresolin} {et~al.}(2012){Bresolin}, {Kennicutt}, \&
  {Ryan-Weber}}]{Bresolin_2012}
{Bresolin}, F., {Kennicutt}, R.~C., \& {Ryan-Weber}, E. 2012, \apj, 750, 122

\bibitem[{{Bruzual} \& {Charlot}(2003)}]{Bruzual_2003}
{Bruzual}, G. \& {Charlot}, S. 2003, \mnras, 344, 1000

\bibitem[{{Bundy} {et~al.}(2015){Bundy}, {Bershady}, {Law}, {Yan}, {Drory},
  {MacDonald}, {Wake}, {Cherinka}, {S{\'a}nchez-Gallego}, {Weijmans}, {Thomas},
  {Tremonti}, {Masters}, {Coccato}, {Diamond-Stanic}, {Arag{\'o}n-Salamanca},
  {Avila-Reese}, {Badenes}, {Falc{\'o}n-Barroso}, {Belfiore}, {Bizyaev},
  {Blanc}, {Bland-Hawthorn}, {Blanton}, {Brownstein}, {Byler}, {Cappellari},
  {Conroy}, {Dutton}, {Emsellem}, {Etherington}, {Frinchaboy}, {Fu}, {Gunn},
  {Harding}, {Johnston}, {Kauffmann}, {Kinemuchi}, {Klaene}, {Knapen},
  {Leauthaud}, {Li}, {Lin}, {Maiolino}, {Malanushenko}, {Malanushenko}, {Mao},
  {Maraston}, {McDermid}, {Merrifield}, {Nichol}, {Oravetz}, {Pan}, {Parejko},
  {Sanchez}, {Schlegel}, {Simmons}, {Steele}, {Steinmetz}, {Thanjavur},
  {Thompson}, {Tinker}, {van den Bosch}, {Westfall}, {Wilkinson}, {Wright},
  {Xiao}, \& {Zhang}}]{Bundy_2015}
{Bundy}, K., {Bershady}, M.~A., {Law}, D.~R., {et~al.} 2015, \apj, 798, 7

\bibitem[{{Calura} {et~al.}(2008){Calura}, {Pipino}, \&
  {Matteucci}}]{Calura_2008}
{Calura}, F., {Pipino}, A., \& {Matteucci}, F. 2008, \aap, 479, 669

\bibitem[{{Camps-Fari{\~n}a} {et~al.}(2023){Camps-Fari{\~n}a},
  {S{\'a}nchez-Bl{\'a}zquez}, {Roca-F{\`a}brega}, \&
  {S{\'a}nchez}}]{Camps_2023}
{Camps-Fari{\~n}a}, A., {S{\'a}nchez-Bl{\'a}zquez}, P., {Roca-F{\`a}brega}, S.,
  \& {S{\'a}nchez}, S.~F. 2023, \aap, 678, A65

\bibitem[{{Cardoso} {et~al.}(2025){Cardoso}, {Cavichia}, {Moll{\'a}}, \&
  {S{\'a}nchez-Menguiano}}]{Cardoso_2025}
{Cardoso}, A.~F.~S., {Cavichia}, O., {Moll{\'a}}, M., \&
  {S{\'a}nchez-Menguiano}, L. 2025, \apj, 980, 45

\bibitem[{{Carton} {et~al.}(2018){Carton}, {Brinchmann}, {Contini}, {Epinat},
  {Finley}, {Richard}, {Patr{\'\i}cio}, {Schaye}, {Nanayakkara}, {Weilbacher},
  \& {Wisotzki}}]{Carton_2018}
{Carton}, D., {Brinchmann}, J., {Contini}, T., {et~al.} 2018, \mnras, 478, 4293

\bibitem[{{Carvalho} {et~al.}(2020){Carvalho}, {Dors}, {Cardaci}, {H{\"a}gele},
  {Krabbe}, {P{\'e}rez-Montero}, {Monteiro}, {Armah}, \&
  {Freitas-Lemes}}]{Carvalho_2020}
{Carvalho}, S.~P., {Dors}, O.~L., {Cardaci}, M.~V., {et~al.} 2020, \mnras, 492,
  5675

\bibitem[{{Cazzoli} {et~al.}(2022){Cazzoli}, {Hermosa Mu{\~n}oz},
  {M{\'a}rquez}, {Masegosa}, {Castillo-Morales}, {Gil de Paz},
  {Hern{\'a}ndez-Garc{\'\i}a}, {La Franca}, \& {Ramos Almeida}}]{Cazzoli_2022}
{Cazzoli}, S., {Hermosa Mu{\~n}oz}, L., {M{\'a}rquez}, I., {et~al.} 2022, \aap,
  664, A135

\bibitem[{{Cazzoli} {et~al.}(2018){Cazzoli}, {M{\'a}rquez}, {Masegosa}, {del
  Olmo}, {Povi{\'c}}, {Gonz{\'a}lez-Mart{\'\i}n}, {Balmaverde},
  {Hern{\'a}ndez-Garc{\'\i}a}, \& {Garc{\'\i}a-Burillo}}]{Cazzoli_2018}
{Cazzoli}, S., {M{\'a}rquez}, I., {Masegosa}, J., {et~al.} 2018, \mnras, 480,
  1106

\bibitem[{{Chabrier}(2003)}]{Chabrier_2003}
{Chabrier}, G. 2003, \apjl, 586, L133

\bibitem[{{Cid Fernandes} {et~al.}(2005){Cid Fernandes}, {Mateus}, {Sodr{\'e}},
  {Stasi{\'n}ska}, \& {Gomes}}]{Cid-Fernandes_2005}
{Cid Fernandes}, R., {Mateus}, A., {Sodr{\'e}}, L., {Stasi{\'n}ska}, G., \&
  {Gomes}, J.~M. 2005, \mnras, 358, 363

\bibitem[{{Cid Fernandes} {et~al.}(2011){Cid Fernandes}, {Stasi{\'n}ska},
  {Mateus}, \& {Vale Asari}}]{Cid-Fernandes_2011}
{Cid Fernandes}, R., {Stasi{\'n}ska}, G., {Mateus}, A., \& {Vale Asari}, N.
  2011, \mnras, 413, 1687

\bibitem[{{Cid Fernandes} {et~al.}(2010){Cid Fernandes}, {Stasi{\'n}ska},
  {Schlickmann}, {Mateus}, {Vale Asari}, {Schoenell}, \&
  {Sodr{\'e}}}]{Cid-Fernandes_2010}
{Cid Fernandes}, R., {Stasi{\'n}ska}, G., {Schlickmann}, M.~S., {et~al.} 2010,
  \mnras, 403, 1036

\bibitem[{{Coziol} {et~al.}(1999){Coziol}, {Reyes}, {Consid{\`e}re}, {Davoust},
  \& {Contini}}]{Coziol_1999}
{Coziol}, R., {Reyes}, R.~E.~C., {Consid{\`e}re}, S., {Davoust}, E., \&
  {Contini}, T. 1999, \aap, 345, 733

\bibitem[{{Cresci} {et~al.}(2019){Cresci}, {Mannucci}, \&
  {Curti}}]{Cresci_2019}
{Cresci}, G., {Mannucci}, F., \& {Curti}, M. 2019, \aap, 627, A42

\bibitem[{{Cresci} {et~al.}(2010){Cresci}, {Mannucci}, {Maiolino}, {Marconi},
  {Gnerucci}, \& {Magrini}}]{Cresci_2010}
{Cresci}, G., {Mannucci}, F., {Maiolino}, R., {et~al.} 2010, \nat, 467, 811

\bibitem[{{Curti} {et~al.}(2017){Curti}, {Cresci}, {Mannucci}, {Marconi},
  {Maiolino}, \& {Esposito}}]{Curti_2017}
{Curti}, M., {Cresci}, G., {Mannucci}, F., {et~al.} 2017, \mnras, 465, 1384

\bibitem[{{Curti} {et~al.}(2020){Curti}, {Mannucci}, {Cresci}, \&
  {Maiolino}}]{Curti_2020}
{Curti}, M., {Mannucci}, F., {Cresci}, G., \& {Maiolino}, R. 2020, \mnras, 491,
  944

\bibitem[{{do Nascimento} {et~al.}(2022){do Nascimento}, {Dors},
  {Storchi-Bergmann}, {Mallmann}, {Riffel}, {Ilha}, {Riffel}, {Rembold},
  {Deconto-Machado}, {da Costa}, \& {Armah}}]{Nascimento_2022}
{do Nascimento}, J.~C., {Dors}, O.~L., {Storchi-Bergmann}, T., {et~al.} 2022,
  \mnras, 513, 807

\bibitem[{{Dors} {et~al.}(2019){Dors}, {Monteiro}, {Cardaci}, {H{\"a}gele}, \&
  {Krabbe}}]{Dors_2019}
{Dors}, O.~L., {Monteiro}, A.~F., {Cardaci}, M.~V., {H{\"a}gele}, G.~F., \&
  {Krabbe}, A.~C. 2019, \mnras, 486, 5853

\bibitem[{{Feltre} {et~al.}(2016){Feltre}, {Charlot}, \&
  {Gutkin}}]{Feltre_2016}
{Feltre}, A., {Charlot}, S., \& {Gutkin}, J. 2016, \mnras, 456, 3354

\bibitem[{{Ferland} {et~al.}(2017){Ferland}, {Chatzikos}, {Guzm{\'a}n},
  {Lykins}, {van Hoof}, {Williams}, {Abel}, {Badnell}, {Keenan}, {Porter}, \&
  {Stancil}}]{Ferland_2017}
{Ferland}, G.~J., {Chatzikos}, M., {Guzm{\'a}n}, F., {et~al.} 2017, \rmxaa, 53,
  385

\bibitem[{{Fern{\'a}ndez-Ontiveros} {et~al.}(2021){Fern{\'a}ndez-Ontiveros},
  {P{\'e}rez-Montero}, {V{\'\i}lchez}, {Amor{\'\i}n}, \&
  {Spinoglio}}]{Fernandez-Ontiveros_2021}
{Fern{\'a}ndez-Ontiveros}, J.~A., {P{\'e}rez-Montero}, E., {V{\'\i}lchez},
  J.~M., {Amor{\'\i}n}, R., \& {Spinoglio}, L. 2021, \aap, 652, A23

\bibitem[{{Friedli} {et~al.}(1994){Friedli}, {Benz}, \&
  {Kennicutt}}]{Friedli_1994}
{Friedli}, D., {Benz}, W., \& {Kennicutt}, R. 1994, \apjl, 430, L105

\bibitem[{{Goddard} {et~al.}(2017){Goddard}, {Thomas}, {Maraston}, {Westfall},
  {Etherington}, {Riffel}, {Mallmann}, {Zheng}, {Argudo-Fern{\'a}ndez}, {Lian},
  {Bershady}, {Bundy}, {Drory}, {Law}, {Yan}, {Wake}, {Weijmans}, {Bizyaev},
  {Brownstein}, {Lane}, {Maiolino}, {Masters}, {Merrifield}, {Nitschelm},
  {Pan}, {Roman-Lopes}, {Storchi-Bergmann}, \& {Schneider}}]{Goddard_2017}
{Goddard}, D., {Thomas}, D., {Maraston}, C., {et~al.} 2017, \mnras, 466, 4731

\bibitem[{{Gonz{\'a}lez Delgado} {et~al.}(2015){Gonz{\'a}lez Delgado},
  {Garc{\'\i}a-Benito}, {P{\'e}rez}, {Cid Fernandes}, {de Amorim},
  {Cortijo-Ferrero}, {Lacerda}, {L{\'o}pez Fern{\'a}ndez}, {Vale-Asari},
  {S{\'a}nchez}, {Moll{\'a}}, {Ruiz-Lara}, {S{\'a}nchez-Bl{\'a}zquez},
  {Walcher}, {Alves}, {Aguerri}, {Bekerait{\'e}}, {Bland-Hawthorn}, {Galbany},
  {Gallazzi}, {Husemann}, {Iglesias-P{\'a}ramo}, {Kalinova},
  {L{\'o}pez-S{\'a}nchez}, {Marino}, {M{\'a}rquez}, {Masegosa}, {Mast},
  {M{\'e}ndez-Abreu}, {Mendoza}, {del Olmo}, {P{\'e}rez}, {Quirrenbach}, \&
  {Zibetti}}]{Gonzalez-Delgado_2015}
{Gonz{\'a}lez Delgado}, R.~M., {Garc{\'\i}a-Benito}, R., {P{\'e}rez}, E.,
  {et~al.} 2015, \aap, 581, A103

\bibitem[{{Grasha} {et~al.}(2022){Grasha}, {Chen}, {Battisti}, {Acharyya},
  {Ridolfo}, {Poehler}, {Mably}, {Verma}, {Hayward}, {Kharbanda},
  {Poetrodjojo}, {Seibert}, {Rich}, {Madore}, \& {Kewley}}]{Grasha_2022}
{Grasha}, K., {Chen}, Q.~H., {Battisti}, A.~J., {et~al.} 2022, \apj, 929, 118

\bibitem[{{H{\"a}gele} {et~al.}(2008){H{\"a}gele}, {D{\'\i}az}, {Terlevich},
  {Terlevich}, {P{\'e}rez-Montero}, \& {Cardaci}}]{Hagele_2008}
{H{\"a}gele}, G.~F., {D{\'\i}az}, {\'A}.~I., {Terlevich}, E., {et~al.} 2008,
  \mnras, 383, 209

\bibitem[{{Henry} {et~al.}(2000){Henry}, {Edmunds}, \&
  {K{\"o}ppen}}]{Henry_2000}
{Henry}, R.~B.~C., {Edmunds}, M.~G., \& {K{\"o}ppen}, J. 2000, \apj, 541, 660

\bibitem[{{Hermosa Mu{\~n}oz} {et~al.}(2024){Hermosa Mu{\~n}oz}, {Cazzoli},
  {M{\'a}rquez}, {Masegosa}, {Chamorro-Cazorla}, {Gil de Paz},
  {Castillo-Morales}, {Gallego}, {Carrasco}, {Iglesias-P{\'a}ramo},
  {Garc{\'\i}a-Vargas}, {G{\'o}mez-{\'A}lvarez}, {Pascual},
  {P{\'e}rez-Calpena}, \& {Cardiel}}]{Hermosa_2024}
{Hermosa Mu{\~n}oz}, L., {Cazzoli}, S., {M{\'a}rquez}, I., {et~al.} 2024, \aap,
  683, A43

\bibitem[{{Hermosa Mu{\~n}oz} {et~al.}(2022){Hermosa Mu{\~n}oz}, {M{\'a}rquez},
  {Cazzoli}, {Masegosa}, \& {Ag{\'\i}s-Gonz{\'a}lez}}]{Hermosa_2022}
{Hermosa Mu{\~n}oz}, L., {M{\'a}rquez}, I., {Cazzoli}, S., {Masegosa}, J., \&
  {Ag{\'\i}s-Gonz{\'a}lez}, B. 2022, \aap, 660, A133

\bibitem[{{Ho}(2008)}]{Ho_2008}
{Ho}, L.~C. 2008, \araa, 46, 475

\bibitem[{Howarth(1983)}]{Howarth_1983}
Howarth, I.~D. 1983, Monthly Notices of the Royal Astronomical Society, 203,
  301

\bibitem[{{Hunter}(2007)}]{Hunter_2007}
{Hunter}, J.~D. 2007, Computing in Science and Engineering, 9, 90

\bibitem[{{Kauffmann} {et~al.}(2003){Kauffmann}, {Heckman}, {Tremonti},
  {Brinchmann}, {Charlot}, {White}, {Ridgway}, {Brinkmann}, {Fukugita}, {Hall},
  {Ivezi{\'c}}, {Richards}, \& {Schneider}}]{Kauffmann_2003}
{Kauffmann}, G., {Heckman}, T.~M., {Tremonti}, C., {et~al.} 2003, \mnras, 346,
  1055

\bibitem[{{Kewley} {et~al.}(2006){Kewley}, {Groves}, {Kauffmann}, \&
  {Heckman}}]{Kewley_2006}
{Kewley}, L.~J., {Groves}, B., {Kauffmann}, G., \& {Heckman}, T. 2006, \mnras,
  372, 961

\bibitem[{{Kobayashi} {et~al.}(2020){Kobayashi}, {Karakas}, \&
  {Lugaro}}]{Kobayashi_2020}
{Kobayashi}, C., {Karakas}, A.~I., \& {Lugaro}, M. 2020, \apj, 900, 179

\bibitem[{{Kobayashi} {et~al.}(2007){Kobayashi}, {Springel}, \&
  {White}}]{Kobayashi_2007}
{Kobayashi}, C., {Springel}, V., \& {White}, S. D.~M. 2007, \mnras, 376, 1465

\bibitem[{{K{\"o}ppen} \& {Edmunds}(1999)}]{Koppen_1999}
{K{\"o}ppen}, J. \& {Edmunds}, M.~G. 1999, \mnras, 306, 317

\bibitem[{{Krabbe} {et~al.}(2021){Krabbe}, {Oliveira}, {Zinchenko},
  {Hern{\'a}ndez-Jim{\'e}nez}, {Dors}, {H{\"a}gele}, {Cardaci}, \&
  {Telles}}]{Krabbe_2021}
{Krabbe}, A.~C., {Oliveira}, C.~B., {Zinchenko}, I.~A., {et~al.} 2021, \mnras,
  505, 2087

\bibitem[{{Kreckel} {et~al.}(2019){Kreckel}, {Ho}, {Blanc}, {Groves},
  {Santoro}, {Schinnerer}, {Bigiel}, {Chevance}, {Congiu}, {Emsellem}, {Faesi},
  {Glover}, {Grasha}, {Kruijssen}, {Lang}, {Leroy}, {Meidt}, {McElroy}, {Pety},
  {Rosolowsky}, {Saito}, {Sandstrom}, {Sanchez-Blazquez}, \&
  {Schruba}}]{Kreckel_2019}
{Kreckel}, K., {Ho}, I.-T., {Blanc}, G.~A., {et~al.} 2019, \apj, 887, 80

\bibitem[{{Law} {et~al.}(2016){Law}, {Cherinka}, {Yan}, {Andrews}, {Bershady},
  {Bizyaev}, {Blanc}, {Blanton}, {Bolton}, {Brownstein}, {Bundy}, {Chen},
  {Drory}, {D'Souza}, {Fu}, {Jones}, {Kauffmann}, {MacDonald}, {Masters},
  {Newman}, {Parejko}, {S{\'a}nchez-Gallego}, {S{\'a}nchez}, {Schlegel},
  {Thomas}, {Wake}, {Weijmans}, {Westfall}, \& {Zhang}}]{Law_2016}
{Law}, D.~R., {Cherinka}, B., {Yan}, R., {et~al.} 2016, \aj, 152, 83

\bibitem[{{Lequeux} {et~al.}(1979){Lequeux}, {Peimbert}, {Rayo}, {Serrano}, \&
  {Torres-Peimbert}}]{Lequeux_1979}
{Lequeux}, J., {Peimbert}, M., {Rayo}, J.~F., {Serrano}, A., \&
  {Torres-Peimbert}, S. 1979, \aap, 500, 145

\bibitem[{{Lia} {et~al.}(2002){Lia}, {Portinari}, \& {Carraro}}]{Lia_2002}
{Lia}, C., {Portinari}, L., \& {Carraro}, G. 2002, \mnras, 330, 821

\bibitem[{{Lilly} {et~al.}(2013){Lilly}, {Carollo}, {Pipino}, {Renzini}, \&
  {Peng}}]{Lilly_2013}
{Lilly}, S.~J., {Carollo}, C.~M., {Pipino}, A., {Renzini}, A., \& {Peng}, Y.
  2013, \apj, 772, 119

\bibitem[{{Maiolino} \& {Mannucci}(2019)}]{Maiolino_2019}
{Maiolino}, R. \& {Mannucci}, F. 2019, \aapr, 27, 3

\bibitem[{{M{\'a}rquez} {et~al.}(2017){M{\'a}rquez}, {Masegosa},
  {Gonz{\'a}lez-Martin}, {Hern{\'a}ndez-Garcia}, {Povi{\'c}}, {Netzer},
  {Cazzoli}, \& {del Olmo}}]{Marquez_2017}
{M{\'a}rquez}, I., {Masegosa}, J., {Gonz{\'a}lez-Martin}, O., {et~al.} 2017,
  Frontiers in Astronomy and Space Sciences, 4, 34

\bibitem[{{Masegosa} {et~al.}(2011){Masegosa}, {M{\'a}rquez}, {Ramirez}, \&
  {Gonz{\'a}lez-Mart{\'\i}n}}]{Masegosa_2011}
{Masegosa}, J., {M{\'a}rquez}, I., {Ramirez}, A., \&
  {Gonz{\'a}lez-Mart{\'\i}n}, O. 2011, \aap, 527, A23

\bibitem[{{Mateus} {et~al.}(2006){Mateus}, {Sodr{\'e}}, {Cid Fernandes},
  {Stasi{\'n}ska}, {Schoenell}, \& {Gomes}}]{Mateus_2006}
{Mateus}, A., {Sodr{\'e}}, L., {Cid Fernandes}, R., {et~al.} 2006, \mnras, 370,
  721

\bibitem[{{Matteucci} \& {Francois}(1989)}]{Matteucci_1989}
{Matteucci}, F. \& {Francois}, P. 1989, \mnras, 239, 885

\bibitem[{{Moll{\'a}} {et~al.}(2009){Moll{\'a}}, {Garc{\'\i}a-Vargas}, \&
  {Bressan}}]{Molla_2009}
{Moll{\'a}}, M., {Garc{\'\i}a-Vargas}, M.~L., \& {Bressan}, A. 2009, \mnras,
  398, 451

\bibitem[{{Moll{\'a}} {et~al.}(2006){Moll{\'a}}, {V{\'\i}lchez}, {Gavil{\'a}n},
  \& {D{\'\i}az}}]{Molla_2006}
{Moll{\'a}}, M., {V{\'\i}lchez}, J.~M., {Gavil{\'a}n}, M., \& {D{\'\i}az},
  A.~I. 2006, \mnras, 372, 1069

\bibitem[{{Montuori} {et~al.}(2010){Montuori}, {Di Matteo}, {Lehnert},
  {Combes}, \& {Semelin}}]{Montuori_2010}
{Montuori}, M., {Di Matteo}, P., {Lehnert}, M.~D., {Combes}, F., \& {Semelin},
  B. 2010, \aap, 518, A56

\bibitem[{{Mosconi} {et~al.}(2001){Mosconi}, {Tissera}, {Lambas}, \&
  {Cora}}]{Mosconi_2001}
{Mosconi}, M.~B., {Tissera}, P.~B., {Lambas}, D.~G., \& {Cora}, S.~A. 2001,
  \mnras, 325, 34

\bibitem[{{Mu{\~n}oz-Mateos} {et~al.}(2007){Mu{\~n}oz-Mateos}, {Gil de Paz},
  {Boissier}, {Zamorano}, {Jarrett}, {Gallego}, \&
  {Madore}}]{Munoz-Mateos_2007}
{Mu{\~n}oz-Mateos}, J.~C., {Gil de Paz}, A., {Boissier}, S., {et~al.} 2007,
  \apj, 658, 1006

\bibitem[{{Oliveira} {et~al.}(2024){Oliveira}, {Krabbe}, {Dors}, {Zinchenko},
  {Hernandez-Jimenez}, {Cardaci}, {H{\"a}gele}, \& {Ilha}}]{Oliveira_2024}
{Oliveira}, C.~B., {Krabbe}, A.~C., {Dors}, O.~L., {et~al.} 2024, \mnras, 531,
  199

\bibitem[{{Osterbrock} \& {Ferland}(2006)}]{Osterbrock_book}
{Osterbrock}, D.~E. \& {Ferland}, G.~J. 2006, {Astrophysics of gaseous nebulae
  and active galactic nuclei}

\bibitem[{{Pagel} \& {Patchett}(1975)}]{Pagel_1975}
{Pagel}, B.~E.~J. \& {Patchett}, B.~E. 1975, \mnras, 172, 13

\bibitem[{{Peimbert}(1967)}]{Peimbert_1967}
{Peimbert}, M. 1967, \apj, 150, 825

\bibitem[{{Peimbert} {et~al.}(2007){Peimbert}, {Luridiana}, \&
  {Peimbert}}]{Peimbert_2007}
{Peimbert}, M., {Luridiana}, V., \& {Peimbert}, A. 2007, \apj, 666, 636

\bibitem[{{Peng} \& {Maiolino}(2014)}]{Peng_2014}
{Peng}, Y.-j. \& {Maiolino}, R. 2014, \mnras, 443, 3643

\bibitem[{{Perez} {et~al.}(2011){Perez}, {Michel-Dansac}, \&
  {Tissera}}]{Perez_2011}
{Perez}, J., {Michel-Dansac}, L., \& {Tissera}, P.~B. 2011, \mnras, 417, 580

\bibitem[{{P{\'e}rez-D{\'\i}az} {et~al.}(2021){P{\'e}rez-D{\'\i}az},
  {Masegosa}, {M{\'a}rquez}, \& {P{\'e}rez-Montero}}]{Perez-Diaz_2021}
{P{\'e}rez-D{\'\i}az}, B., {Masegosa}, J., {M{\'a}rquez}, I., \&
  {P{\'e}rez-Montero}, E. 2021, \mnras, 505, 4289

\bibitem[{{P{\'e}rez-D{\'\i}az} {et~al.}(2022){P{\'e}rez-D{\'\i}az},
  {P{\'e}rez-Montero}, {Fern{\'a}ndez-Ontiveros}, \&
  {V{\'\i}lchez}}]{Perez-Diaz_2022}
{P{\'e}rez-D{\'\i}az}, B., {P{\'e}rez-Montero}, E., {Fern{\'a}ndez-Ontiveros},
  J.~A., \& {V{\'\i}lchez}, J.~M. 2022, \aap, 666, A115

\bibitem[{{P{\'e}rez-D{\'\i}az} {et~al.}(2024){P{\'e}rez-D{\'\i}az},
  {P{\'e}rez-Montero}, {Fern{\'a}ndez-Ontiveros}, {V{\'\i}lchez}, \&
  {Amor{\'\i}n}}]{Perez-Diaz_2024}
{P{\'e}rez-D{\'\i}az}, B., {P{\'e}rez-Montero}, E., {Fern{\'a}ndez-Ontiveros},
  J.~A., {V{\'\i}lchez}, J.~M., \& {Amor{\'\i}n}, R. 2024, Nature Astronomy, 8,
  368

\bibitem[{{P{\'e}rez-D{\'\i}az} {et~al.}(2025){P{\'e}rez-D{\'\i}az},
  {P{\'e}rez-Montero}, {Zinchenko}, \& {V{\'\i}lchez}}]{Perez-Diaz_2025}
{P{\'e}rez-D{\'\i}az}, B., {P{\'e}rez-Montero}, E., {Zinchenko}, I.~A., \&
  {V{\'\i}lchez}, J.~M. 2025, \aap, 694, A18

\bibitem[{{P{\'e}rez-Montero}(2014)}]{Perez-Montero_2014}
{P{\'e}rez-Montero}, E. 2014, \mnras, 441, 2663

\bibitem[{{P{\'e}rez-Montero} \& {Amor{\'\i}n}(2017)}]{Perez-Montero_2017}
{P{\'e}rez-Montero}, E. \& {Amor{\'\i}n}, R. 2017, \mnras, 467, 1287

\bibitem[{{P{\'e}rez-Montero} {et~al.}(2023){P{\'e}rez-Montero}, {Amor{\'\i}n},
  {P{\'e}rez-D{\'\i}az}, {V{\'\i}lchez}, \&
  {Garc{\'\i}a-Benito}}]{Perez-Montero_2023}
{P{\'e}rez-Montero}, E., {Amor{\'\i}n}, R., {P{\'e}rez-D{\'\i}az}, B.,
  {V{\'\i}lchez}, J.~M., \& {Garc{\'\i}a-Benito}, R. 2023, \mnras, 521, 1556

\bibitem[{{P{\'e}rez-Montero} \& {Contini}(2009)}]{Perez-Montero_2009}
{P{\'e}rez-Montero}, E. \& {Contini}, T. 2009, \mnras, 398, 949

\bibitem[{{P{\'e}rez-Montero} \& {D{\'\i}az}(2003)}]{Perez-Montero_2003}
{P{\'e}rez-Montero}, E. \& {D{\'\i}az}, A.~I. 2003, \mnras, 346, 105

\bibitem[{{P{\'e}rez-Montero} {et~al.}(2019){P{\'e}rez-Montero}, {Dors},
  {V{\'\i}lchez}, {Garc{\'\i}a-Benito}, {Cardaci}, \&
  {H{\"a}gele}}]{Perez-Montero_2019}
{P{\'e}rez-Montero}, E., {Dors}, O.~L., {V{\'\i}lchez}, J.~M., {et~al.} 2019,
  \mnras, 489, 2652

\bibitem[{{P{\'e}rez-Montero} {et~al.}(2025){P{\'e}rez-Montero},
  {Fern{\'a}ndez-Ontiveros}, {P{\'e}rez-D{\'\i}az}, {V{\'\i}lchez}, \&
  {Amor{\'\i}n}}]{Perez-Montero_2025}
{P{\'e}rez-Montero}, E., {Fern{\'a}ndez-Ontiveros}, J.~A.,
  {P{\'e}rez-D{\'\i}az}, B., {V{\'\i}lchez}, J.~M., \& {Amor{\'\i}n}, R. 2025,
  arXiv e-prints, arXiv:2503.09267

\bibitem[{{P{\'e}rez-Montero} {et~al.}(2016){P{\'e}rez-Montero},
  {Garc{\'\i}a-Benito}, {V{\'\i}lchez}, {S{\'a}nchez}, {Kehrig}, {Husemann},
  {Duarte Puertas}, {Iglesias-P{\'a}ramo}, {Galbany}, {Moll{\'a}}, {Walcher},
  {Ascas{\'\i}bar}, {Gonz{\'a}lez Delgado}, {Marino}, {Masegosa}, {P{\'e}rez},
  {Rosales-Ortega}, {S{\'a}nchez-Bl{\'a}zquez}, {Bland-Hawthorn}, {Bomans},
  {L{\'o}pez-S{\'a}nchez}, {Ziegler}, \& {CALIFA
  Collaboration}}]{Perez-Montero_2016}
{P{\'e}rez-Montero}, E., {Garc{\'\i}a-Benito}, R., {V{\'\i}lchez}, J.~M.,
  {et~al.} 2016, \aap, 595, A62

\bibitem[{{Pilyugin} \& {Grebel}(2016)}]{Pilyugin_2016}
{Pilyugin}, L.~S. \& {Grebel}, E.~K. 2016, \mnras, 457, 3678

\bibitem[{{Pilyugin} \& {Tautvai{\v{s}}ien{\.{e}}}(2024)}]{Pilyugin_2024}
{Pilyugin}, L.~S. \& {Tautvai{\v{s}}ien{\.{e}}}, G. 2024, \aap, 682, A41

\bibitem[{{Pilyugin} {et~al.}(2004){Pilyugin}, {V{\'\i}lchez}, \&
  {Contini}}]{Pilyugin_2004}
{Pilyugin}, L.~S., {V{\'\i}lchez}, J.~M., \& {Contini}, T. 2004, \aap, 425, 849

\bibitem[{{Poetrodjojo} {et~al.}(2018){Poetrodjojo}, {Groves}, {Kewley},
  {Medling}, {Sweet}, {van de Sande}, {Sanchez}, {Bland-Hawthorn}, {Brough},
  {Bryant}, {Cortese}, {Croom}, {L{\'o}pez-S{\'a}nchez}, {Richards}, {Zafar},
  {Lawrence}, {Lorente}, {Owers}, \& {Scott}}]{Poetrodjojo_2018}
{Poetrodjojo}, H., {Groves}, B., {Kewley}, L.~J., {et~al.} 2018, \mnras, 479,
  5235

\bibitem[{{Rich} {et~al.}(2012){Rich}, {Torrey}, {Kewley}, {Dopita}, \&
  {Rupke}}]{Rich_2012}
{Rich}, J.~A., {Torrey}, P., {Kewley}, L.~J., {Dopita}, M.~A., \& {Rupke},
  D.~S.~N. 2012, \apj, 753, 5

\bibitem[{{Rupke} {et~al.}(2010){Rupke}, {Kewley}, \& {Barnes}}]{Rupke_2010}
{Rupke}, D. S.~N., {Kewley}, L.~J., \& {Barnes}, J.~E. 2010, \apjl, 710, L156

\bibitem[{{Ryder}(1995)}]{Ryder_1995}
{Ryder}, S.~D. 1995, \apj, 444, 610

\bibitem[{{S{\'a}nchez} {et~al.}(2015){S{\'a}nchez}, {Galbany}, {P{\'e}rez},
  {S{\'a}nchez-Bl{\'a}zquez}, {Falc{\'o}n-Barroso}, {Rosales-Ortega},
  {S{\'a}nchez-Menguiano}, {Marino}, {Kuncarayakti}, {Anderson}, {Kruehler},
  {Cano-D{\'\i}az}, {Barrera-Ballesteros}, \&
  {Gonz{\'a}lez-Gonz{\'a}lez}}]{Sanchez_2015}
{S{\'a}nchez}, S.~F., {Galbany}, L., {P{\'e}rez}, E., {et~al.} 2015, \aap, 573,
  A105

\bibitem[{{S{\'a}nchez} {et~al.}(2014){S{\'a}nchez}, {Rosales-Ortega},
  {Iglesias-P{\'a}ramo}, {Moll{\'a}}, {Barrera-Ballesteros}, {Marino},
  {P{\'e}rez}, {S{\'a}nchez-Blazquez}, {Gonz{\'a}lez Delgado}, {Cid Fernandes},
  {de Lorenzo-C{\'a}ceres}, {Mendez-Abreu}, {Galbany}, {Falcon-Barroso},
  {Miralles-Caballero}, {Husemann}, {Garc{\'\i}a-Benito}, {Mast}, {Walcher},
  {Gil de Paz}, {Garc{\'\i}a-Lorenzo}, {Jungwiert}, {V{\'\i}lchez},
  {J{\'\i}lkov{\'a}}, {Lyubenova}, {Cortijo-Ferrero}, {D{\'\i}az}, {Wisotzki},
  {M{\'a}rquez}, {Bland-Hawthorn}, {Ellis}, {van de Ven}, {Jahnke},
  {Papaderos}, {Gomes}, {Mendoza}, \& {L{\'o}pez-S{\'a}nchez}}]{Sanchez_2014}
{S{\'a}nchez}, S.~F., {Rosales-Ortega}, F.~F., {Iglesias-P{\'a}ramo}, J.,
  {et~al.} 2014, \aap, 563, A49

\bibitem[{{S{\'a}nchez} {et~al.}(2013){S{\'a}nchez}, {Rosales-Ortega},
  {Jungwiert}, {Iglesias-P{\'a}ramo}, {V{\'\i}lchez}, {Marino}, {Walcher},
  {Husemann}, {Mast}, {Monreal-Ibero}, {Cid Fernandes}, {P{\'e}rez},
  {Gonz{\'a}lez Delgado}, {Garc{\'\i}a-Benito}, {Galbany}, {van de Ven},
  {Jahnke}, {Flores}, {Bland-Hawthorn}, {L{\'o}pez-S{\'a}nchez}, {Stanishev},
  {Miralles-Caballero}, {D{\'\i}az}, {S{\'a}nchez-Blazquez}, {Moll{\'a}},
  {Gallazzi}, {Papaderos}, {Gomes}, {Gruel}, {P{\'e}rez}, {Ruiz-Lara},
  {Florido}, {de Lorenzo-C{\'a}ceres}, {Mendez-Abreu}, {Kehrig}, {Roth},
  {Ziegler}, {Alves}, {Wisotzki}, {Kupko}, {Quirrenbach}, {Bomans}, \& {CALIFA
  Collaboration}}]{Sanchez_2013}
{S{\'a}nchez}, S.~F., {Rosales-Ortega}, F.~F., {Jungwiert}, B., {et~al.} 2013,
  \aap, 554, A58

\bibitem[{{S{\'a}nchez-Bl{\'a}zquez} {et~al.}(2009){S{\'a}nchez-Bl{\'a}zquez},
  {Courty}, {Gibson}, \& {Brook}}]{Sanchez-Blazquez_2009}
{S{\'a}nchez-Bl{\'a}zquez}, P., {Courty}, S., {Gibson}, B.~K., \& {Brook},
  C.~B. 2009, \mnras, 398, 591

\bibitem[{{S{\'a}nchez-Menguiano} {et~al.}(2016){S{\'a}nchez-Menguiano},
  {S{\'a}nchez}, {P{\'e}rez}, {Garc{\'\i}a-Benito}, {Husemann}, {Mast},
  {Mendoza}, {Ruiz-Lara}, {Ascasibar}, {Bland-Hawthorn}, {Cavichia},
  {D{\'\i}az}, {Florido}, {Galbany}, {G{\'o}nzalez Delgado}, {Kehrig},
  {Marino}, {M{\'a}rquez}, {Masegosa}, {M{\'e}ndez-Abreu}, {Moll{\'a}}, {Del
  Olmo}, {P{\'e}rez}, {S{\'a}nchez-Bl{\'a}zquez}, {Stanishev}, {Walcher},
  {L{\'o}pez-S{\'a}nchez}, \& {CALIFA Collaboration}}]{Sanchez-Menguiano_2016}
{S{\'a}nchez-Menguiano}, L., {S{\'a}nchez}, S.~F., {P{\'e}rez}, I., {et~al.}
  2016, \aap, 587, A70

\bibitem[{{S{\'a}nchez-Menguiano} {et~al.}(2018){S{\'a}nchez-Menguiano},
  {S{\'a}nchez}, {P{\'e}rez}, {Ruiz-Lara}, {Galbany}, {Anderson},
  {Kr{\"u}hler}, {Kuncarayakti}, \& {Lyman}}]{Sanchez-Menguiano_2018}
{S{\'a}nchez-Menguiano}, L., {S{\'a}nchez}, S.~F., {P{\'e}rez}, I., {et~al.}
  2018, \aap, 609, A119

\bibitem[{{S{\'a}nchez-Menguiano} {et~al.}(2024){S{\'a}nchez-Menguiano},
  {S{\'a}nchez Almeida}, {S{\'a}nchez}, \&
  {Mu{\~n}oz-Tu{\~n}{\'o}n}}]{Sanchez-Menguiano_2024}
{S{\'a}nchez-Menguiano}, L., {S{\'a}nchez Almeida}, J., {S{\'a}nchez}, S.~F.,
  \& {Mu{\~n}oz-Tu{\~n}{\'o}n}, C. 2024, \aap, 681, A121

\bibitem[{{Sharda} {et~al.}(2024){Sharda}, {Ginzburg}, {Krumholz}, {Forbes},
  {Wisnioski}, {Mingozzi}, {Zovaro}, \& {Dekel}}]{Sharda_2024}
{Sharda}, P., {Ginzburg}, O., {Krumholz}, M.~R., {et~al.} 2024, \mnras, 528,
  2232

\bibitem[{{Sharda} {et~al.}(2021){Sharda}, {Krumholz}, {Wisnioski}, {Forbes},
  {Federrath}, \& {Acharyya}}]{Sharda_2021}
{Sharda}, P., {Krumholz}, M.~R., {Wisnioski}, E., {et~al.} 2021, \mnras, 502,
  5935

\bibitem[{{Sillero} {et~al.}(2017){Sillero}, {Tissera}, {Lambas}, \&
  {Michel-Dansac}}]{Sillero_2017}
{Sillero}, E., {Tissera}, P.~B., {Lambas}, D.~G., \& {Michel-Dansac}, L. 2017,
  \mnras, 472, 4404

\bibitem[{{Spindler} {et~al.}(2018){Spindler}, {Wake}, {Belfiore}, {Bershady},
  {Bundy}, {Drory}, {Masters}, {Thomas}, {Westfall}, \& {Wild}}]{Spindler_2021}
{Spindler}, A., {Wake}, D., {Belfiore}, F., {et~al.} 2018, \mnras, 476, 580

\bibitem[{{Spitoni} {et~al.}(2021){Spitoni}, {Calura}, {Silva Aguirre}, \&
  {Gilli}}]{Spitoni_2021}
{Spitoni}, E., {Calura}, F., {Silva Aguirre}, V., \& {Gilli}, R. 2021, \aap,
  648, L5

\bibitem[{{Spitoni} {et~al.}(2019){Spitoni}, {Cescutti}, {Minchev},
  {Matteucci}, {Silva Aguirre}, {Martig}, {Bono}, \&
  {Chiappini}}]{Spitoni_2019}
{Spitoni}, E., {Cescutti}, G., {Minchev}, I., {et~al.} 2019, \aap, 628, A38

\bibitem[{{Spitoni} {et~al.}(2013){Spitoni}, {Matteucci}, \&
  {Marcon-Uchida}}]{Spitoni_2013}
{Spitoni}, E., {Matteucci}, F., \& {Marcon-Uchida}, M.~M. 2013, \aap, 551, A123

\bibitem[{{Tapia-Contreras} {et~al.}(2025){Tapia-Contreras}, {Tissera},
  {Sillero}, {Gonzalez-Jara}, {Casanueva-Villarreal}, {Pedrosa}, {Bignone},
  {Padilla}, \& {Dom{\'\i}nguez-Tenreiro}}]{Tapia-Contreras_2025}
{Tapia-Contreras}, B., {Tissera}, P.~B., {Sillero}, E., {et~al.} 2025, arXiv
  e-prints, arXiv:2502.02080

\bibitem[{{Taylor} \& {Kobayashi}(2017)}]{Taylor_2017}
{Taylor}, P. \& {Kobayashi}, C. 2017, \mnras, 471, 3856

\bibitem[{{Taylor} {et~al.}(2005){Taylor}, {Jansen}, {Windhorst}, {Odewahn}, \&
  {Hibbard}}]{Taylor_2005}
{Taylor}, V.~A., {Jansen}, R.~A., {Windhorst}, R.~A., {Odewahn}, S.~C., \&
  {Hibbard}, J.~E. 2005, \apj, 630, 784

\bibitem[{{Thomas} {et~al.}(2018){Thomas}, {Dopita}, {Kewley}, {Groves},
  {Sutherland}, {Hopkins}, \& {Blanc}}]{Thomas_2018}
{Thomas}, A.~D., {Dopita}, M.~A., {Kewley}, L.~J., {et~al.} 2018, \apj, 856, 89

\bibitem[{{Thuan} {et~al.}(1995){Thuan}, {Izotov}, \&
  {Lipovetsky}}]{Thuan_1995}
{Thuan}, T.~X., {Izotov}, Y.~I., \& {Lipovetsky}, V.~A. 1995, \apj, 445, 108

\bibitem[{{Tinsley}(1980)}]{Tinsley_1980}
{Tinsley}, B.~M. 1980, \fcp, 5, 287

\bibitem[{{Tissera} {et~al.}(2025){Tissera}, {Bignone}, {Gonzalez-Jara},
  {Mu{\~n}oz-Escobar}, {Cataldi}, {Miranda}, {Barrientos-Acevedo},
  {Tapia-Contreras}, {Pedrosa}, {Padilla}, {Dominguez-Tenreiro},
  {Casanueva-Villarreal}, {Sillero}, {Silva-Mella}, {Shailesh}, \&
  {Jara-Ferreira}}]{Tissera_2025}
{Tissera}, P.~B., {Bignone}, L., {Gonzalez-Jara}, J., {et~al.} 2025, \aap, 697,
  A134

\bibitem[{{Tissera} {et~al.}(2022){Tissera}, {Rosas-Guevara}, {Sillero},
  {Pedrosa}, {Theuns}, \& {Bignone}}]{Tissera_2022}
{Tissera}, P.~B., {Rosas-Guevara}, Y., {Sillero}, E., {et~al.} 2022, \mnras,
  511, 1667

\bibitem[{{van der Walt} {et~al.}(2011){van der Walt}, {Colbert}, \&
  {Varoquaux}}]{Walt_2011}
{van der Walt}, S., {Colbert}, S.~C., \& {Varoquaux}, G. 2011, Computing in
  Science and Engineering, 13, 22

\bibitem[{{Veilleux} {et~al.}(2005){Veilleux}, {Cecil}, \&
  {Bland-Hawthorn}}]{Veilleux_2005}
{Veilleux}, S., {Cecil}, G., \& {Bland-Hawthorn}, J. 2005, \araa, 43, 769

\bibitem[{{Veilleux} {et~al.}(2020){Veilleux}, {Maiolino}, {Bolatto}, \&
  {Aalto}}]{Veilleux_2020}
{Veilleux}, S., {Maiolino}, R., {Bolatto}, A.~D., \& {Aalto}, S. 2020, \aapr,
  28, 2

\bibitem[{{Vila-Costas} \& {Edmunds}(1992)}]{Vila-Costas_1992}
{Vila-Costas}, M.~B. \& {Edmunds}, M.~G. 1992, \mnras, 259, 121

\bibitem[{{Vilchez} {et~al.}(1988){Vilchez}, {Pagel}, {Diaz}, {Terlevich}, \&
  {Edmunds}}]{Vilchez_1988}
{Vilchez}, J.~M., {Pagel}, B.~E.~J., {Diaz}, A.~I., {Terlevich}, E., \&
  {Edmunds}, M.~G. 1988, \mnras, 235, 633

\bibitem[{{Villar-Martin} {et~al.}(2024){Villar-Martin}, {L{\'o}pez Cob{\'a}},
  {Cazzoli}, {P{\'e}rez Montero}, \& {Cabrera Lavers}}]{Villar-Martin_2024}
{Villar-Martin}, M., {L{\'o}pez Cob{\'a}}, C., {Cazzoli}, S., {P{\'e}rez
  Montero}, E., \& {Cabrera Lavers}, A. 2024, arXiv e-prints, arXiv:2407.02115

\bibitem[{{Vincenzo} {et~al.}(2016){Vincenzo}, {Belfiore}, {Maiolino},
  {Matteucci}, \& {Ventura}}]{Vincenzo_2016}
{Vincenzo}, F., {Belfiore}, F., {Maiolino}, R., {Matteucci}, F., \& {Ventura},
  P. 2016, \mnras, 458, 3466

\bibitem[{{Virtanen} {et~al.}(2020){Virtanen}, {Gommers}, {Oliphant},
  {Haberland}, {Reddy}, {Cournapeau}, {Burovski}, {Peterson}, {Weckesser},
  {Bright}, {van der Walt}, {Brett}, {Wilson}, {Millman}, {Mayorov}, {Nelson},
  {Jones}, {Kern}, {Larson}, {Carey}, {Polat}, {Feng}, {Moore}, {VanderPlas},
  {Laxalde}, {Perktold}, {Cimrman}, {Henriksen}, {Quintero}, {Harris},
  {Archibald}, {Ribeiro}, {Pedregosa}, {van Mulbregt}, \& {SciPy 1. 0
  Contributors}}]{Virtanen_2020}
{Virtanen}, P., {Gommers}, R., {Oliphant}, T.~E., {et~al.} 2020, Nature
  Methods, 17, 261

\bibitem[{{Vogt} {et~al.}(2017){Vogt}, {P{\'e}rez}, {Dopita},
  {Verdes-Montenegro}, \& {Borthakur}}]{Vogt_2017}
{Vogt}, F.~P.~A., {P{\'e}rez}, E., {Dopita}, M.~A., {Verdes-Montenegro}, L., \&
  {Borthakur}, S. 2017, \aap, 601, A61

\bibitem[{{Yuan} {et~al.}(2022){Yuan}, {Wang}, \& {Yang}}]{Yuan_2022}
{Yuan}, F., {Wang}, H., \& {Yang}, H. 2022, \apj, 924, 124

\bibitem[{{Zaritsky} {et~al.}(1994){Zaritsky}, {Kennicutt}, \&
  {Huchra}}]{Zaristky_1994}
{Zaritsky}, D., {Kennicutt}, Robert~C., J., \& {Huchra}, J.~P. 1994, \apj, 420,
  87

\bibitem[{{Zinchenko} {et~al.}(2019){Zinchenko}, {Just}, {Pilyugin}, \&
  {Lara-Lopez}}]{Zinchenko_2019}
{Zinchenko}, I.~A., {Just}, A., {Pilyugin}, L.~S., \& {Lara-Lopez}, M.~A. 2019,
  \aap, 623, A7

\bibitem[{{Zinchenko} {et~al.}(2016){Zinchenko}, {Pilyugin}, {Grebel},
  {S{\'a}nchez}, \& {V{\'\i}lchez}}]{Zinchenko_2016}
{Zinchenko}, I.~A., {Pilyugin}, L.~S., {Grebel}, E.~K., {S{\'a}nchez}, S.~F.,
  \& {V{\'\i}lchez}, J.~M. 2016, \mnras, 462, 2715

\bibitem[{{Zinchenko} {et~al.}(2021){Zinchenko}, {V{\'\i}lchez},
  {P{\'e}rez-Montero}, {Sukhorukov}, {Sobolenko}, \& {Duarte
  Puertas}}]{Zinchenko_2021}
{Zinchenko}, I.~A., {V{\'\i}lchez}, J.~M., {P{\'e}rez-Montero}, E., {et~al.}
  2021, \aap, 655, A58

\bibitem[{{Zurita} {et~al.}(2021{\natexlab{a}}){Zurita}, {Florido}, {Bresolin},
  {P{\'e}rez}, \& {P{\'e}rez-Montero}}]{Zurita_2021b}
{Zurita}, A., {Florido}, E., {Bresolin}, F., {P{\'e}rez}, I., \&
  {P{\'e}rez-Montero}, E. 2021{\natexlab{a}}, \mnras, 500, 2380

\bibitem[{{Zurita} {et~al.}(2021{\natexlab{b}}){Zurita}, {Florido}, {Bresolin},
  {P{\'e}rez-Montero}, \& {P{\'e}rez}}]{Zurita_2021}
{Zurita}, A., {Florido}, E., {Bresolin}, F., {P{\'e}rez-Montero}, E., \&
  {P{\'e}rez}, I. 2021{\natexlab{b}}, \mnras, 500, 2359

\end{thebibliography}

\appendix
\section{Abundance radial gradients}
\label{A1}
We present in this appendix the abundance radial gradients, as traced by 12+log(O/H) and log(N/O), for our sample of LINER-like galaxies. We only display those regions classified as HII regions according to the diagnostic diagrams \citep{Baldwin_1981, Kauffmann_2003, Kewley_2006}, and we normalized their distances to the galaxy center by \ensuremath{R_{e}}. For the nuclear estimations, we show the estimations discussed in \citetalias{Perez-Diaz_2025}. 

\begin{figure*}
\centering
   \includegraphics[width=16cm]{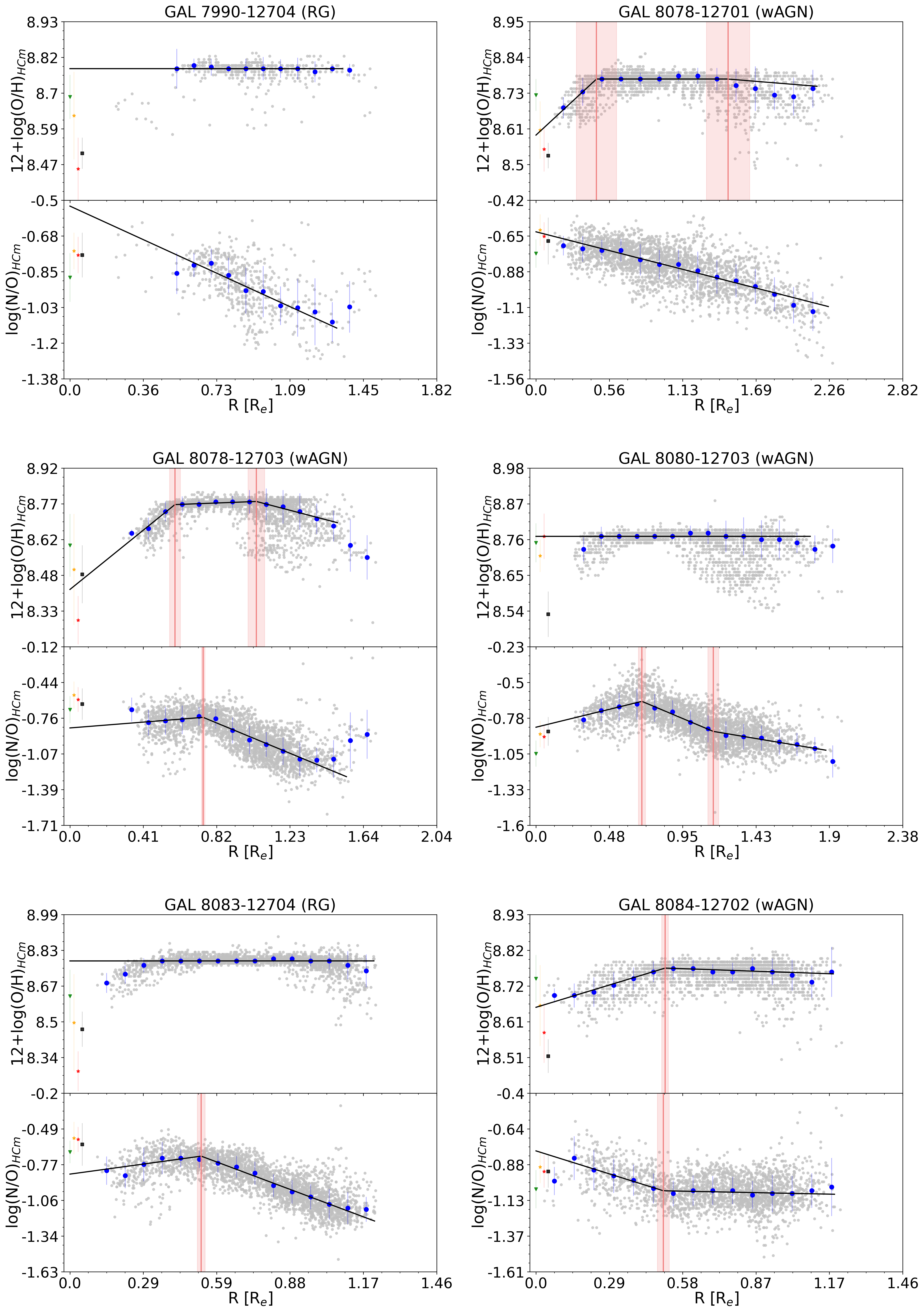}
   \caption{Metallicity gradients, 12+log(O/H) and log(N/O), in our sample of LINER-like galaxies. Nuclear estimations of the corresponding chemical abundance ratio are represented as follows: green triangles are the estimations from AGN models with \ensuremath{\alpha_{OX} = -1.6}; orange and red stars are the estimations from pAGB models with T\ensuremath{_{eff} = 10^{5}} K and T\ensuremath{_{eff} = 1.5\cdot10^{5}} K respectively; and black squares are the estimations from ADAF models. Red vertical lines mark the break point, and red shaded areas their corresponding uncertainty.}
	\label{Gradients_1}
\end{figure*}

The results are showed in Fig. \ref{Gradients_1}, where we show all HII regions considered (gray dots) and the median value (blue) at a given distance considering 15 different intervals in distance and ensuring that each segment contains at least 10 HII regions. We also represent the fit (solid black line) obtained from the piecewise algorithm. We only show in the manuscript Fig. \ref{Gradients_1} as an example. The rest of the figures are available on Zenodo.

\section{log(N/O) vs 12+log(O/H)}
\label{A2}
\begin{figure*}
\centering
   \includegraphics[width=16cm]{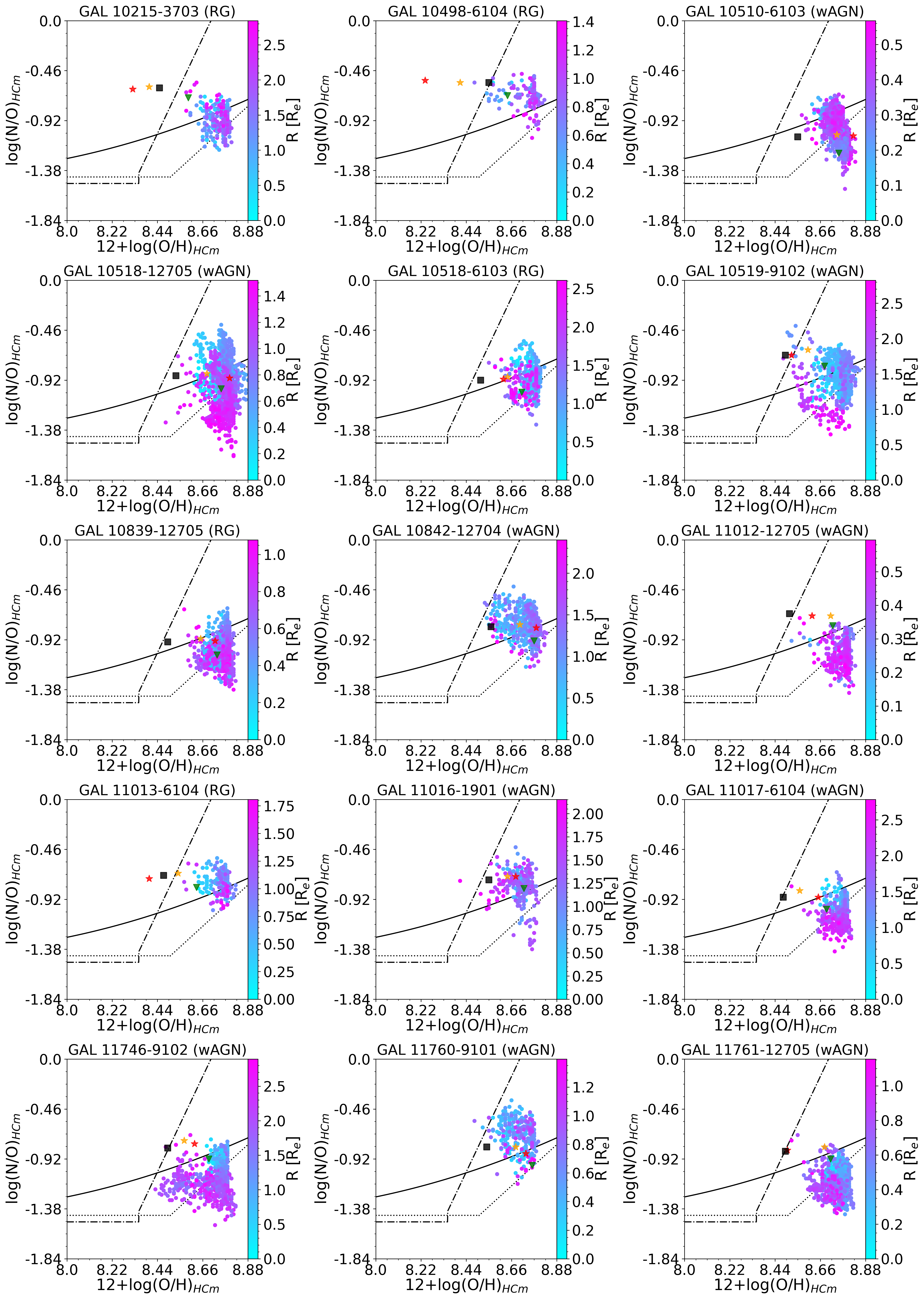}
   \caption{log(N/O) vs 12+log(O/H) diagram for the HII regions in our sample of LINERs. The colorbar shows the distance to the galactic center in terms of R\ensuremath{_{50}}. Nuclear estimations of the corresponding chemical abundance ratio are represented as follows: green triangles are the estimations from AGN models with \ensuremath{\alpha_{OX} = -1.6}; orange and red stars are the estimations from pAGB models with T\ensuremath{_{eff} = 10^{5}} K and T\ensuremath{_{eff} = 1.5\cdot10^{5}} K respectively; and black squares are the estimations from ADAF models. The solid back line represents the fit provided by \citet{Coziol_1999}, the dotted line shows the fit by \citet{Andrews_2013}, and the dash-dotted line shows the fit by \citep{Belfiore_2015}.}
	\label{NOOH_1}
\end{figure*}

We present in this appendix the log(N/O) vs 12+log(O/H) diagram for the HII regions selected in each galaxy in our sample. We also present the estimations of the chemical abundances of the nuclear region that comprehends the LINER-like emission. We only show in the manuscript Fig. \ref{NOOH_1} as an example. The rest of the figures are available on Zenodo.
\end{document}